\documentclass[structabstract]{aa}
\usepackage{txfonts}
\usepackage{graphicx}
\usepackage{longtable} 
\usepackage{subfig}  
\usepackage{float}

\begin{document}

\title{Exploring the bulk of the BL Lac object population:\\
1. parsec-scale radio structures.
  }

\subtitle{}

\author{E. Liuzzo\inst{1}, M. Giroletti\inst{1} , G. Giovannini\inst{1,2}, B. Boccardi\inst{3}, S. Tamburri \inst{4,5}, G.B. Taylor\inst{6, 7}, C. Casadio\inst{8}, M. Kadler\inst{9}, G. Tosti\inst{10}, A. Mignano\inst{1}.}

\offprints{E.Liuzzo, \email{liuzzo@ira.inaf.it}}

\institute{INAF - Istituto di Radioastronomia, via Gobetti 101, 40129 Bologna, Italy.
  \and Dipartimento di Fisica e Astronomia, Universit\`a di Bologna, via Ranzani 1 , 40127 Bologna, Italy.
  \and Max Planck Institute for Radioastronomy, Auf dem H\"ugel 69  53121 Bonn, Germany.
  \and INAF-Osservatorio Astronomico di Brera, via Brera 28 20121 Milano, Italy.
  \and Universit\`a degli Studi dell'Insubria, via Valleggio 11 22100 Como, Italy.  
  \and Department of Physics and Astronomy, University of New Mexico, Albuquerque NM 87131, USA.
  \and also Adjunct Astronomer at the National Radio Astronomy Observatory, USA.
  \and Instituto de Astrofisica de Andalucia, CSIC, Apartado 3004, 18080, Spain.
  \and Universit\"at W\"urzbur\"g, Lehrstuhl f\"ur Astronomie, 
Emil-Fischer-Str. 31, 97074 W\"urzburg, Germany.
  \and Dipartimento di Fisica, Universit\`a degli Studi di Perugia, 06123 Perugia, Italy.}

\date{Received ........ / Accepted .........}

\abstract  
 {The advent of \textit{Fermi} is changing our understanding on the radio and $\gamma$-ray emission in Active Galactic Nuclei. Contrary to pre-\textit{Fermi} ideas, BL Lac objects are found to be the most abundant emitters in the $\gamma$-ray band. However, since they are relatively weak radio sources, most of their parsec-scale structure and their multi-frequency properties are poorly understood and/or have not been investigated in a systematically fashion.}
   {Our main goal is to analyze the radio and $\gamma$-ray emission properties of a sample of 42 BL Lacs selected, for the first time in the literature, with no constraint on their radio and $\gamma$-ray flux densities/emission.}
   {Thanks to new Very Long Baseline Array observations at 8 and 15 GHz for the whole sample, we present here fundamental parameters such as radio flux densities, spectral index information, and parsec-scale structure. Moreover, we search for $\gamma$-ray counterparts using data reported in the Second Catalog of Fermi Gamma-ray sources.}
   {Parsec-scale radio emission is observed in the majority of the sources at both frequencies. Gamma-ray counterparts are found for 14/42 sources.}
{ The comparison between our results in radio and gamma-ray bands points out the presence of a large number of faint BL Lacs showing ``non classical" properties such as low source compactness, core dominance, no gamma-ray emission and steep radio spectral indexes. A deeper multiwavelength analysis will be needed.}


\maketitle

\section{Introduction} \label{intro}

Based on results from the Energetic Gamma-Ray Experiment Telescope
(EGRET) on the Compton Gamma-Ray Observatory (CGRO), it is well known
that the main contribution to the $\gamma$-ray sky comes from the Galactic
plane emission, pulsars, and blazars (Hartman et al. 1999). Among the
latter class of objects, Flat Spectrum Radio Quasars (FSRQs) were the
most numerous compared to the BL Lac objects, making up 77$\%$ of the
high confidence blazar associations.

With the advent of the {\it Fermi} mission, studies of a large number
of $\gamma$-ray sources have become possible thanks to the
unprecedented sensitivity of the Large Area Telescope (LAT, Atwood et
al. 2009). Moreover, the improved Point Spread Function of the LAT
significantly improved the localization of the $\gamma$-ray sources
with respect to the past $\gamma$-ray missions. In contrast with the
previous EGRET results, the LAT has shown that the BL Lacs, and not
the FSRQs, are now the most common $\gamma$-ray emitters (1LAC, Abdo
et al. 2010; 2LAC, Ackermann et al. 2011a, 2011b): in the
Clean 2 LAC Sample, 395 are BL Lac objects, 310 FSRQs, 157 sources
of unknown type, 22 other AGNs, and 2 starburst galaxies (Ackermann et
al. 2011a, 2011b). Moreover, the two
blazar sub-classes are markedly distinct in the average photon
indexes, redshift and flux density distributions.  Big questions are
nevertheless remaining open, such as the $\gamma$-ray origin and location,
the distance of the main energy dissipation site from the nucleus, and the
relation between the $\gamma$-ray and radio emission.

BL Lac objects belong to the blazar population showing a high core
dominance, large degrees of variability and polarization, and
one-sided Doppler beamed parsec-scale jets. However, they present
properties quite different from those of FRSQs: lower apparent
velocities (Gabuzda et al. 1994, Jorstad et al. 2001, Kellermann et
al. 2004, Lister et al. 2009a, 2009b), also in the most extreme BL Lacs, 
the TeV BL Lacs (e.g. MrK 501, Giroletti et al. 2004a, 2004b; MrK 412,
Lico et al. 2012; Piner et al. 2004, 2008, 2010); differences
in polarization degree (Gabuzda et al. 1999) and rotation measure (Hovatta et al. 2012) 
values in their cores at mas resolution.

Thanks to {\it Fermi} findings, the high energy characteristics of BL
Lacs can be investigated, but to understand the origin and the nuclear
properties of these sources more information at different frequencies
are necessary. Very Long Baseline Interferometer (VLBI) campaigns are
one of the most incisive observational tools to address the
questions opened by the \textit{Fermi} results. From a detailed
literature review (Ros et al. 2011), it is evident that previous VLBI
surveys have looked at the parcsec scale properties of the brightest
BL Lac objects, while the majority of the BL Lac population is below
their flux density limits (Wu et al. 2007; Giroletti et al. 2004b,
2006; Rector et al. 2003; Cassaro et al. 2002). In fact, many LAT BL
Lacs are high-frequency--synchrotron-peaked (HSP) sources, discovered at X-rays
and generally they are faint radio sources, seldom studied with
VLBI.  To gain a deeper understanding of BL Lac properties, we selected a
sample of low redshift BL Lacs sources with no selection limits on
their radio flux density or high frequency emission.

In this work, we present our analysis on the radio and $\gamma$-ray
emission of this sample of nearby BL Lacs.  In following papers (Rain\`o et
al. in preparation, Giroletti et al. in preparation), we will discuss
these results at multiple-wavebands.

The paper is organized as following: in Sect. 2, we introduce our new
sample of low redshift BL Lacs; in Sect. 3, we describe the new high
resolution radio observations performed with VLBI; in
Sect. 4, we show the results on the radio properties of the sample; in
Sect. 5, we compare the radio and $\gamma$-ray emission for these
objects.

Throughout this paper, we assume h = 0.71, $\Omega_{m}$ = 0.27 and  $\Omega_{\lambda}$= 0.73, where the Hubble constant H$_{0}$ = 100h km s$^{-1}$ Mpc$^{-1}$. We define radio spectral index $\alpha$ such that the flux density S($\nu$) $\sim$ $\nu^{-\alpha}$ and the gamma photon index $\Gamma$ is defined as N(E) $\propto$ E$^{-\Gamma}$.

\section{The sample} \label{sample}

Recently we began a project aimed to improve our knowledge of the BL
Lac nuclear region, through VLBI observations of a sample of BL Lacs,
independent of their radio flux densities and their
$\gamma$-ray emission, and with multi-frequency information available
for all targets.  We selected our sample of BL Lacs from the Roma
BZCat Catalog of the known blazars (Massaro et al. 2009)\footnote{The
  BZ Cat is a catalogue of blazars based on multifrequency surveys and
  on an extensive review of the literature. The adopted acceptance
  criteria are: 1. detection in the radio band, down to mJy flux
  densities at 1.4 GHz (NVSS, Condon et al. 1998 or FIRST, White et
  al. 1997) or 0.84 GHz (Sydney University Molonglo Sky Survey, SUMSS,
  Mauch et al. 2003); 2.optical identification and knowledge of the
  optical spectrum to establish the type of blazar; 3.isotropic X-ray
  luminosity close to or higher than 10$^{43}$ erg s$^{-1}$; 4. for
  FSRQs a spectral index, measured
  between 1.4 GHz (or 0.843 GHz) and 5 GHz, $\alpha < 0.5$,
  but not for BL Lacs even though most of them have flat spectra;
  5. compact radio morphology, or, when extended, with one dominant
  core and a one-sided jet.  For more details on the Roma BZ Cat see
  Massaro et al. 2009.}  with two constraints:
\begin{itemize}
\item a measured redshift z $<$ 0.2, 
\item BL Lacs located within the sky area covered by the Sloan Digital Sky Survey (SDSS, Abazajian et al. 2009). 
\end{itemize}

We extracted our sample from the BZ Cat as this catalog is the most
complete list of published blazars, and so well suitable for the aims
of our project, which requires a sample unbiased with respect to the
the radio flux density and gamma-ray emission\footnote{The
  completeness of BZ Cat for BL Lacs with z$<$0.2 is however still an
  open question, as the completeness of the whole BZ Cat (Massaro et
  al. 2009). There is in fact a deficiency of blazars in the southern
  sky due to the smaller number of surveys in comparison to the
  Northern hemisphere, even if many new discovered blazars and
  candidates have been identified using spectroscopic observations
  available in the SDSS.}.  These two selection criteria on the BZ Cat
allow us 1) to investigate also the least powerful sources, such as
the weak population of HSP BL Lacs, with a good linear resolution (1
pc $\sim$ 0.5 mas at z = 0.1); 2) a good characterization not only of
the optical properties, but also of their extended radio
characteristics as the Faint Images of the Radio Sky at
Twenty-Centimeters (FIRST) cover the same SDSS field.

\begin{figure}[t!]
\centering \includegraphics[width=0.5\textwidth]{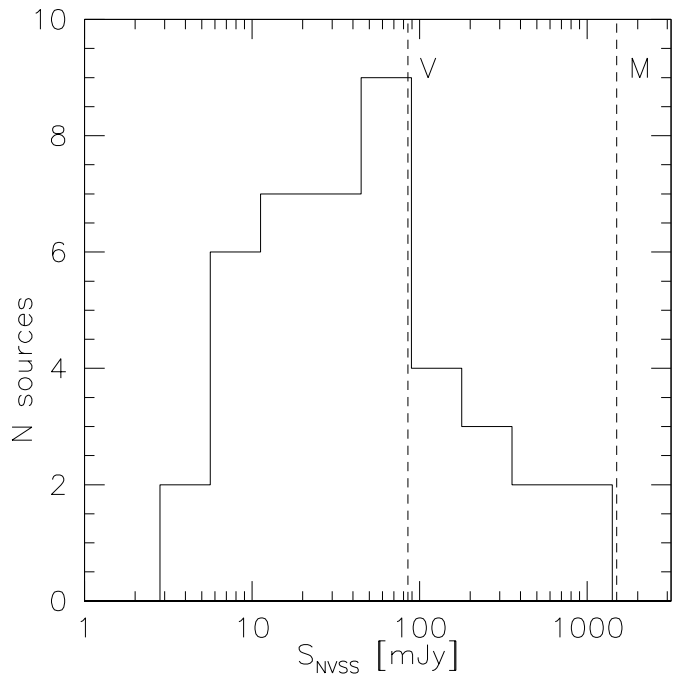}
\caption{Distribution of the NVSS flux density for the sources in our sample. The dashed lines show the flux density limits of the MOJAVE-1 (M) and VIPS surveys
(V), extrapolated assuming $\alpha$ = 0.0. }  \label{fig_flux}
\end{figure}

The total number of BL Lacs in our sample is 42. The National Radio
Astronomy Observatory (NRAO) VLA Sky Survey Catalog (NVSS) flux
density distribution for the sources in the sample is shown in
Fig. \ref{fig_flux}. The flux density limits of the Monitoring of jets in
active galactic nuclei (AGN) with Very Long Baseline Array (VLBA)
Experiments (MOJAVE-1) (correlated flux $S$ $>$ 1.5 Jy at 2 cm, Lister
et al. 2009a, 2009b) and VLBA Imaging and Polarimetry Survey (VIPS) ($S  >$ 85
mJy at 5 GHz, Helmboldt et al. 2007) surveys are also plotted
(assuming $\alpha$= 0.0). As Fig. \ref{fig_flux} points out, our sample is representative of a population
which is the majority of the whole BL Lacs but it is unexplored by two
of the most complete previous VLBI surveys.

\section{VLBA Observations and data reduction.} \label{observations}

We observed with the VLBA all sources of our sample (Table
\ref{tab_oss}, project codes BG197A, B and BG204A, B, C). Each target
was observed at 8 and 15 GHz with the aim of obtaining simultaneous
spectral information.  The observing time was about one hour per
source, with roughly a 1:3 ratio between the low and high frequency
total integration time. Targets weaker than 50 mJy in NVSS images have
been observed in phase reference mode at 15 GHz, while sources weaker
than 30 mJy have been observed in phase reference mode at both 8 and
15 GHz. Indeed, most sources had never been observed before with VLBI
and the phase referencing technique has also provided the possibility
to obtain absolute coordinates for them. We switched between the two
frequencies and moved from different targets within each session to
obtain good ($u, v$)-coverage, necessary to properly map complex faint
structures.  Observations were made with the entire VLBA in full
polarization with 4 intermediate frequencies (IFs). The VLBA
correlator produced 16 frequency channels per IF/polarization for a
total aggregate bit rate of 256 Mbs.  Calibrators were chosen from the
VLBA calibrators list to be bright and close to the target sources: in Table
\ref{tab_oss}, we report the list of the selected calibrators.  Short
scans on strong sources were interspersed with the targets and the
calibrators as fringe finders. The observations were correlated in
Socorro, New Mexico, USA.

Post correlation processing and imaging used the NRAO Astronomical
Image Processing System ({\it AIPS}) package (Cotton 1993) and the
Caltech {\it Difmap} packages (Pearson et al. 1994).  The signal to
noise ratio (SNR) achieved allowed for phase self-calibration on a few
sources. For detected sources observed in phase reference mode, we
give the absolute core position ($^{*}$ in Col.3 in Table
\ref{tab_oss}). Otherwise the nuclear coordinates are taken from Very
Large Array (VLA) data (see Table \ref{tab_oss}).  We show in Fig. \ref{fig_images}
contour clean maps for resolved sources in our VLBA data at least at
one frequency. Contour levels are 3$\times$rms
$\times$ (-1, 1, 2, 4, 8, 16, 32, 64, 128, 256, ..), where rms is
the noise level as measured from uniformly weighted images. We report the rms values for each target images
in Table \ref{tab_vlbaImage}, together with
the beam size and the flux density peak S$_{peak}$ .  For undetected
sources, not observed in phase reference mode, it was not possible
to properly calibrate the visibility phases and we are not able to
give image parameters (see Table \ref{tab_vlbaImage}). As a reference,
the upper flux density limits for non detections are the lowest flux
density revealed without phase referencing in the whole sample,
i.e. 15.3 mJy at 8 GHz (corresponding to J1221+0821) and 14.7 mJy at
15 GHz (corresponding to J1428+4240).  For undetected sources observed
in phase reference mode, upper limits are 3$\times$rms.

To derive an appropriate description of the the parsec-scale structure
of each source components, we applied {\it Difmap} model-fit to 8 and
15 GHz data of all detected sources using elliptical (e), circular (c)
or delta components ($\delta$) chosen on the base of the lower Reduced
Chi-squared values and the most reasonable parameters derived
(e.g. major and minor axis, position) for each component. The {\it
  Difmap} model-fit program fits aggregates of various forms of model
components, fitting directly to the real and imaginary parts of the
observed visibilities using the powerful Levenberg-Marquardt nonlinear
least squares minimization technique. From the model-fits, we derived the total VLBA flux densities, which are in agreement with those of the cleaned images. In Table
\ref{tab_parsecMorphology}, we report the parsec-scale morphology
(point-like  ``p" or core-jet  ``1s") and spectral indexes for detected BL Lacs, while in Table \ref{tab_modelfit} we summarize the model-fit
results for detected BL Lacs. All the calibrated uv and fits files are available in the following link http://www.ira.inaf.it/progetti/bllacs/

\begin{figure*}
\centering
\includegraphics[width=4cm, angle=0]{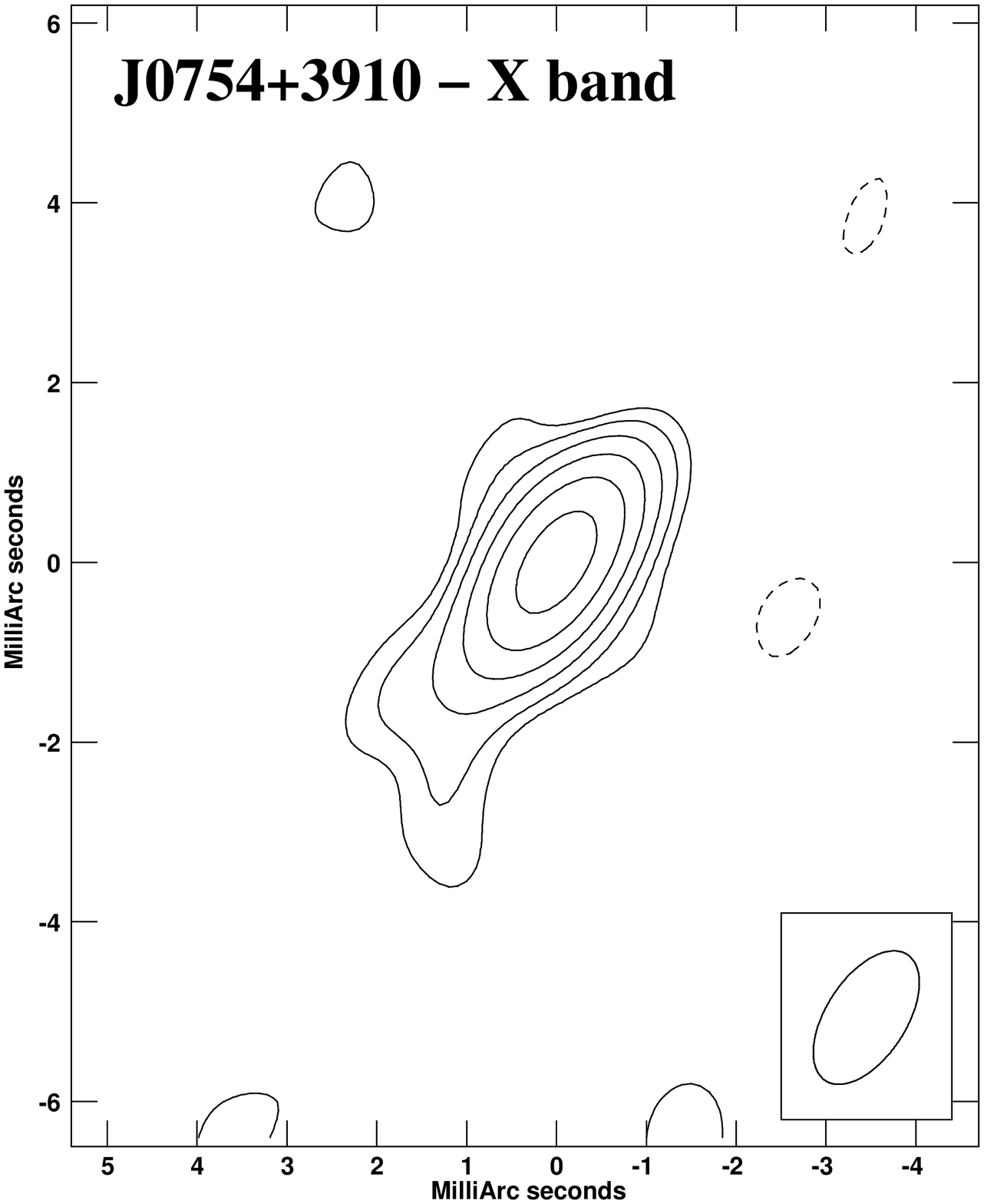}
\includegraphics[width=4cm, angle=0]{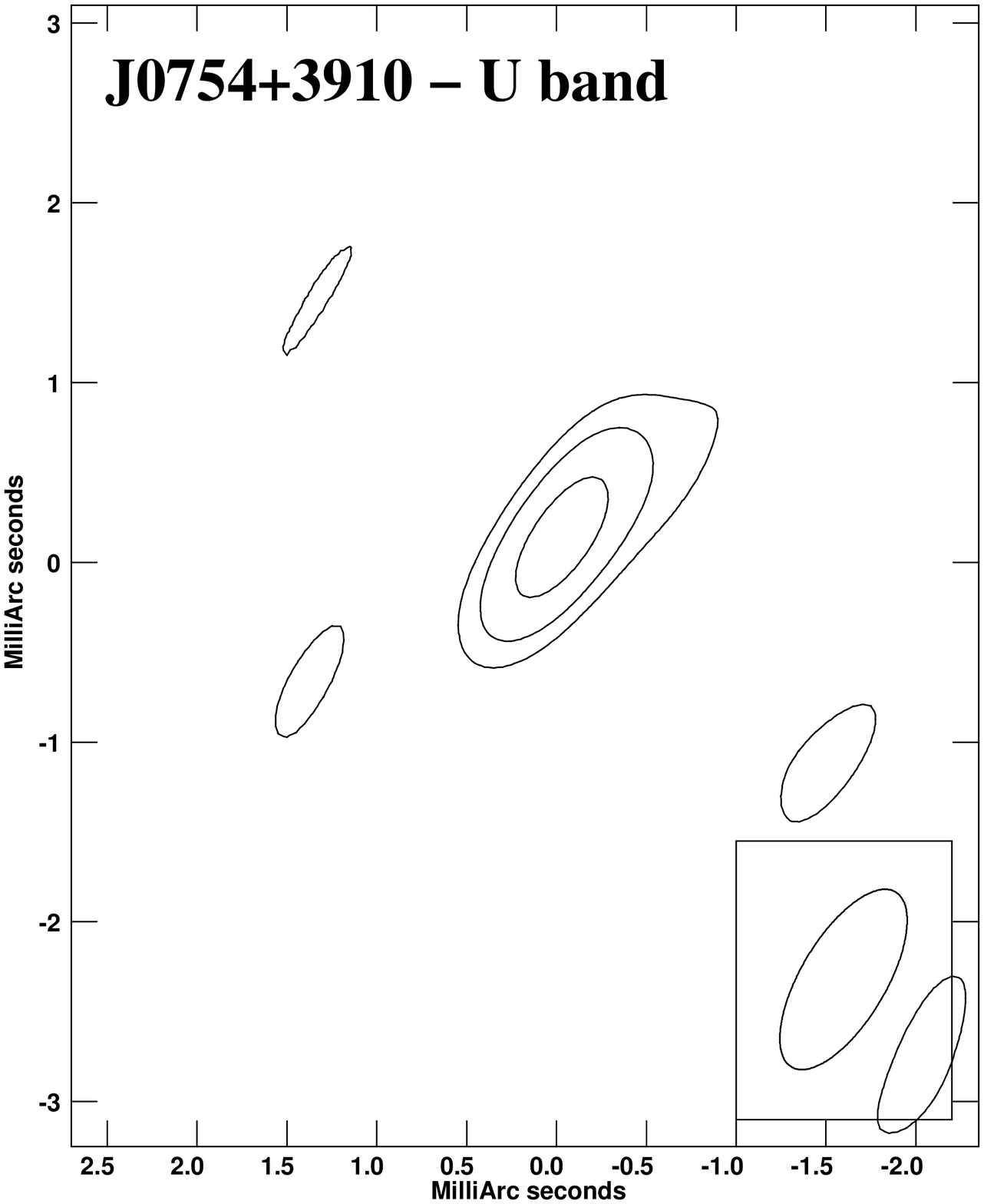}
\includegraphics[width=4.5cm, angle=0]{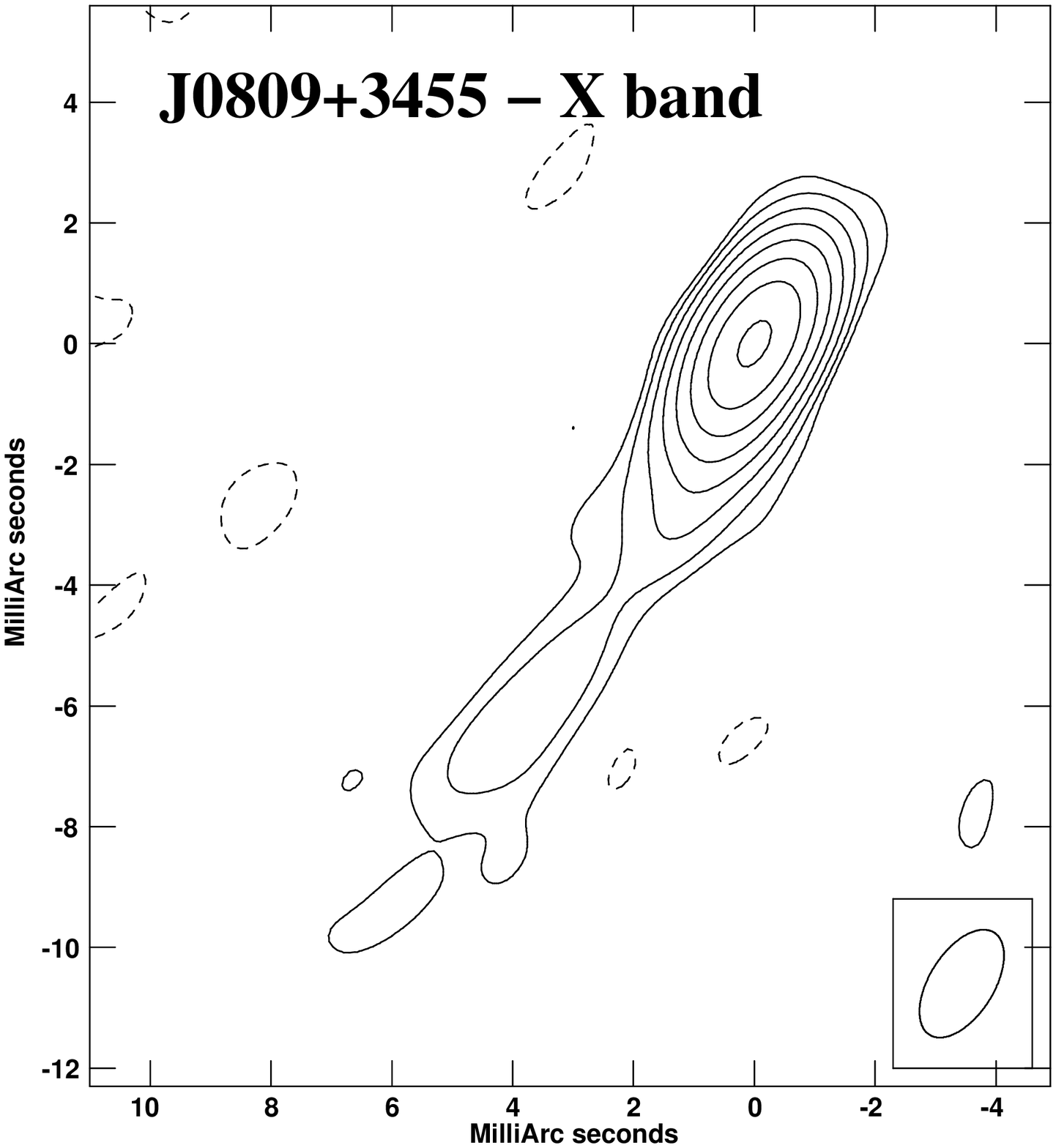}
\includegraphics[width=4.5cm, angle=0]{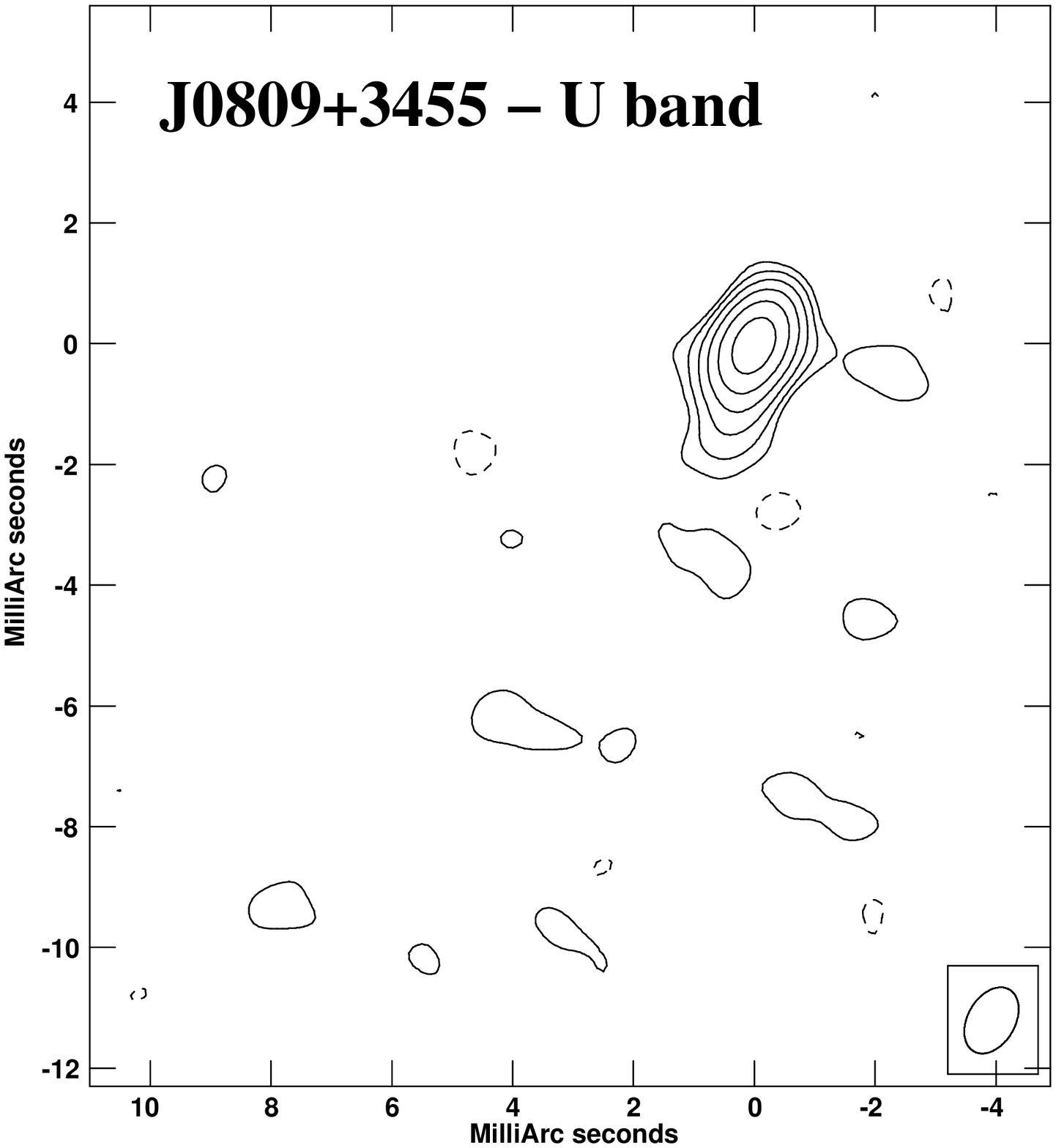}
\includegraphics[width=4cm, angle=0]{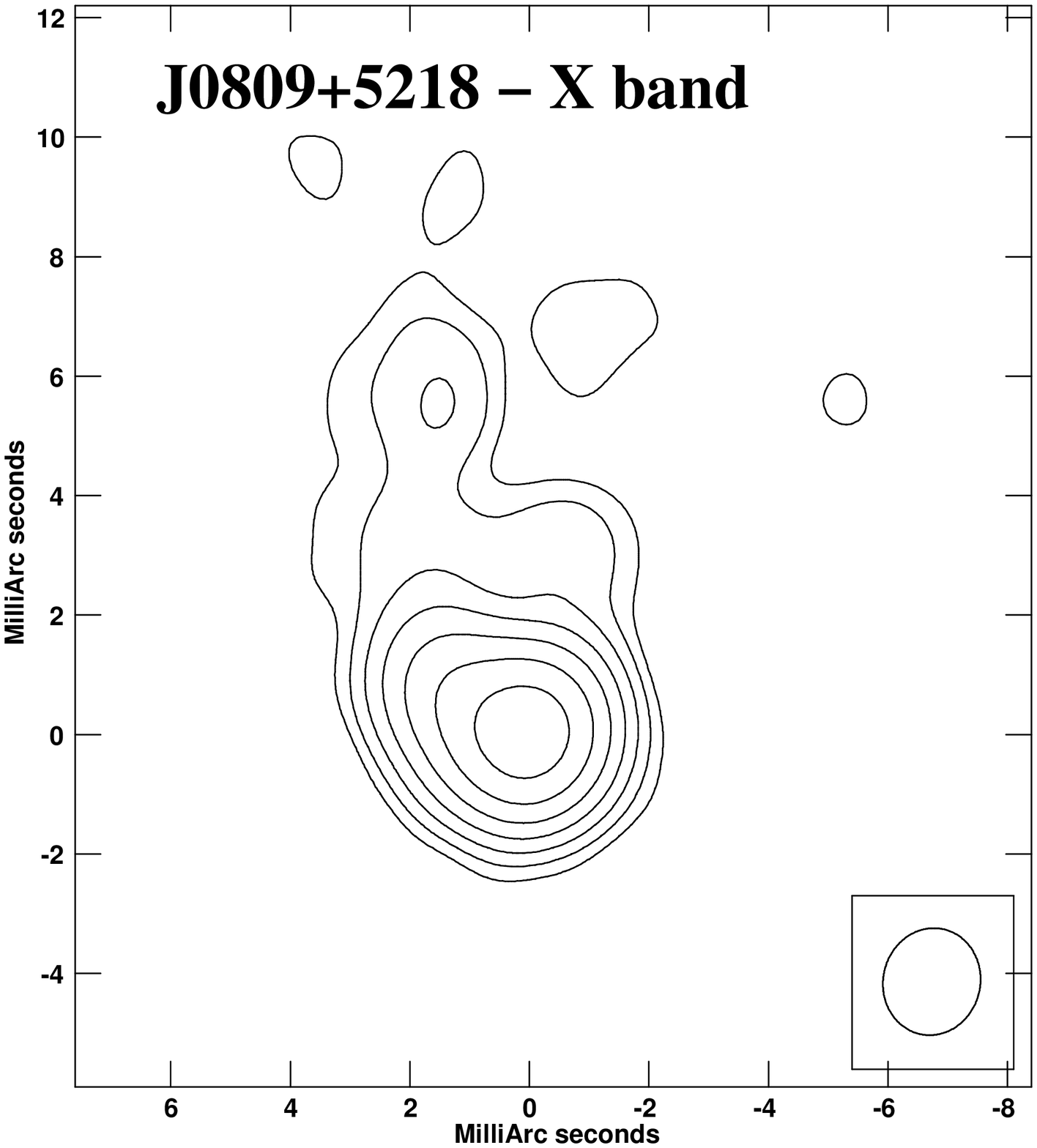}
\includegraphics[width=4cm, angle=0]{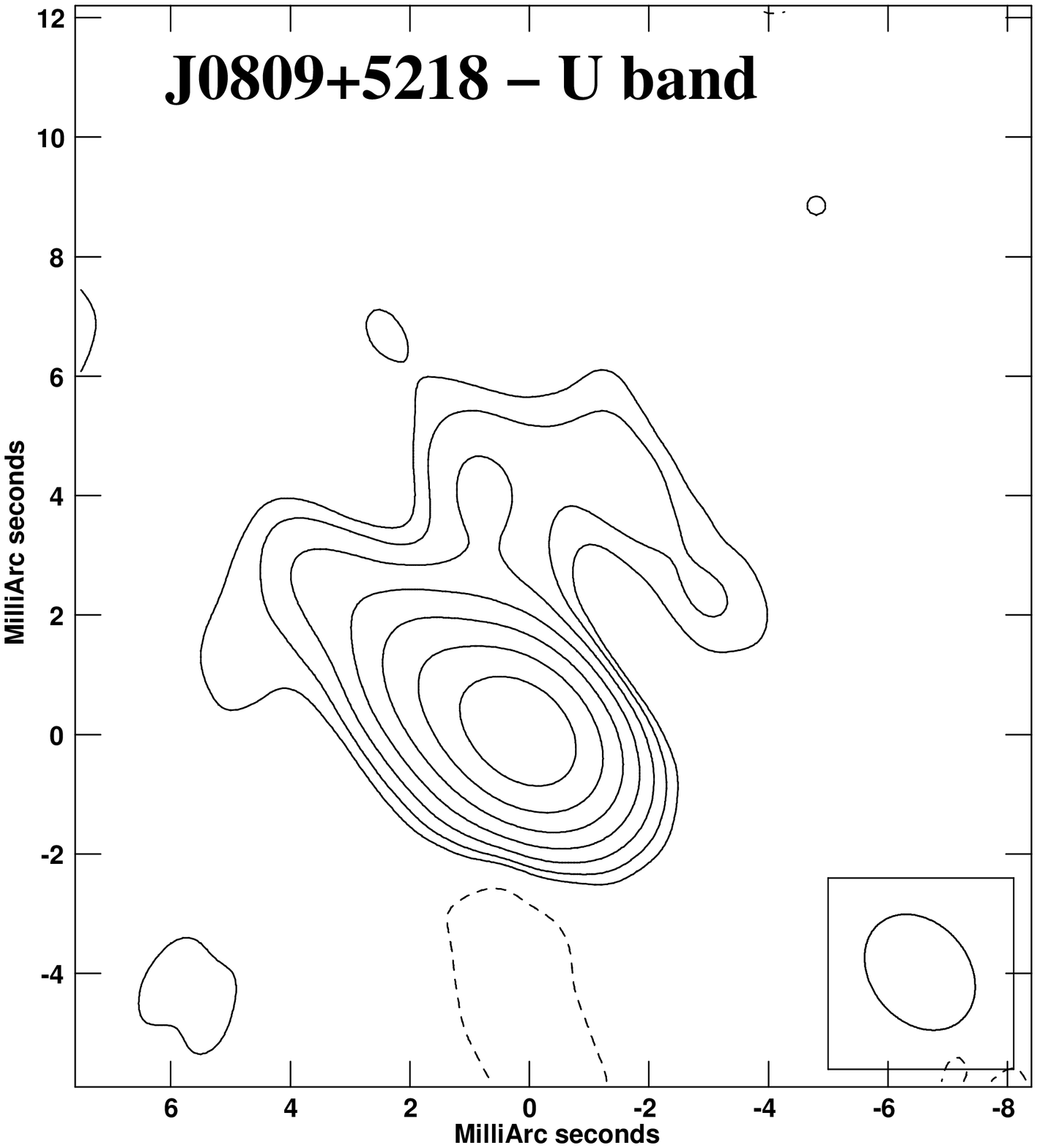}
\includegraphics[width=4.5cm, angle=0]{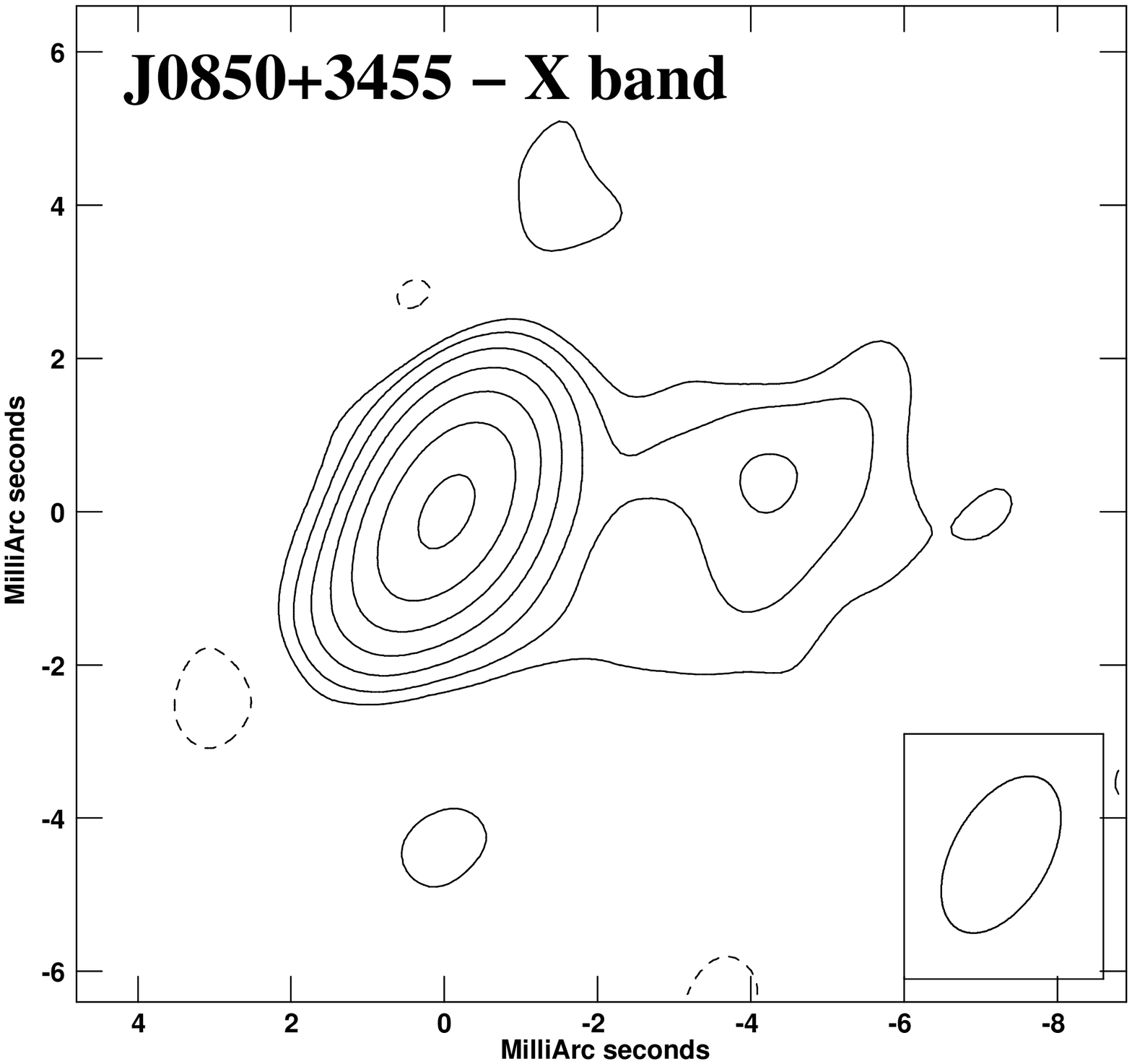}
\includegraphics[width=4.5cm, angle=0]{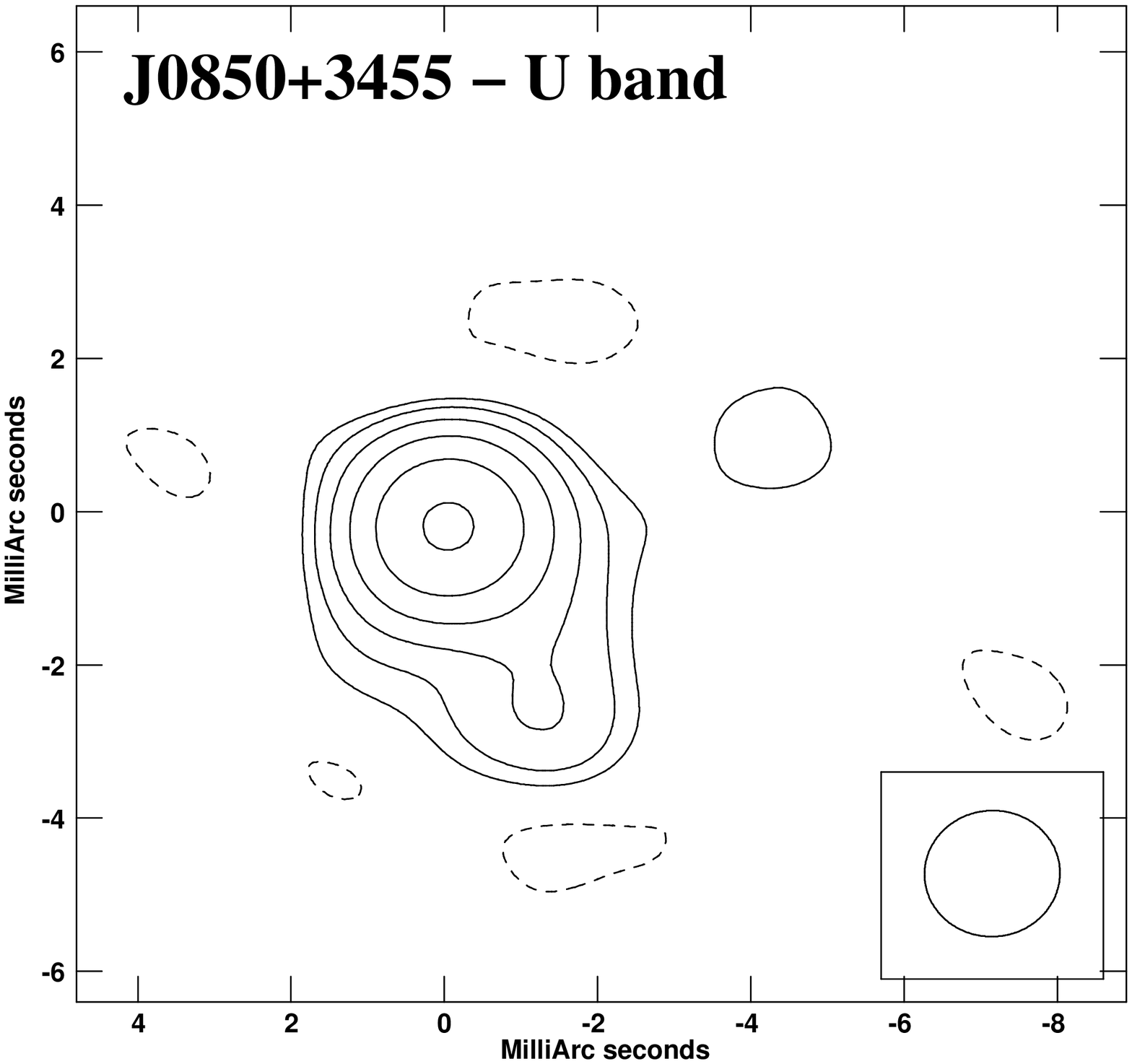}
\includegraphics[width=4cm, angle=0]{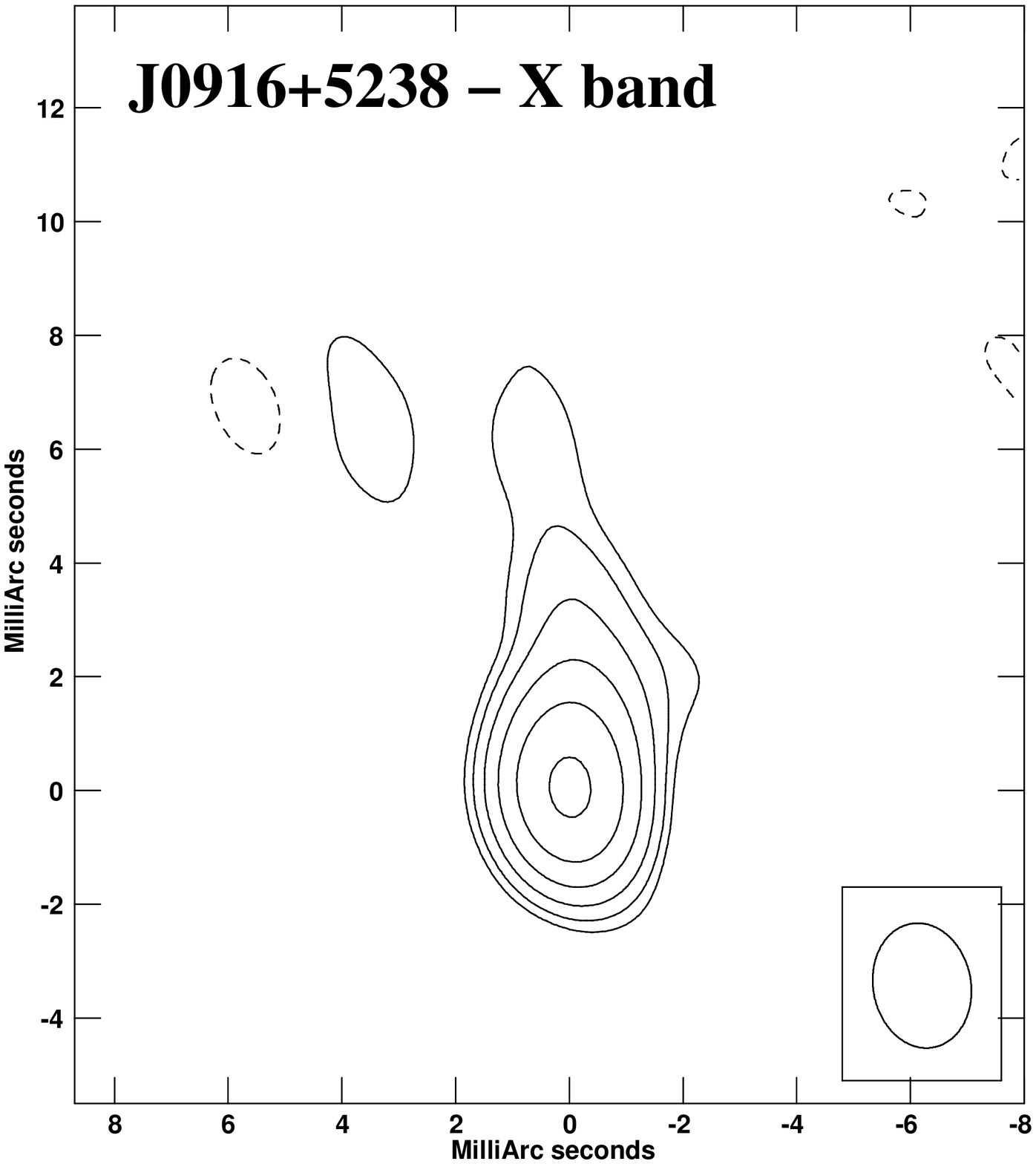}
\includegraphics[width=4cm, angle=0]{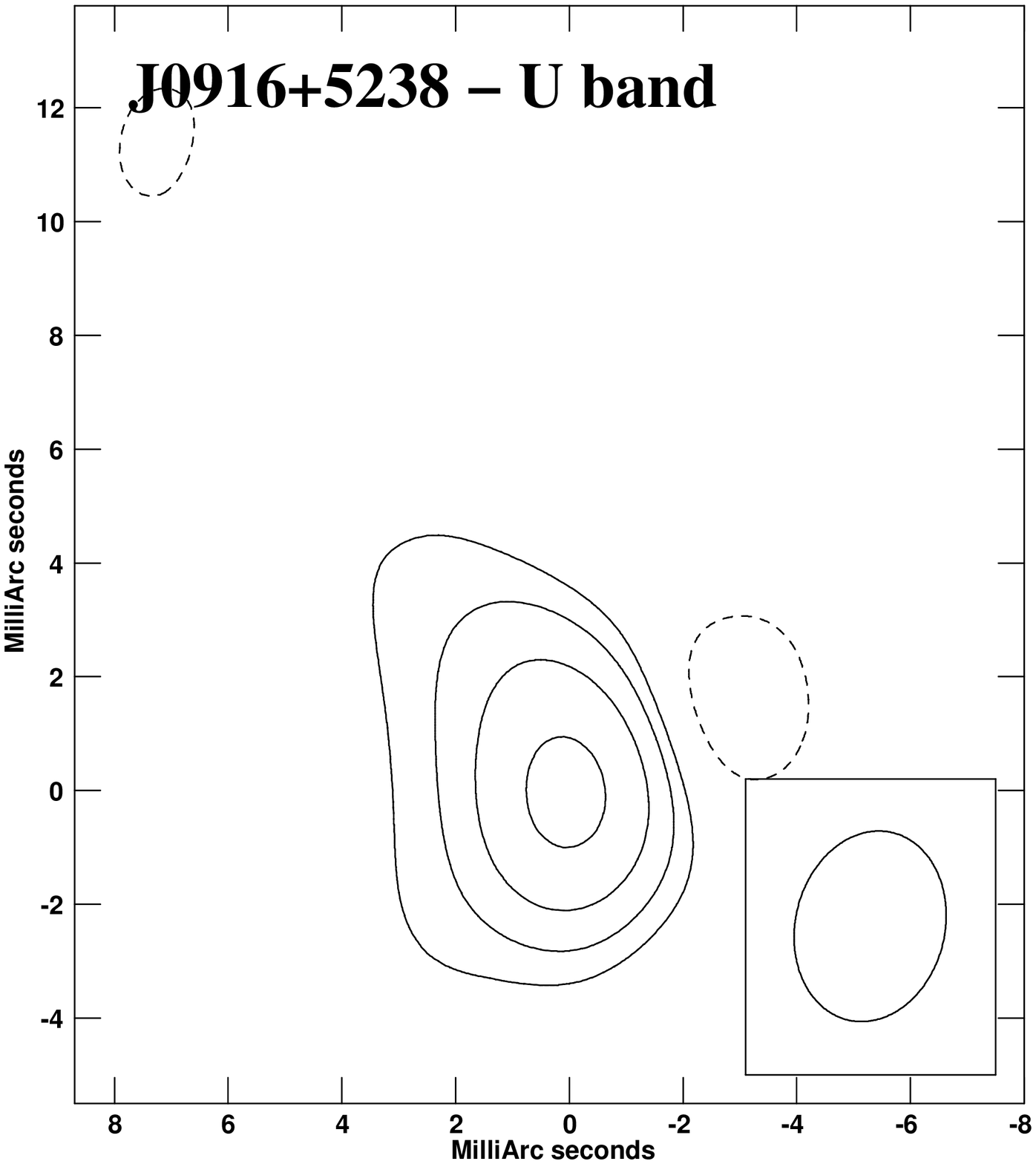}
\includegraphics[width=4.5cm, angle=0]{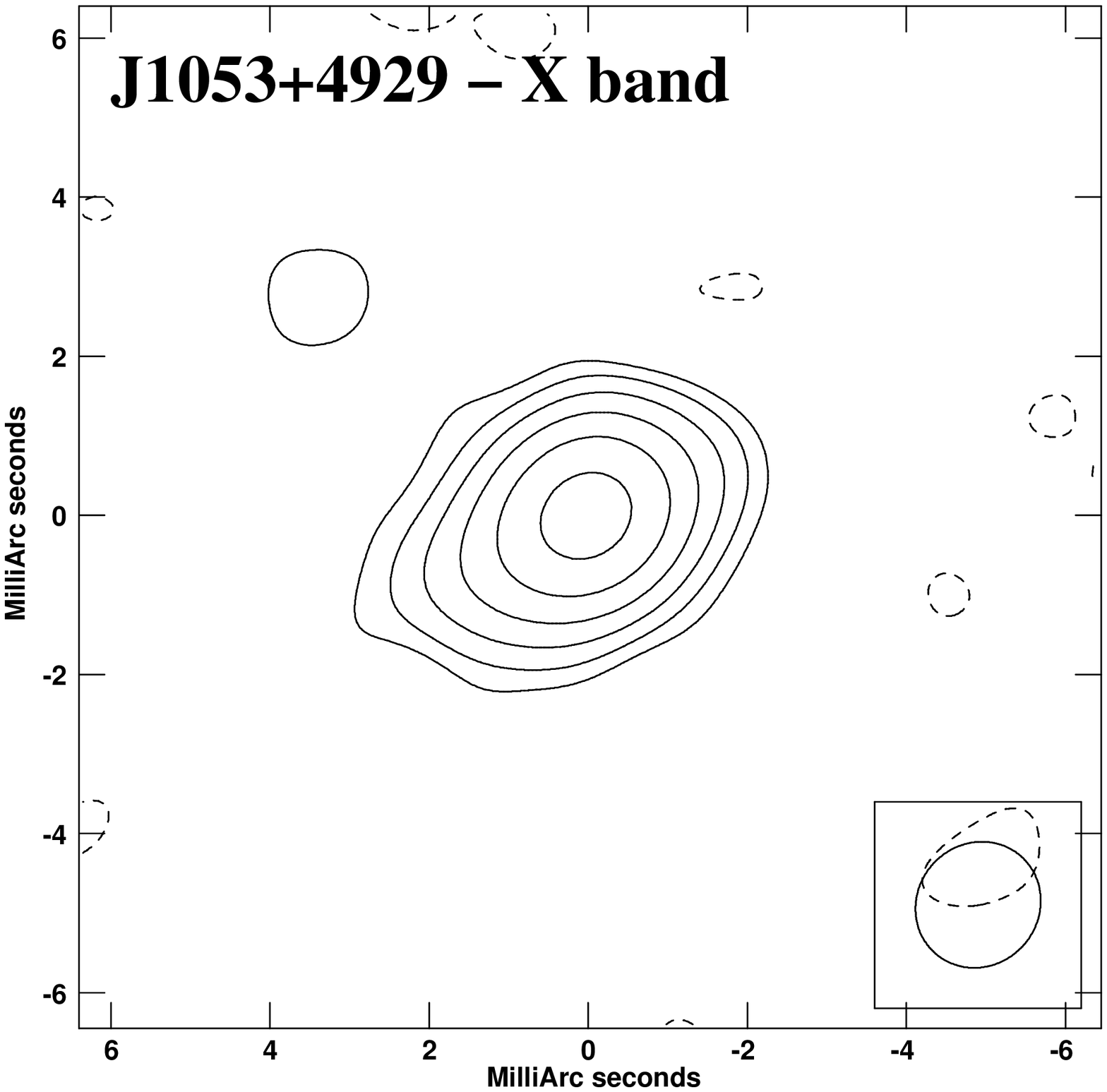}
\includegraphics[width=4.5cm, angle=0]{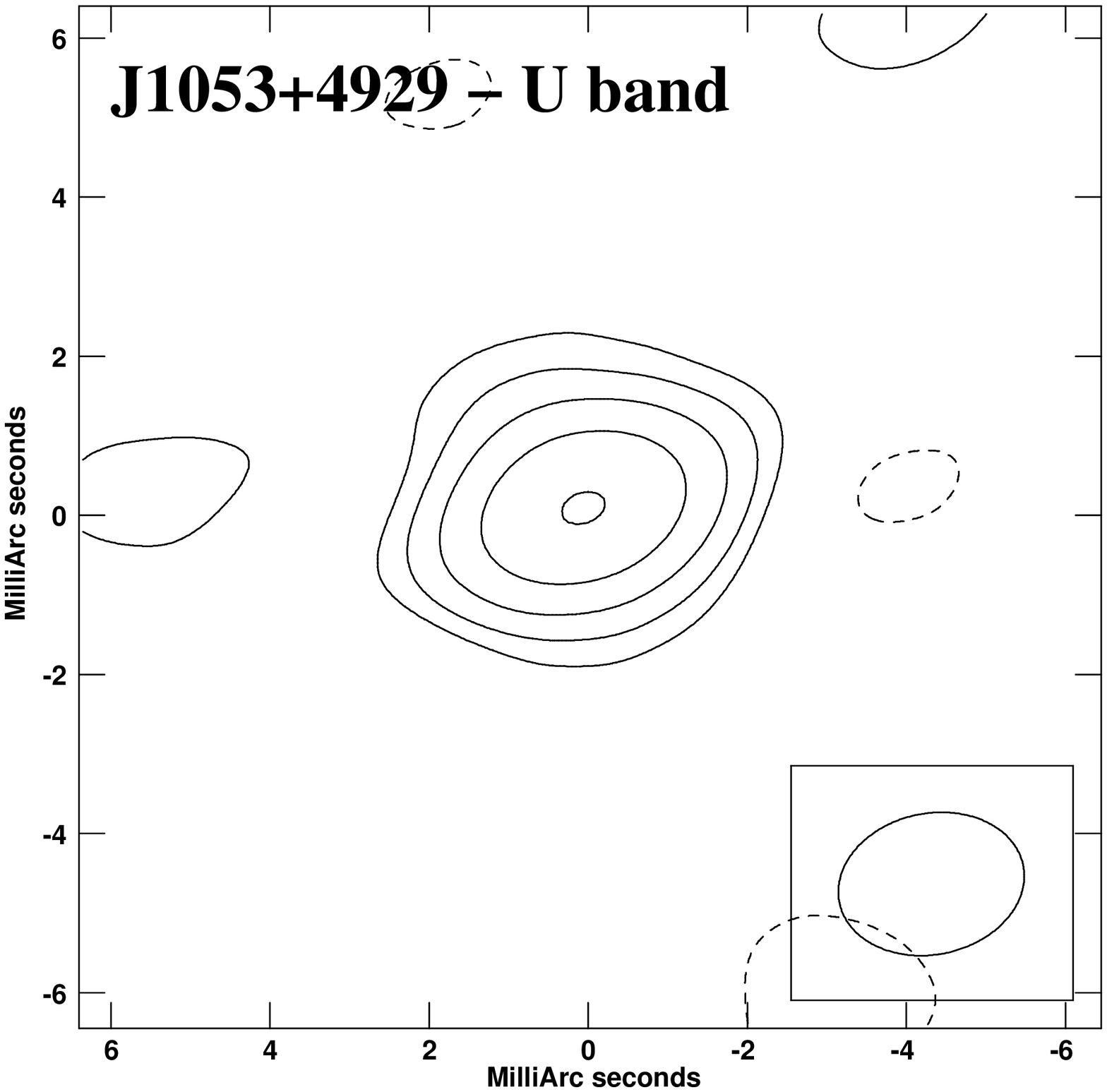}
\includegraphics[width=4.5cm, angle=0]{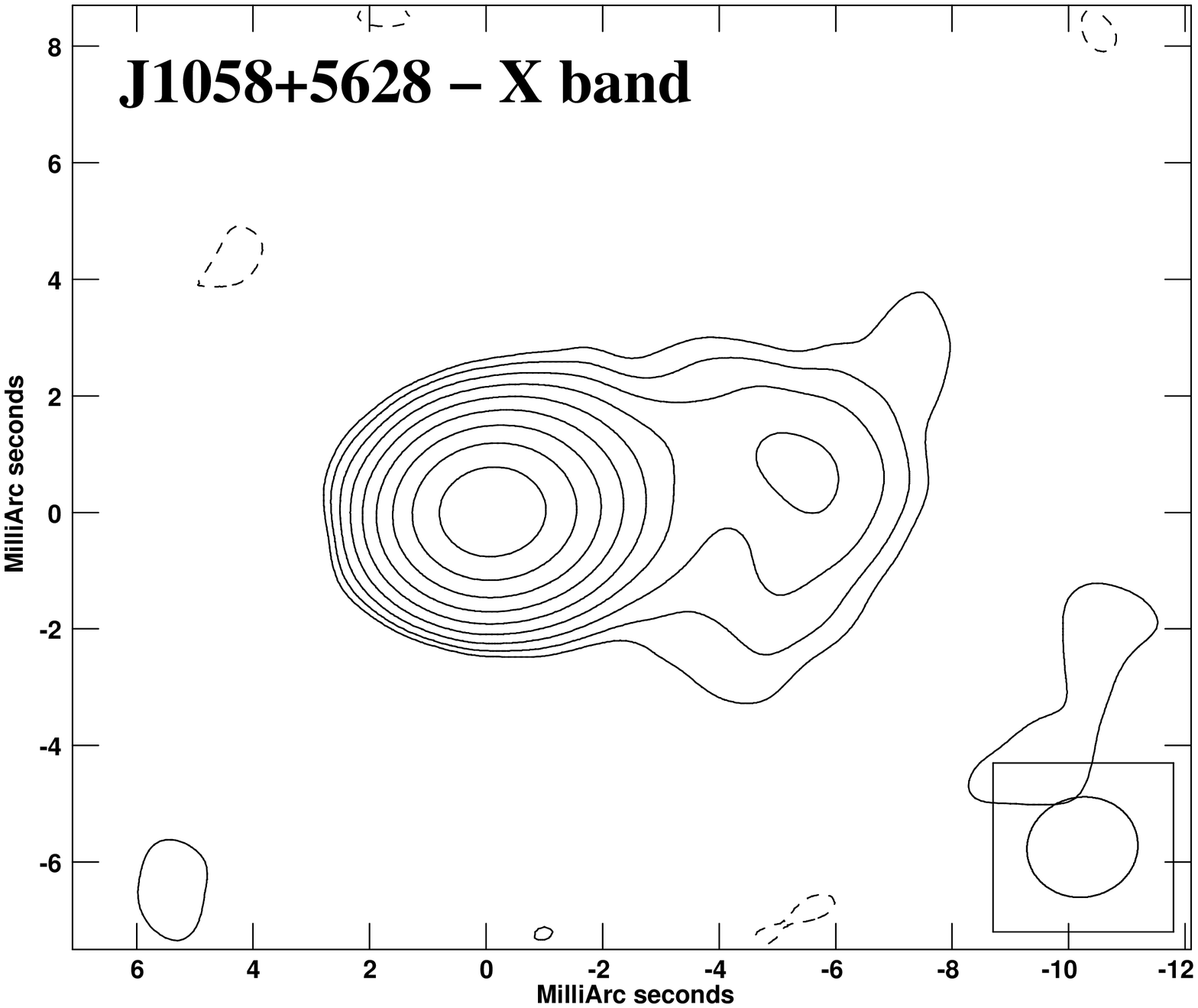}
\includegraphics[width=4.5cm, angle=0]{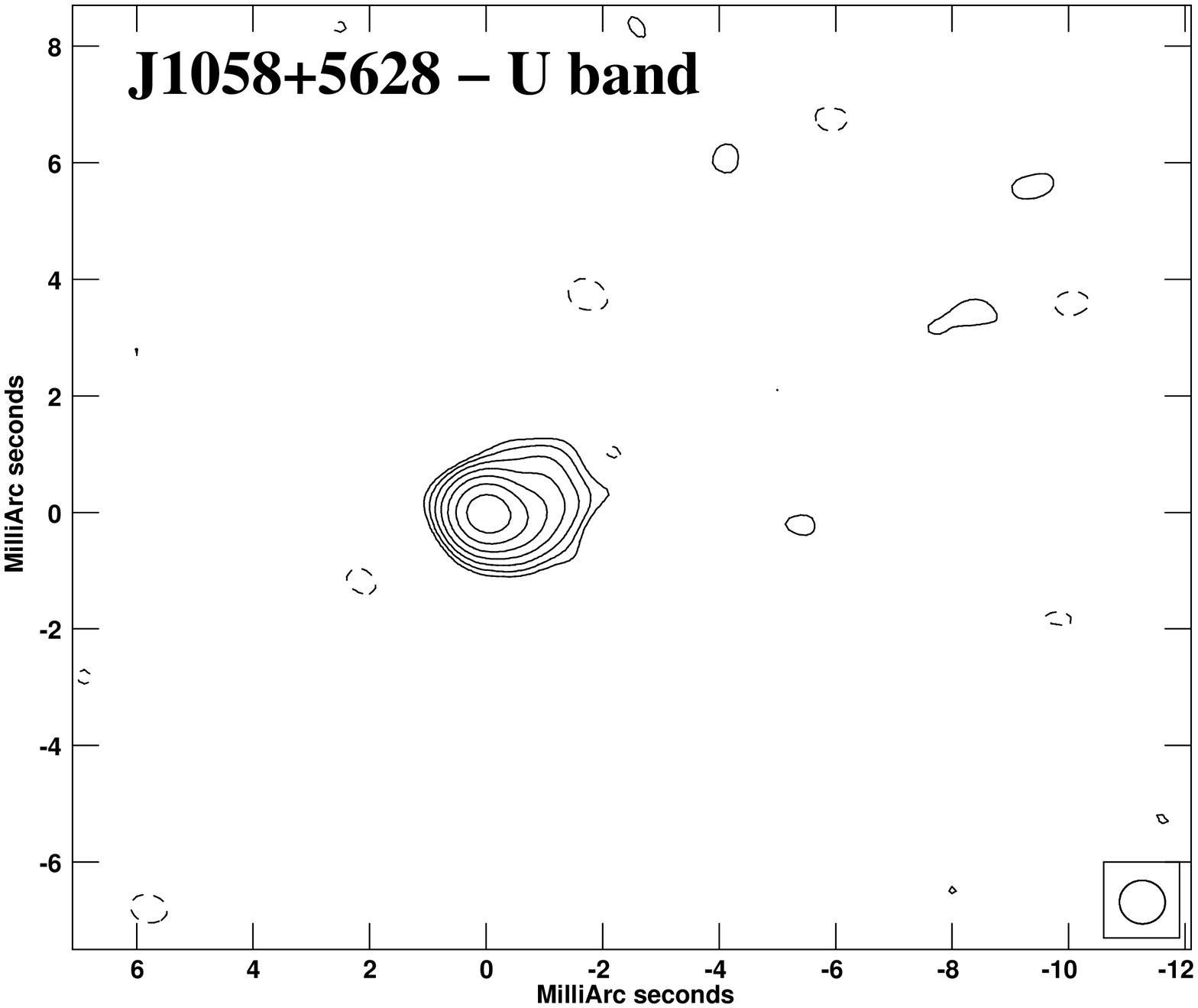}
\includegraphics[width=4cm, angle=0]{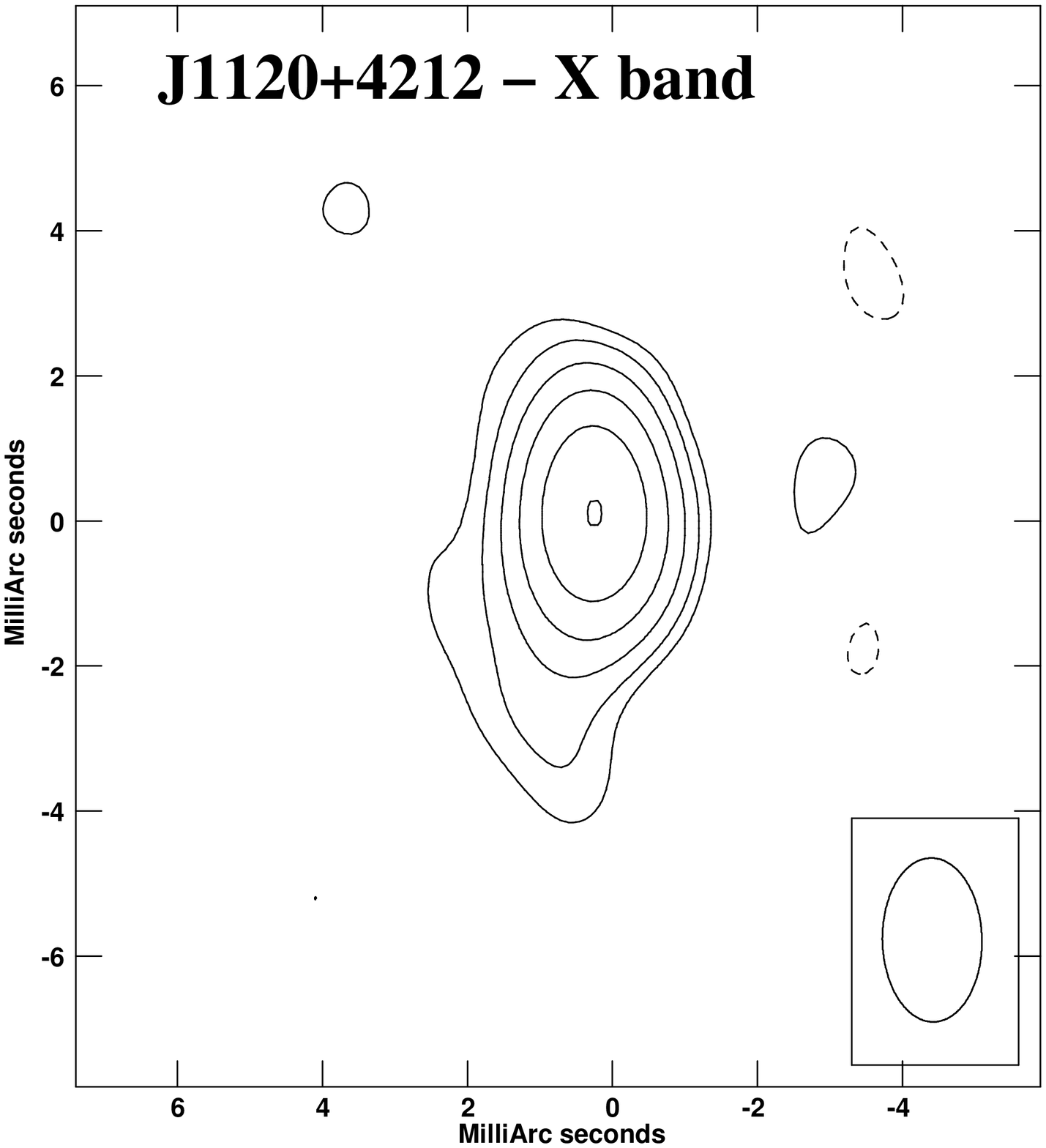}
\includegraphics[width=4cm, angle=0]{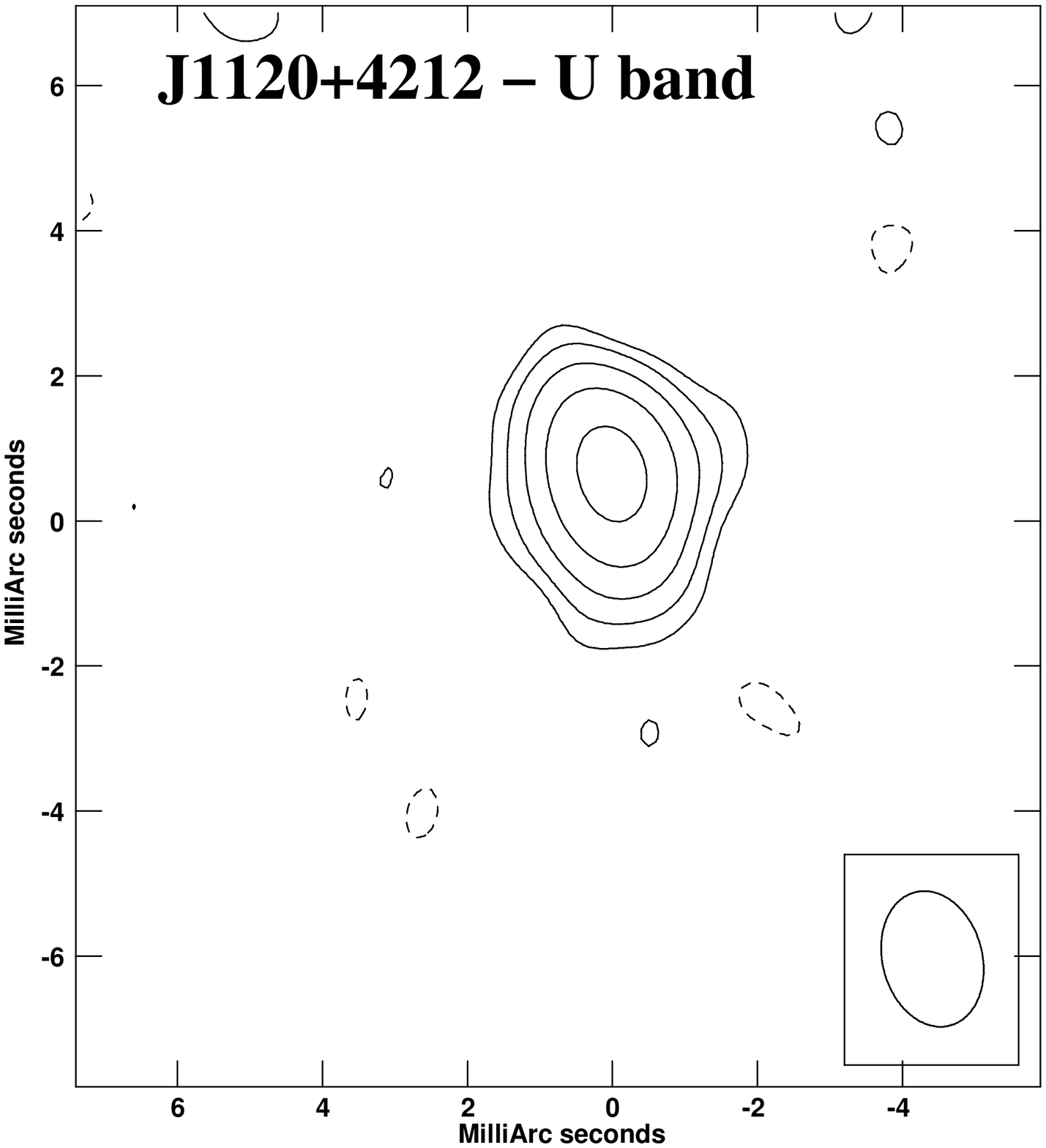}
\includegraphics[width=4.15cm, angle=0]{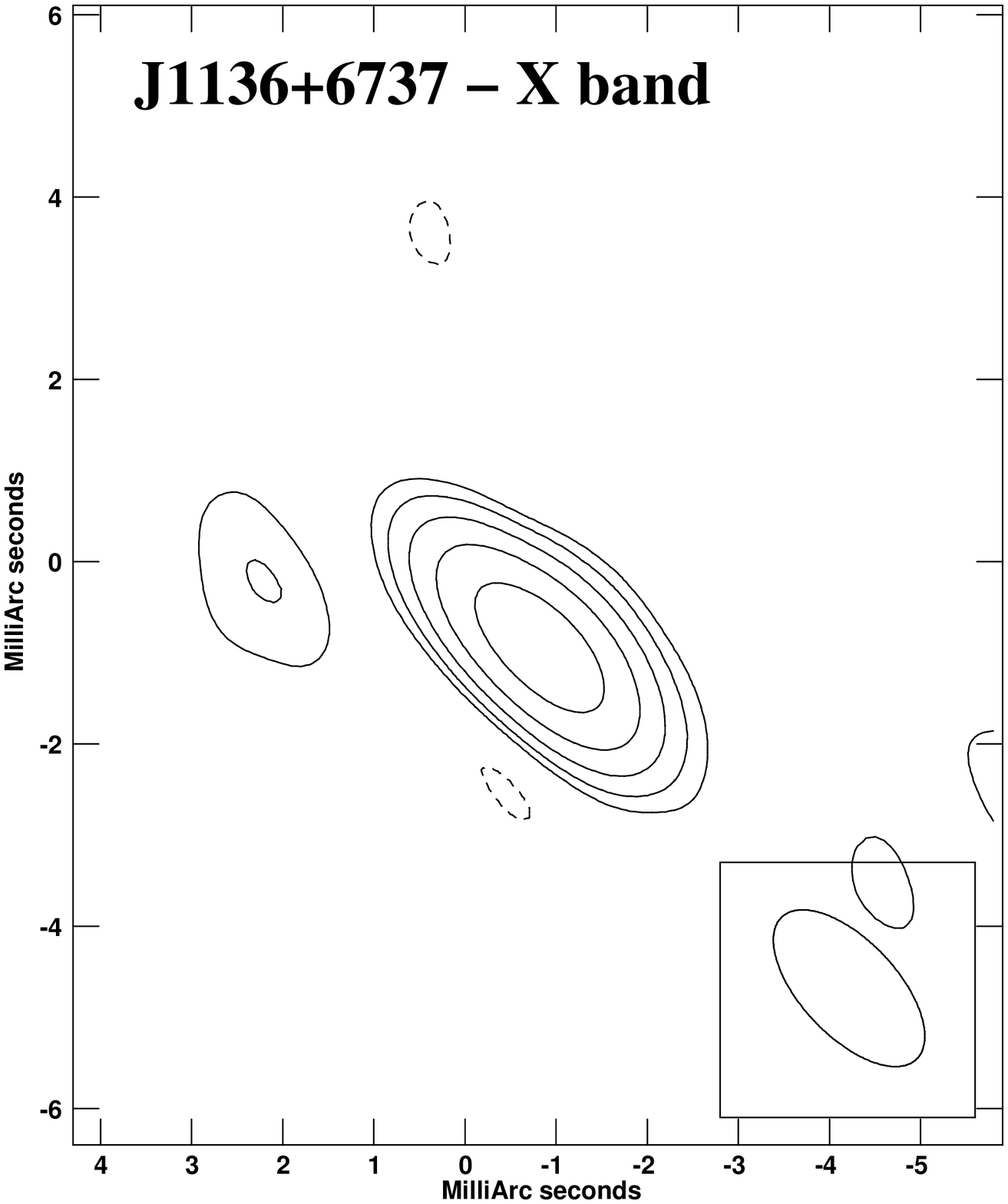}
\includegraphics[width=4.15cm, angle=0]{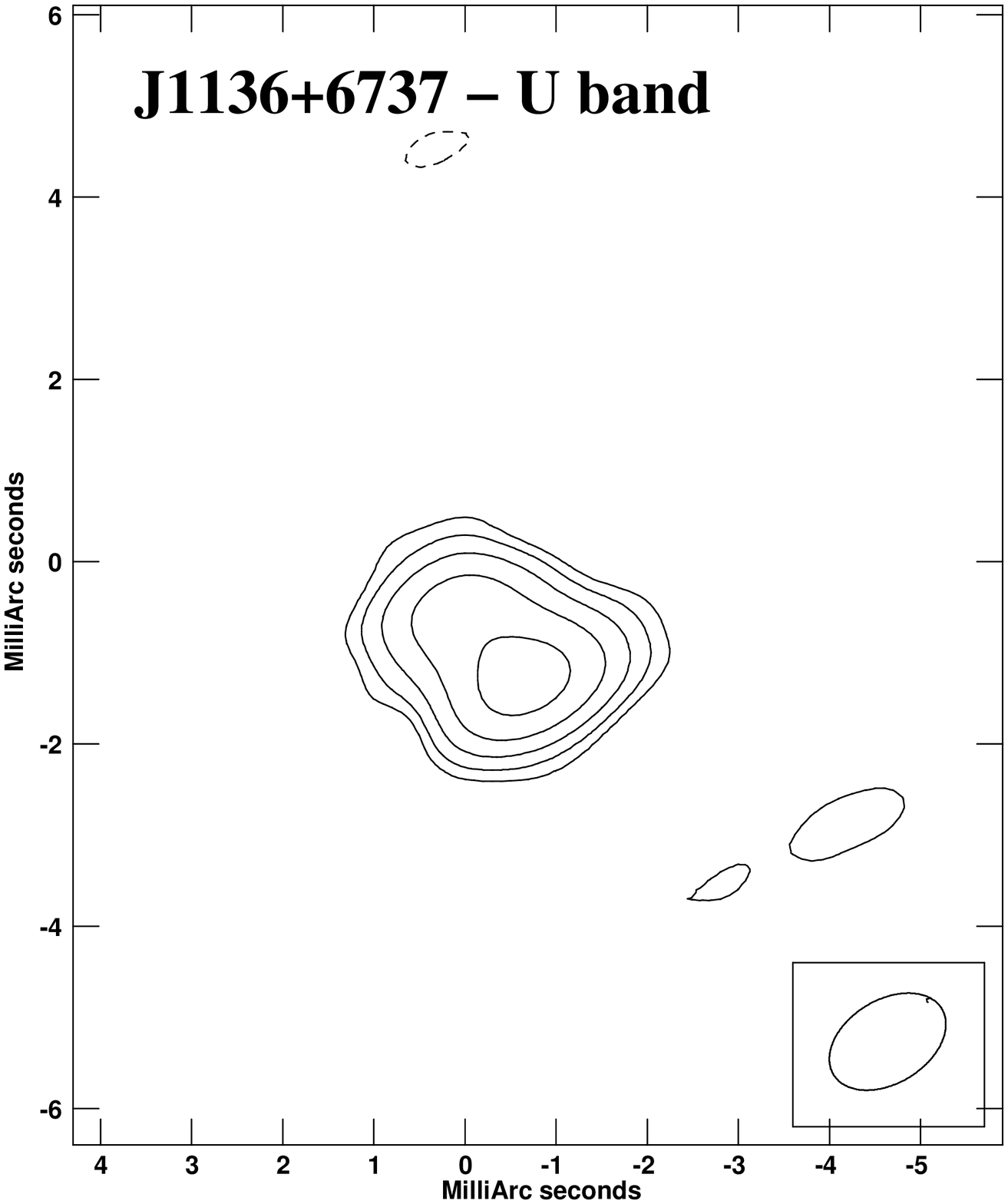}
\includegraphics[width=4.5cm, angle=0]{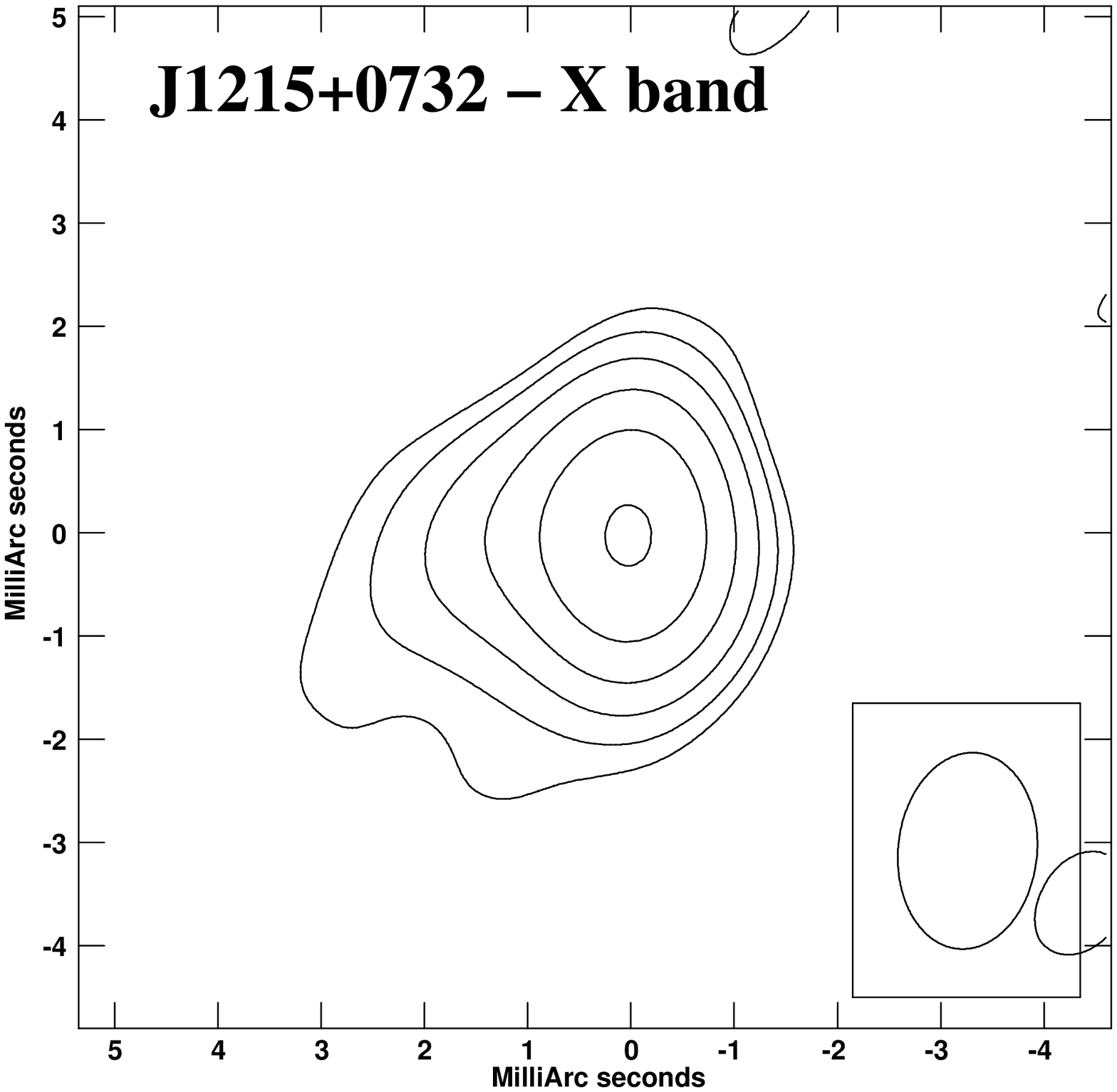}
\includegraphics[width=4.2cm, angle=0]{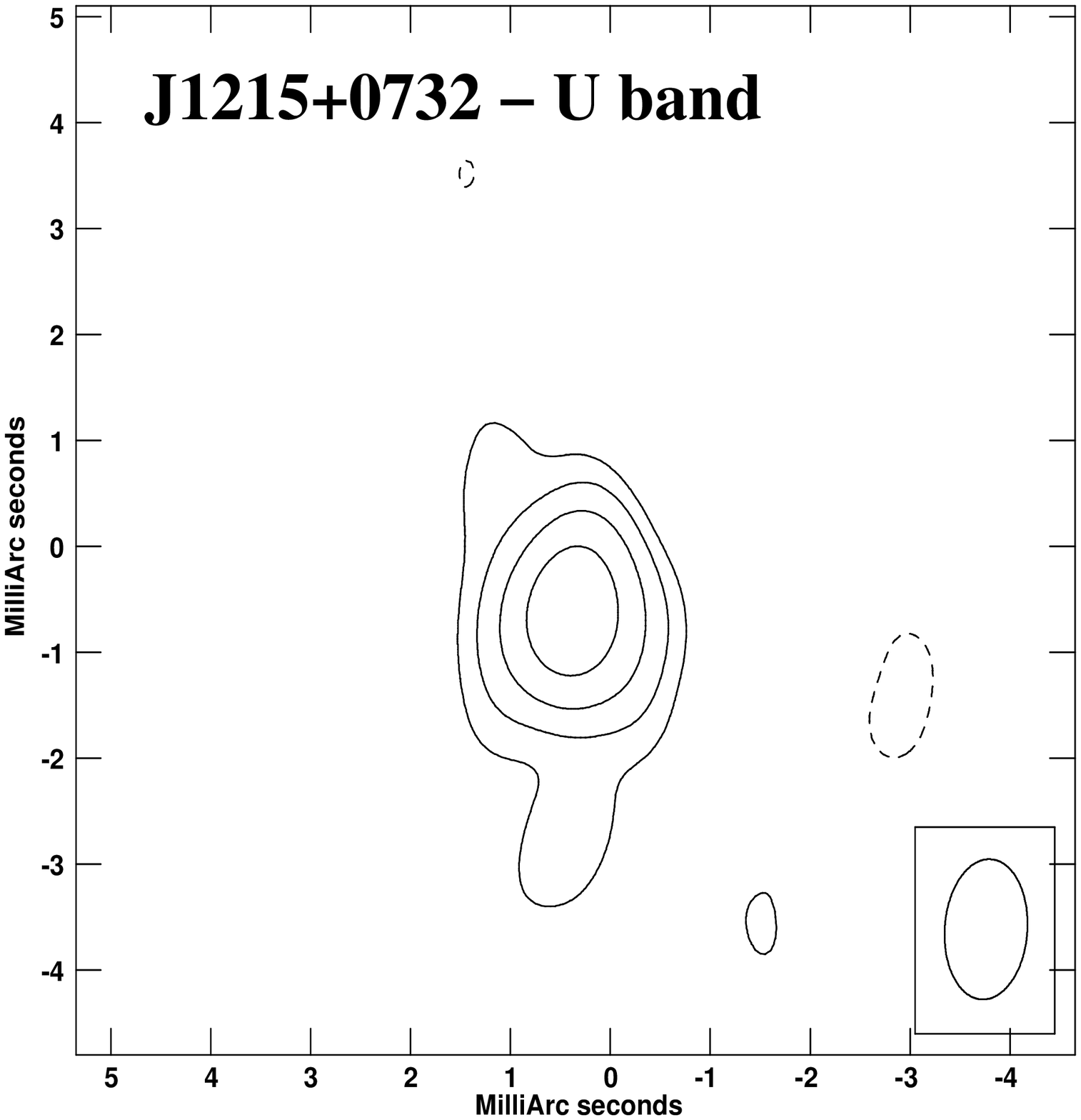}
\caption{\footnotesize {\bf 8.4 and 15 GHz VLBA images of resolved BL Lacs.}}
\label{fig_images}
\end{figure*}

\begin{figure*}
  \ContinuedFloat 
\includegraphics[width=4.25cm, angle=0]{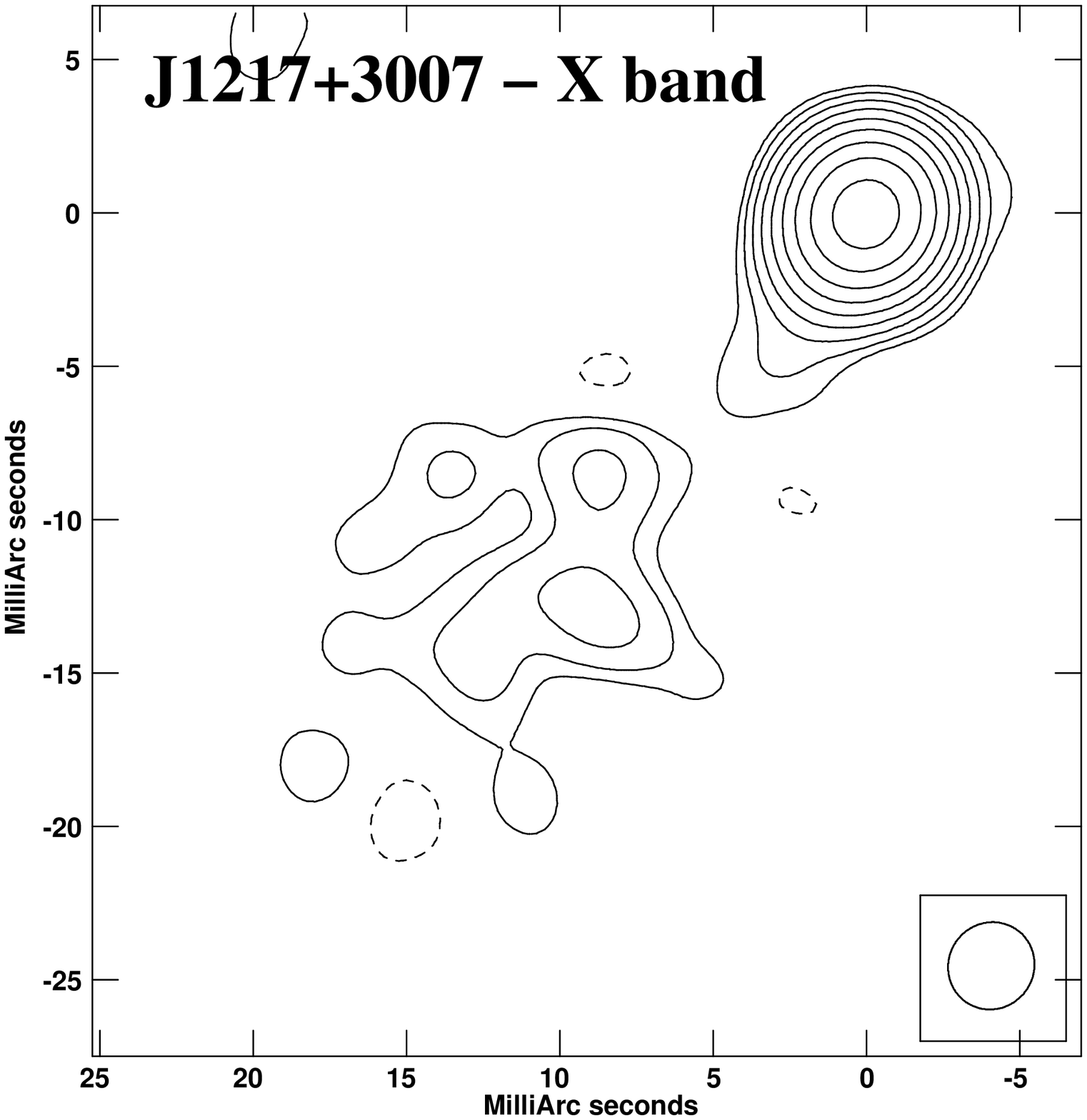}
\includegraphics[width=4.25cm, angle=0]{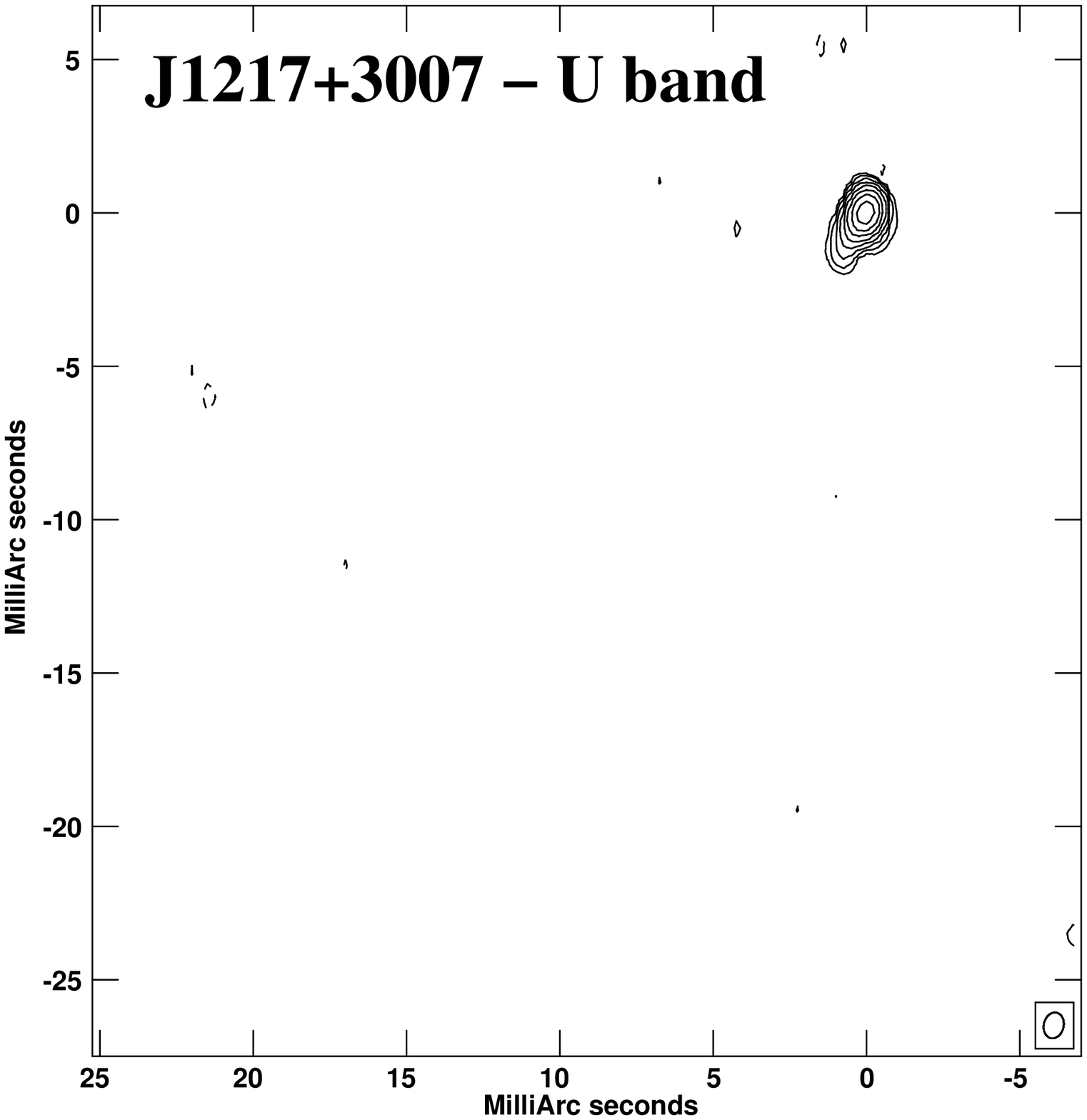}
\includegraphics[width=4.25cm, angle=0]{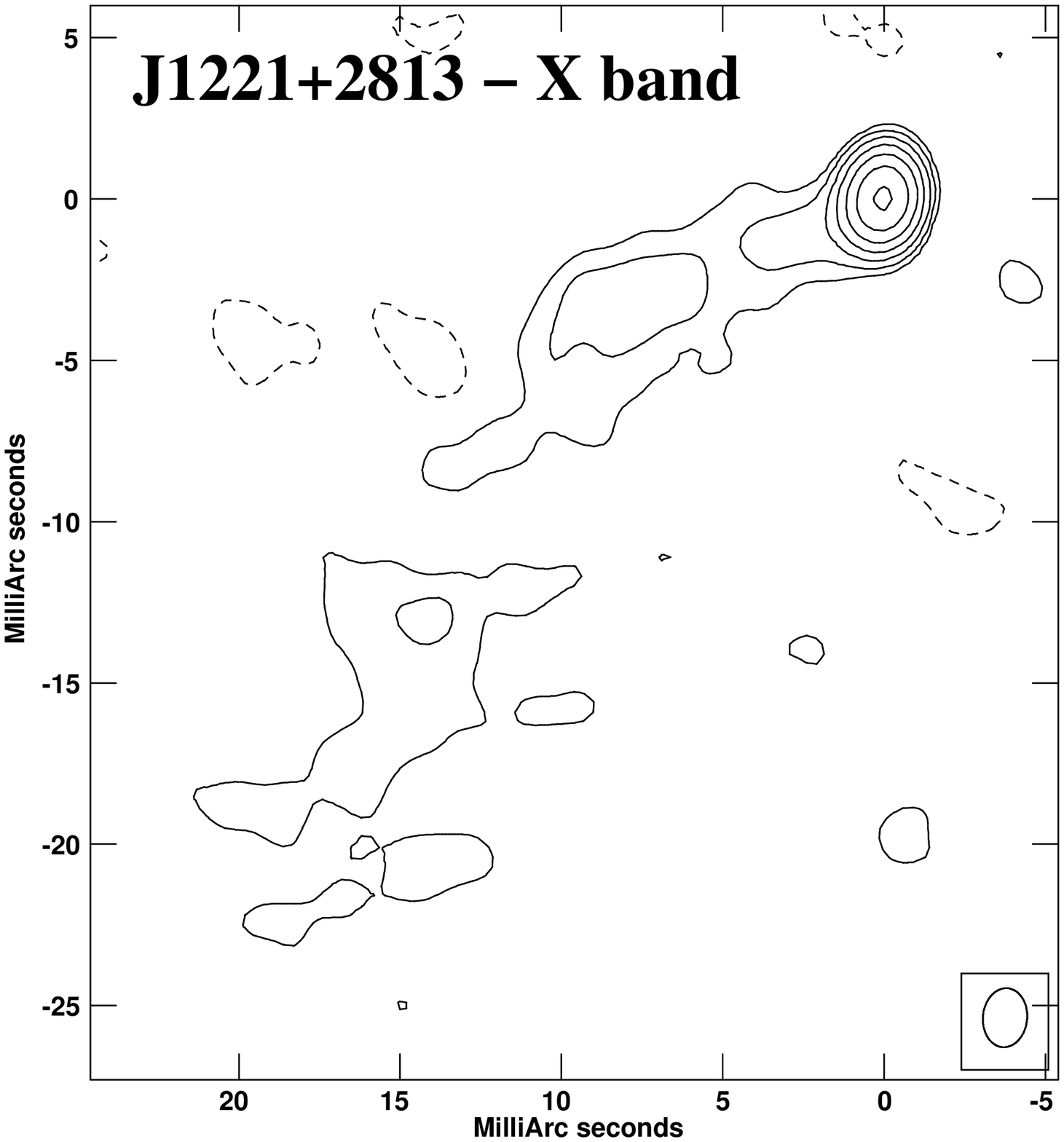}
\includegraphics[width=4.25cm, angle=0]{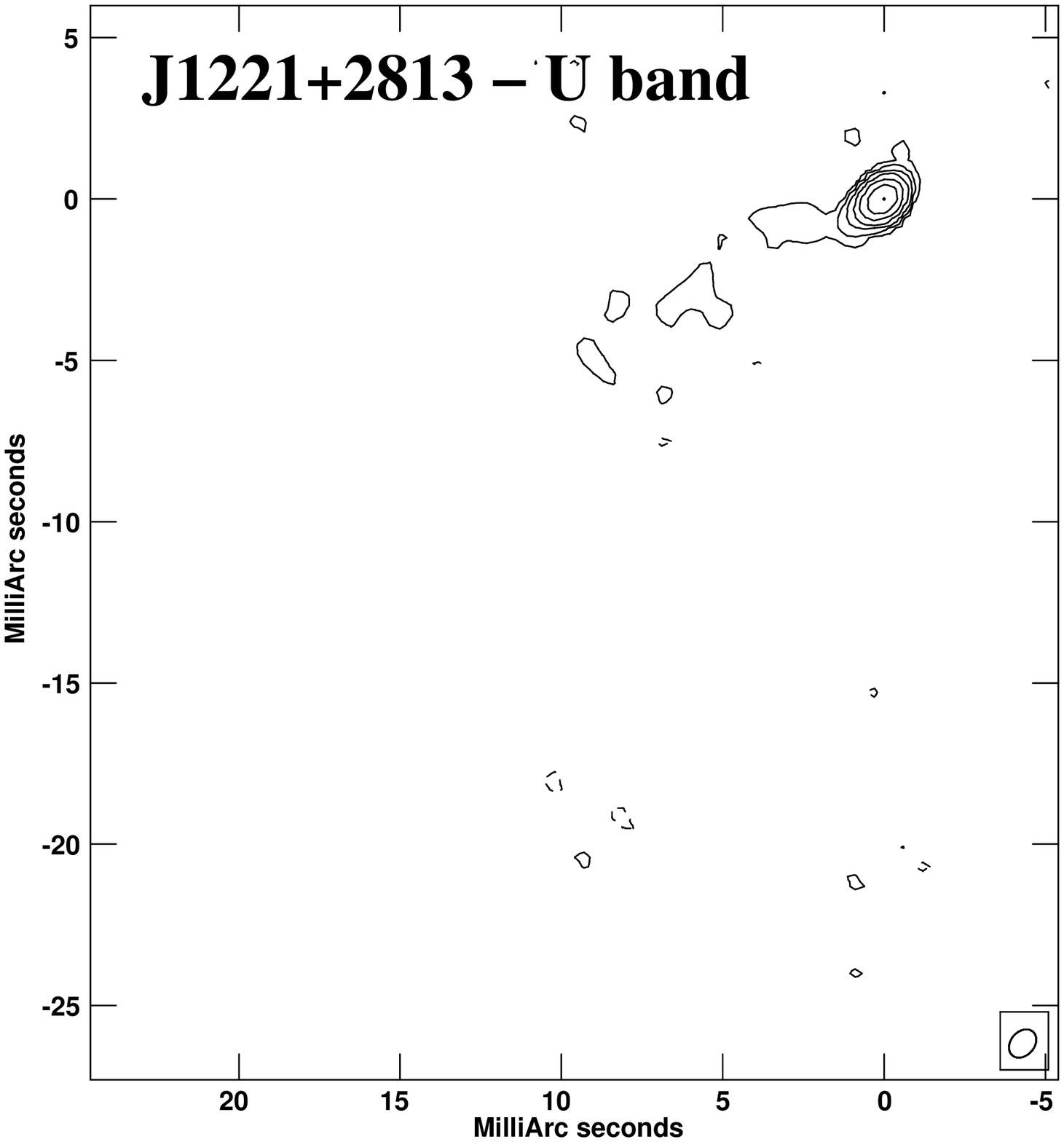}
\includegraphics[width=4.5cm, angle=0]{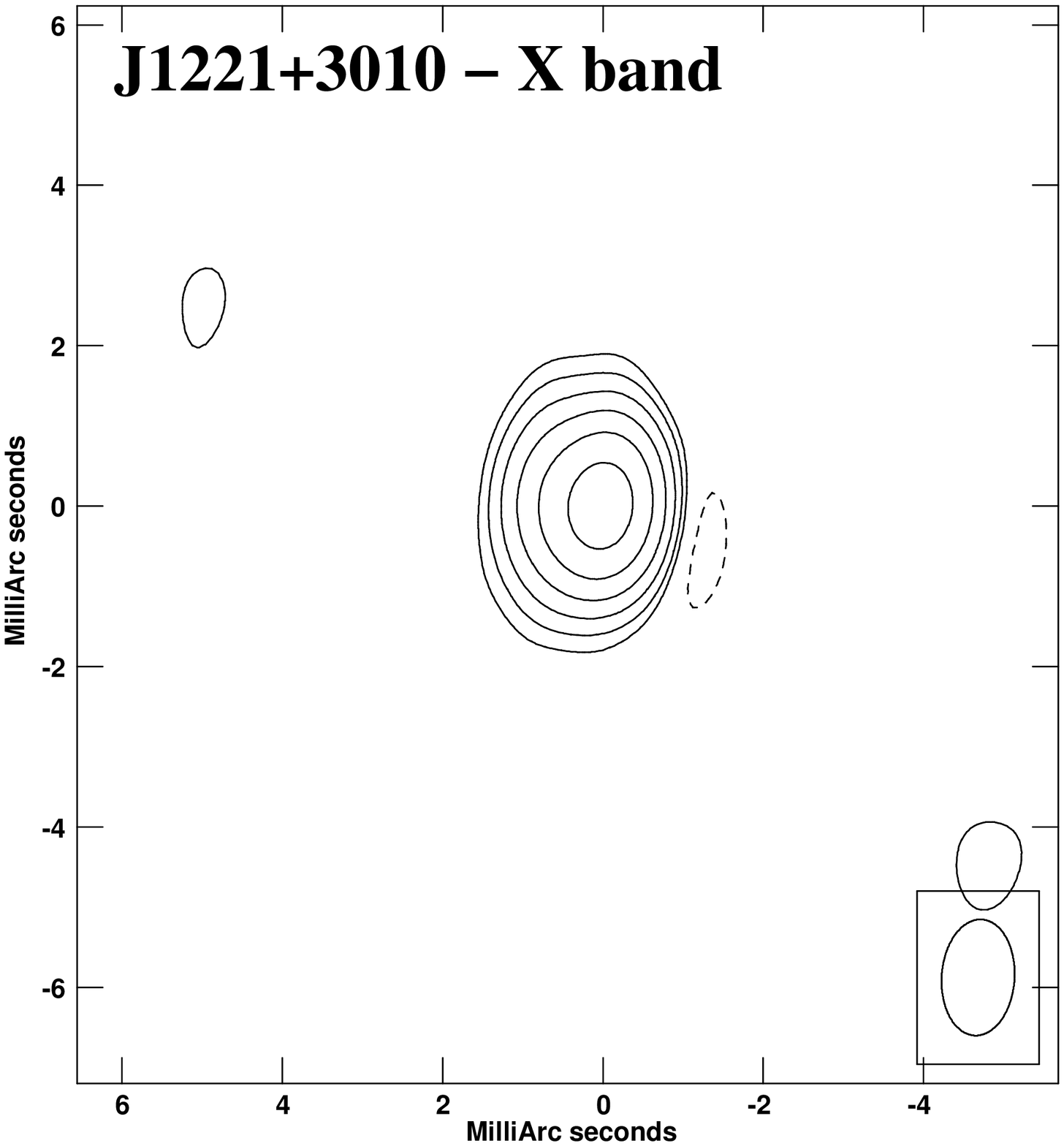}
\includegraphics[width=4.5cm, angle=0]{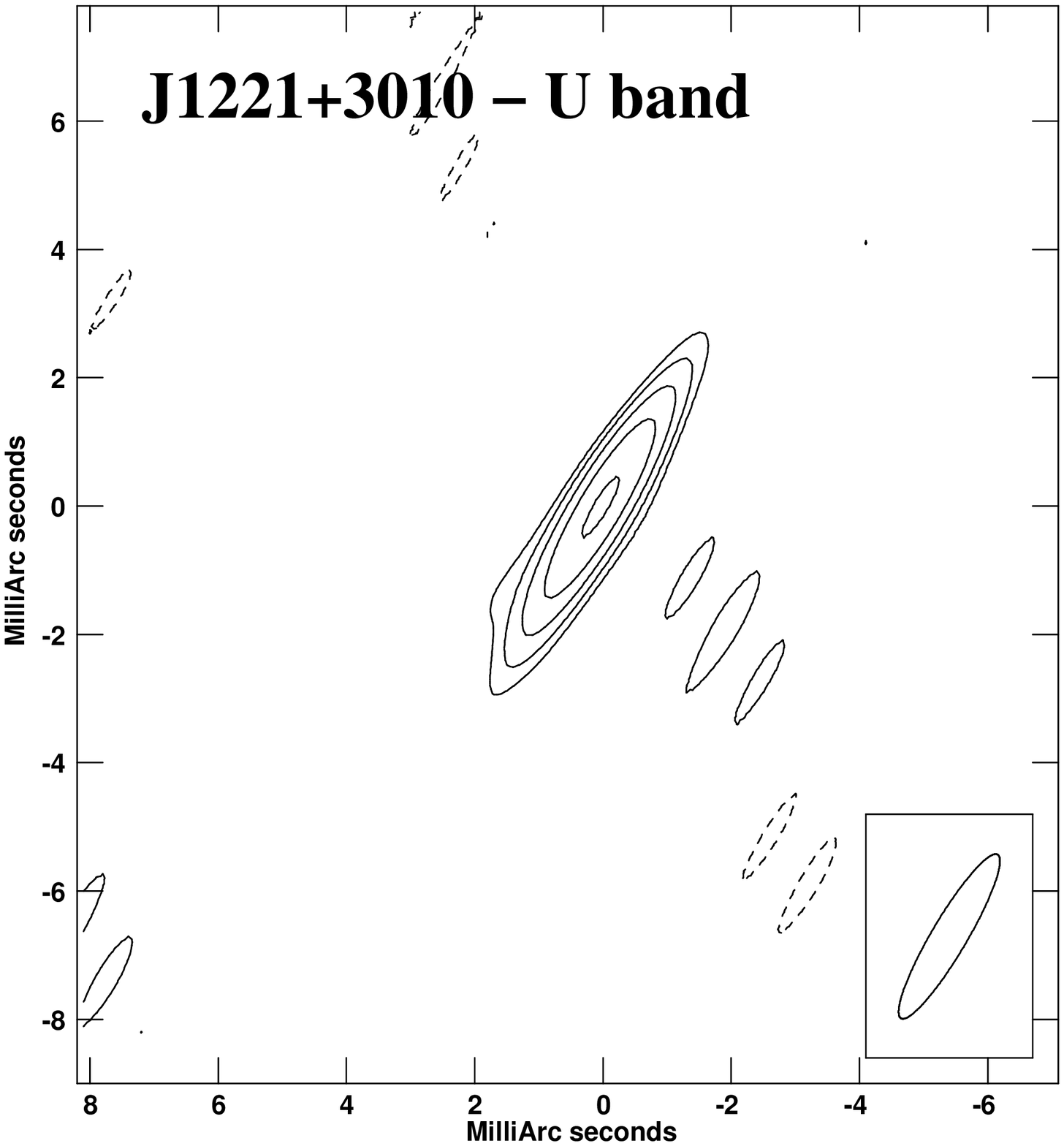}
\includegraphics[width=4cm, angle=0]{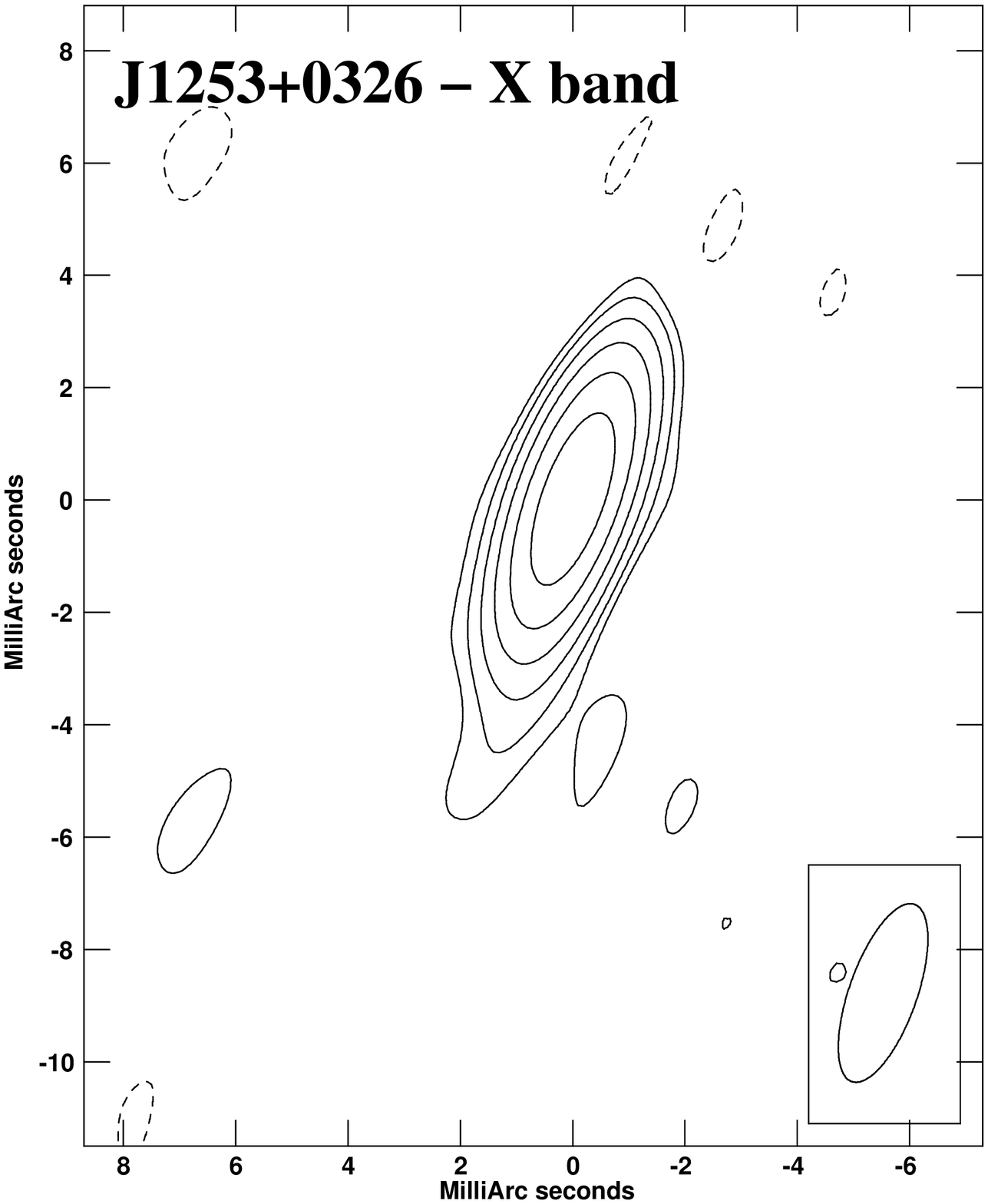}
\includegraphics[width=4cm, angle=0]{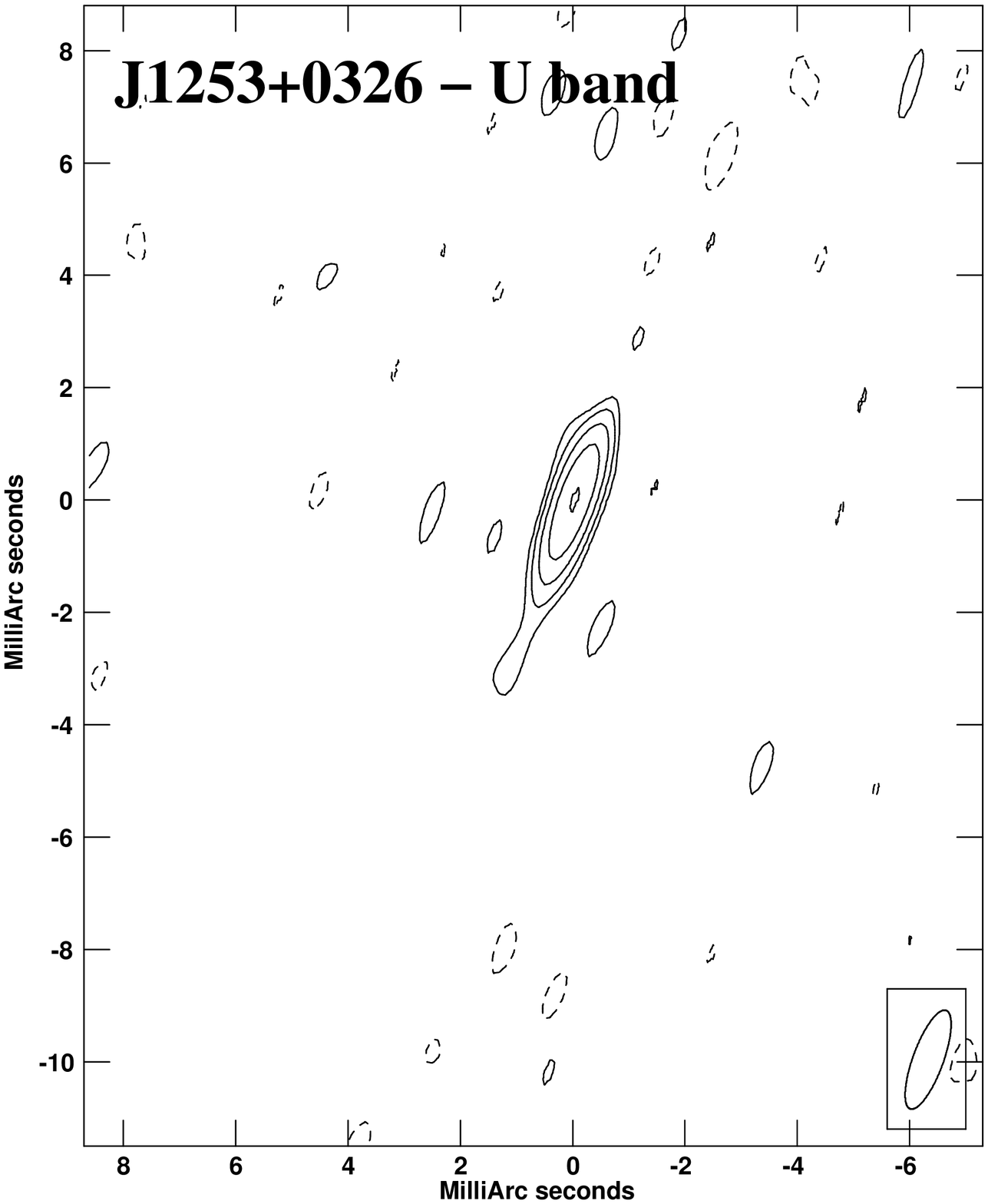}
\includegraphics[width=4.25cm, angle=0]{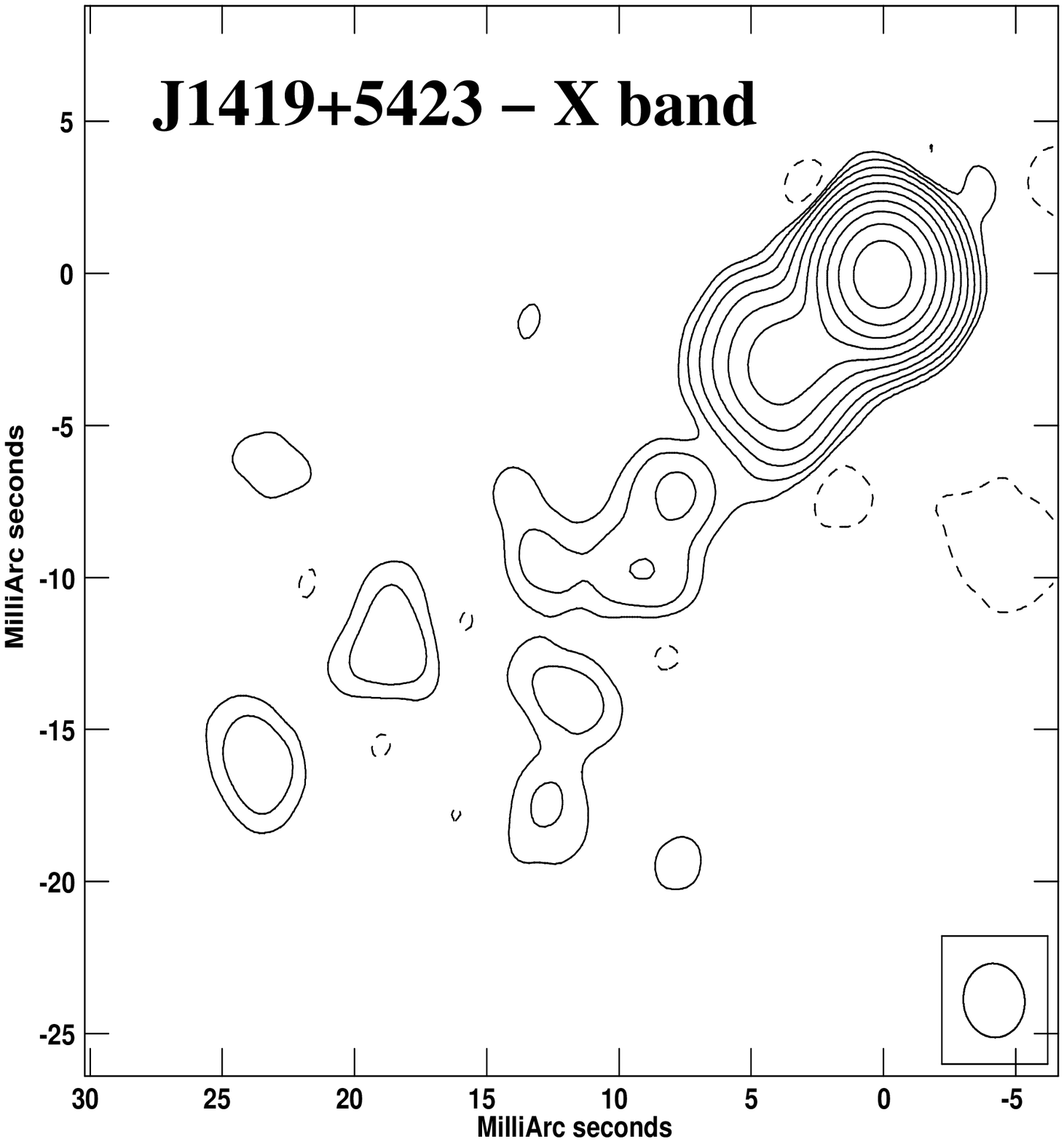}
\includegraphics[width=4.25cm, angle=0]{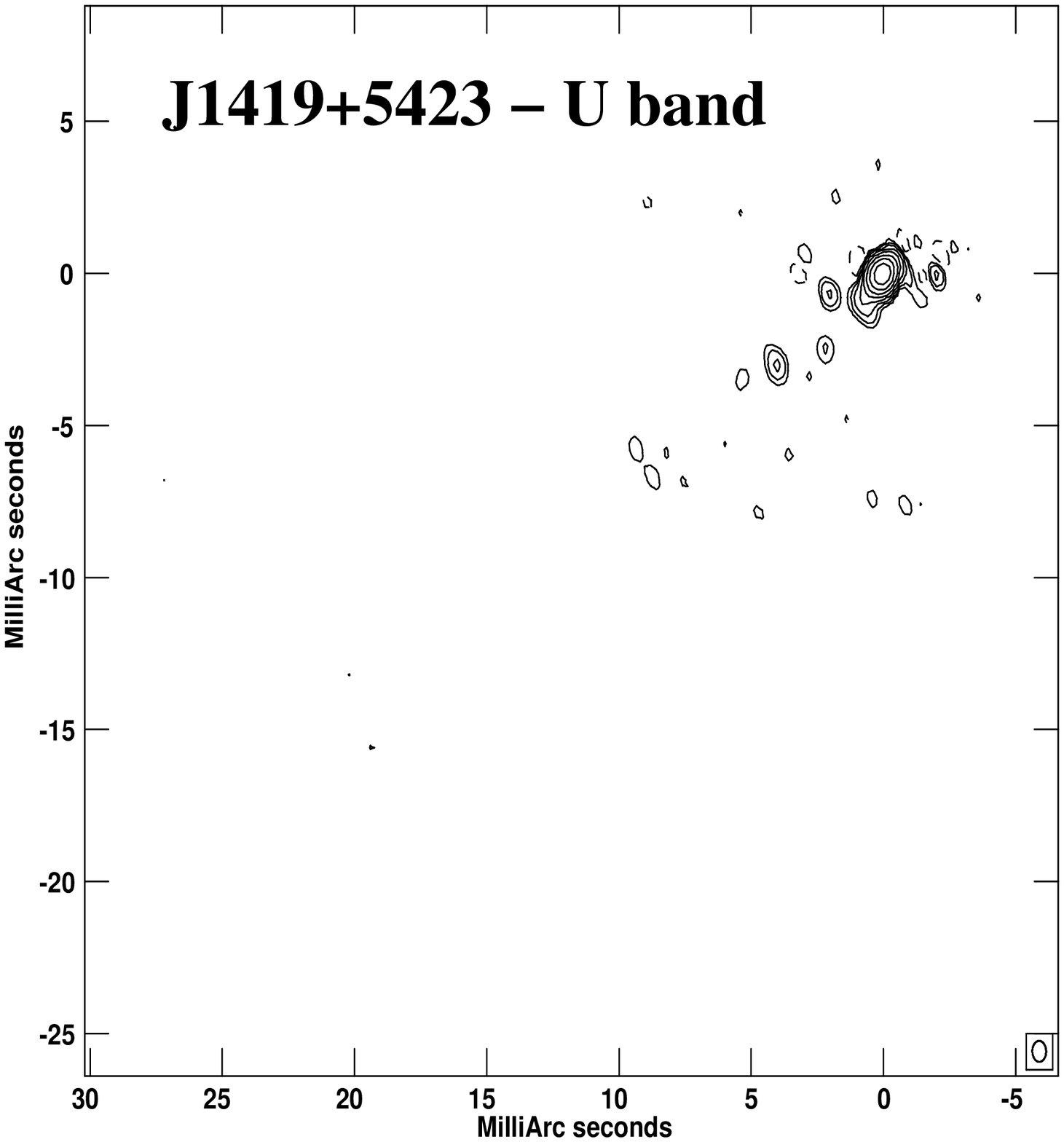}
\includegraphics[width=4.25cm, angle=0]{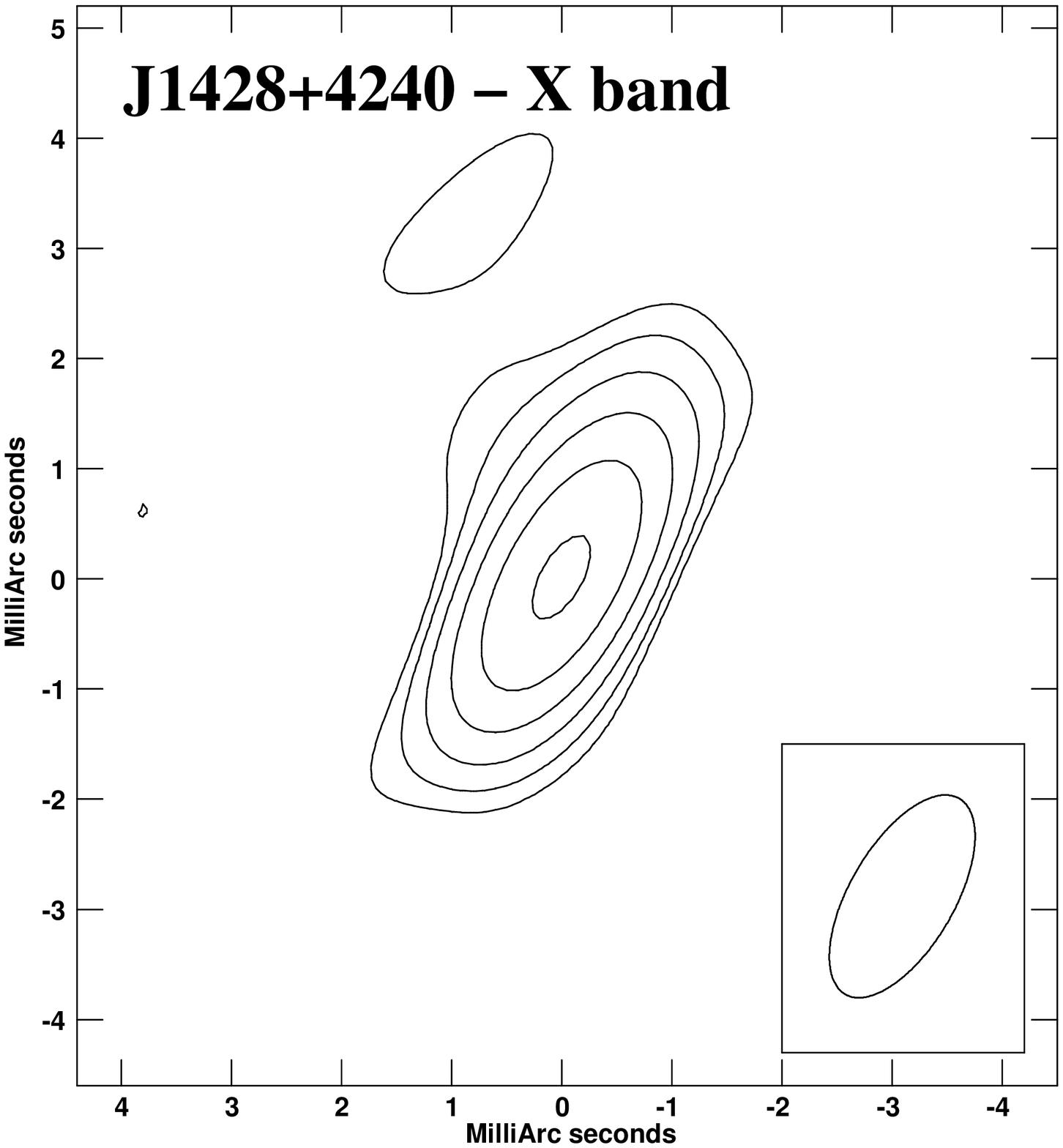}
\includegraphics[width=4.25cm, angle=0]{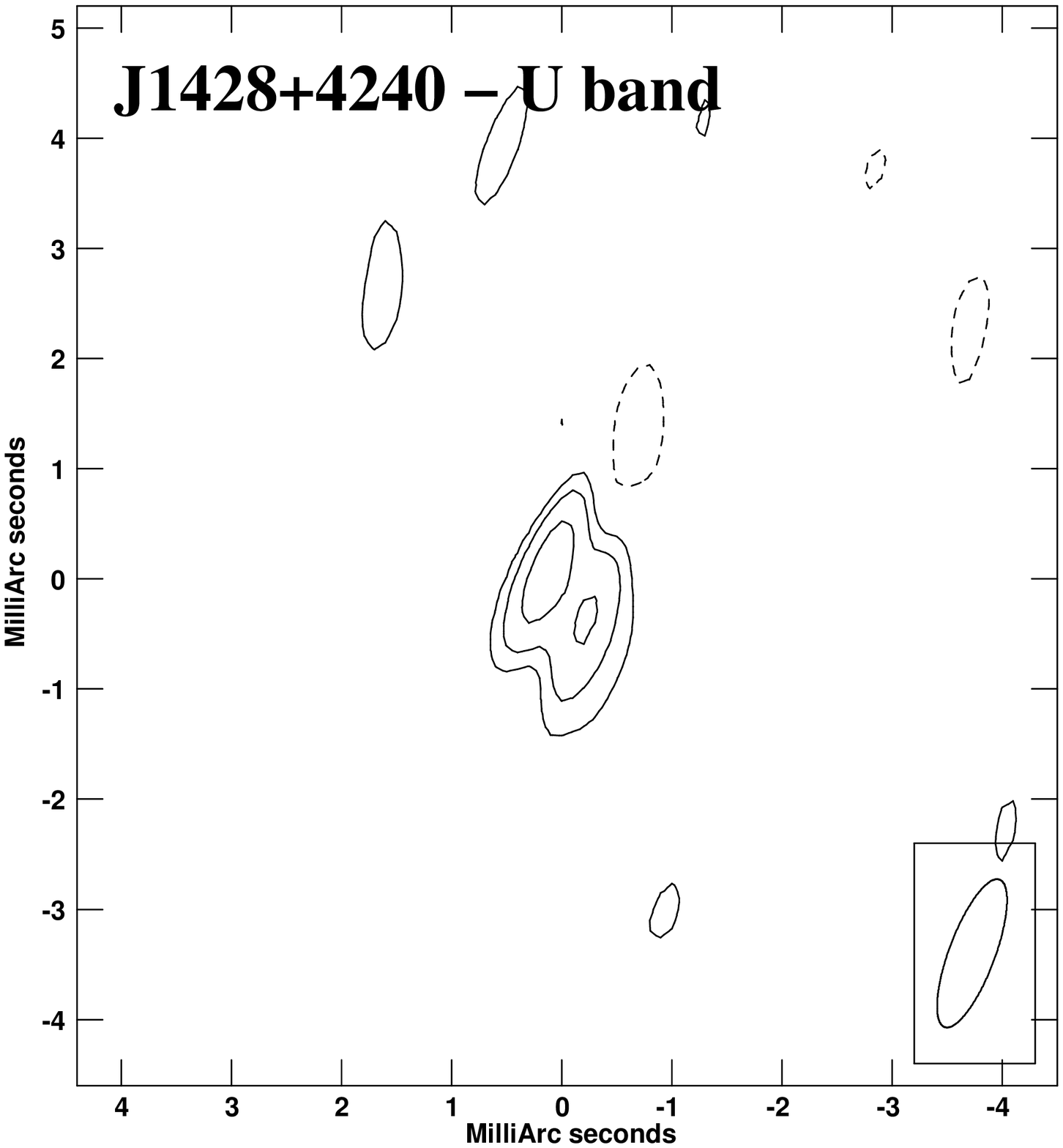}
\includegraphics[width=4.25cm, angle=0]{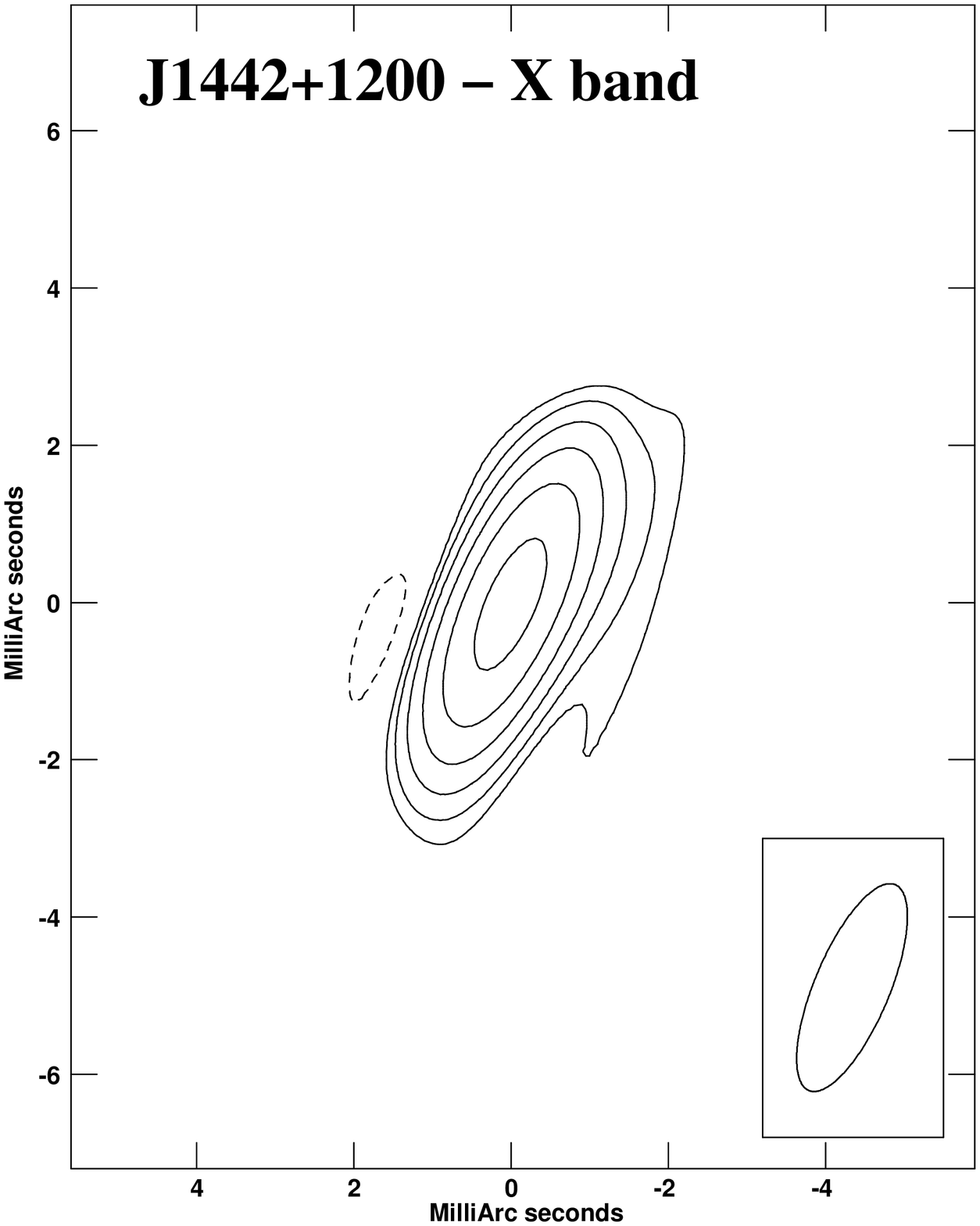}
\includegraphics[width=4.25cm, angle=0]{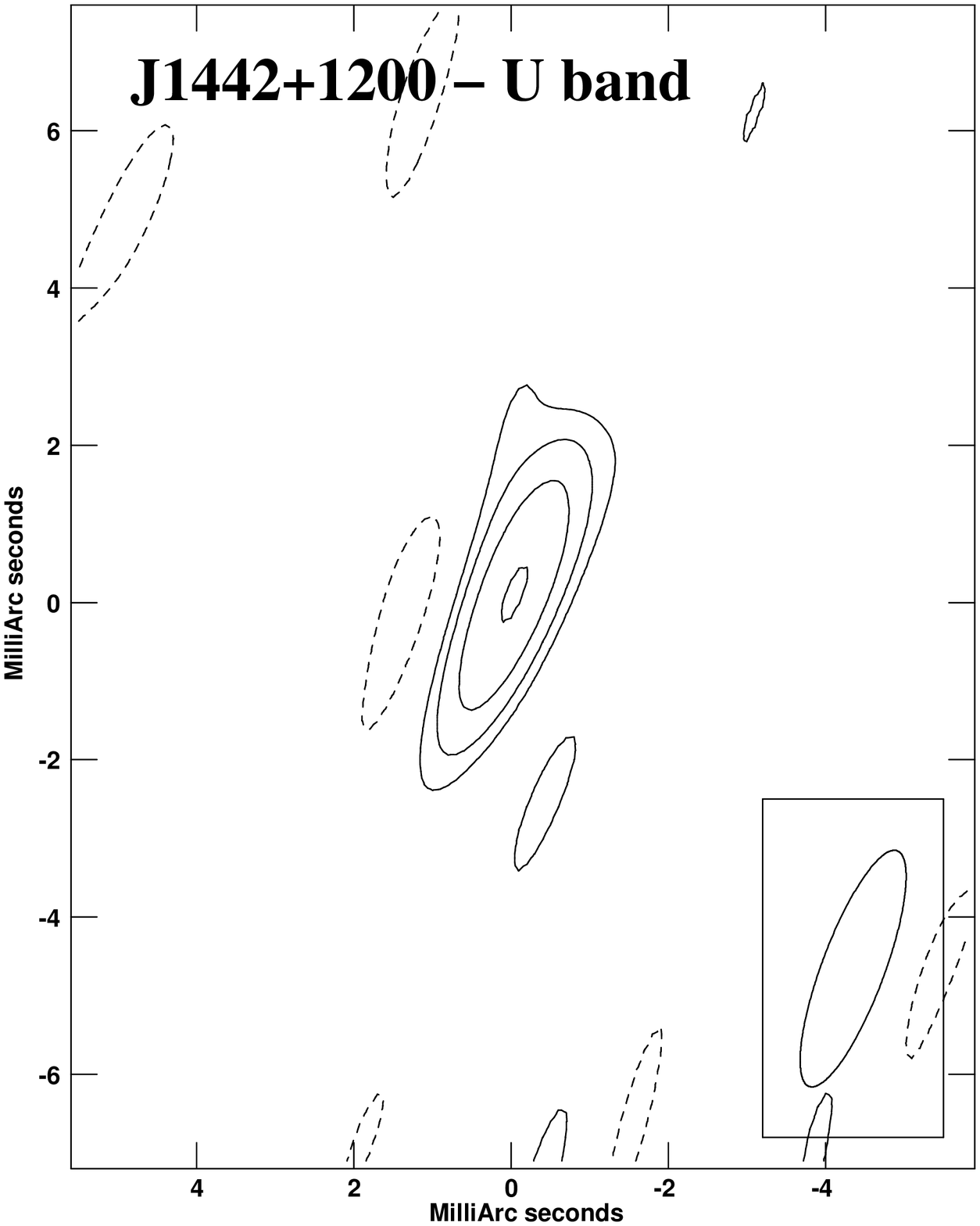}
\includegraphics[width=4.25cm, angle=0]{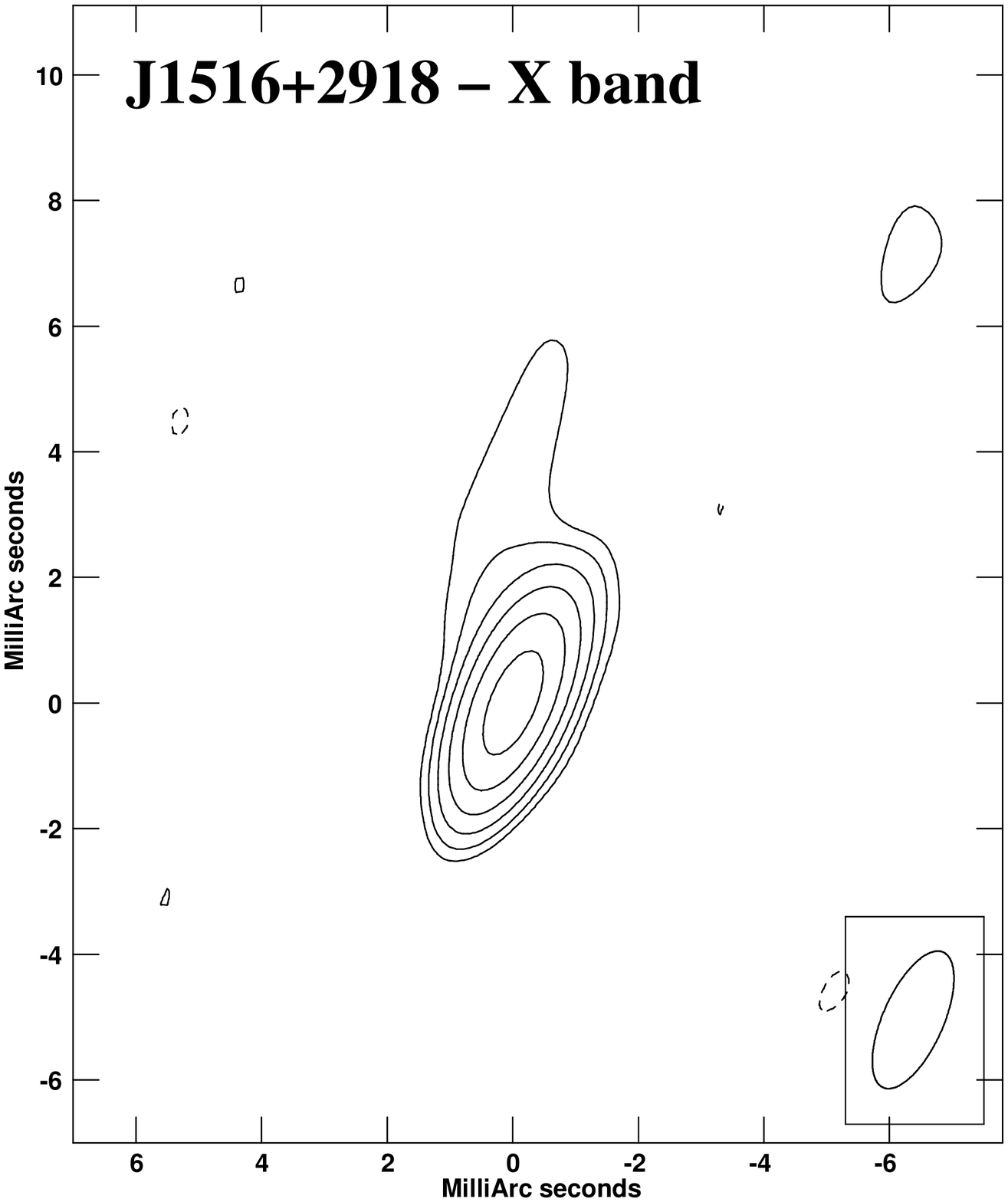}
\includegraphics[width=4.25cm, angle=0]{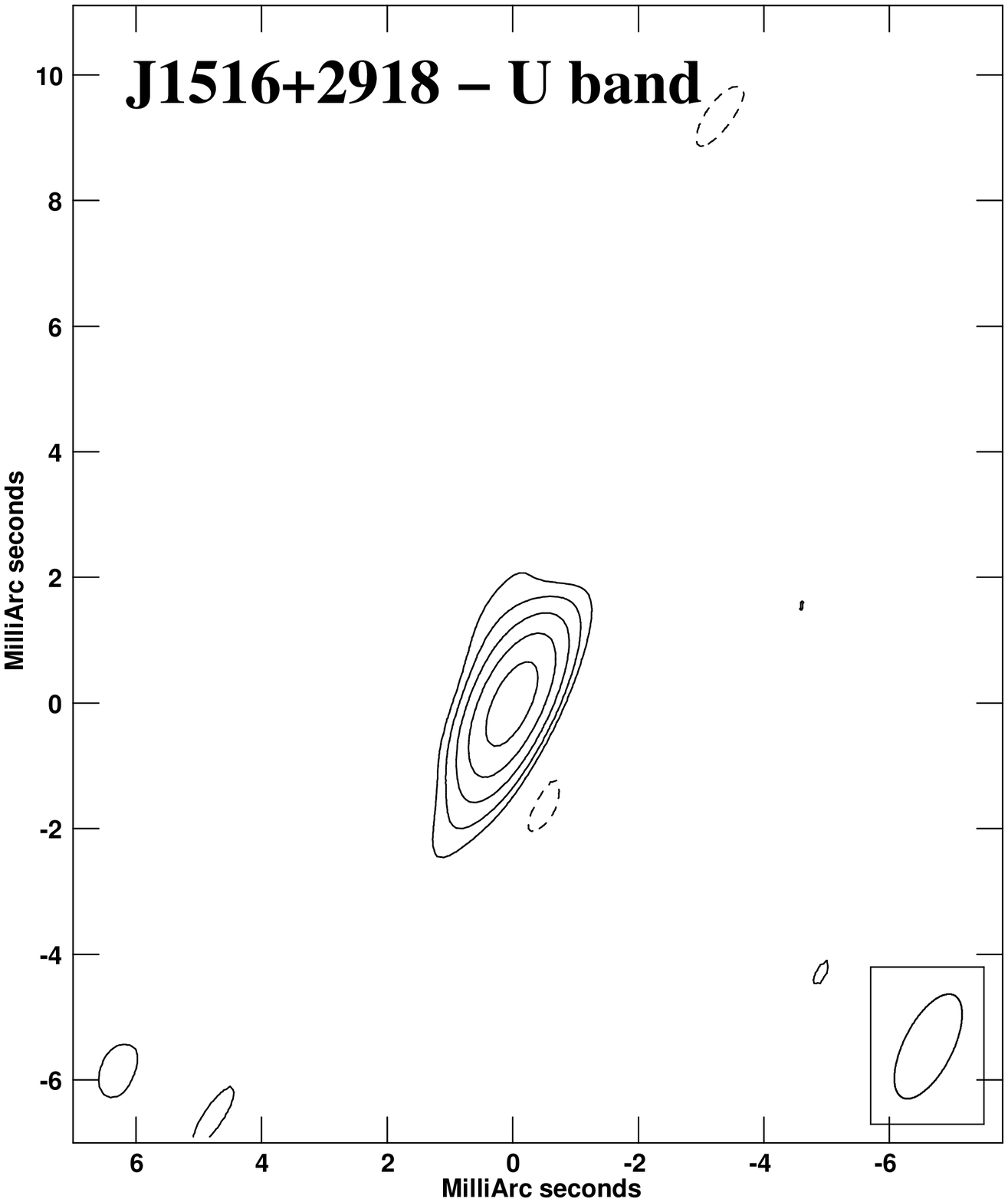}
\includegraphics[width=4.25cm, angle=0]{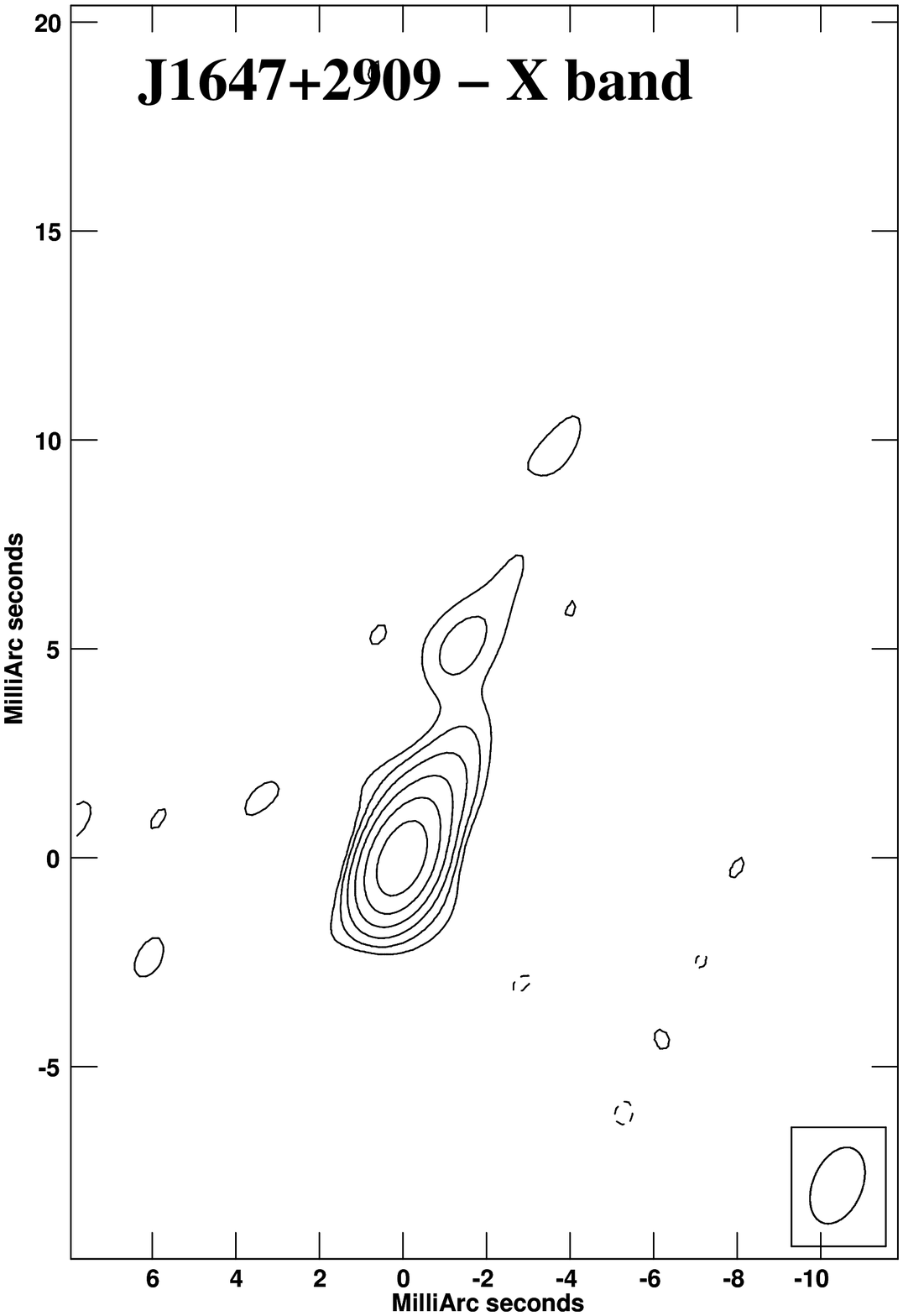}
\includegraphics[width=4.25cm, angle=0]{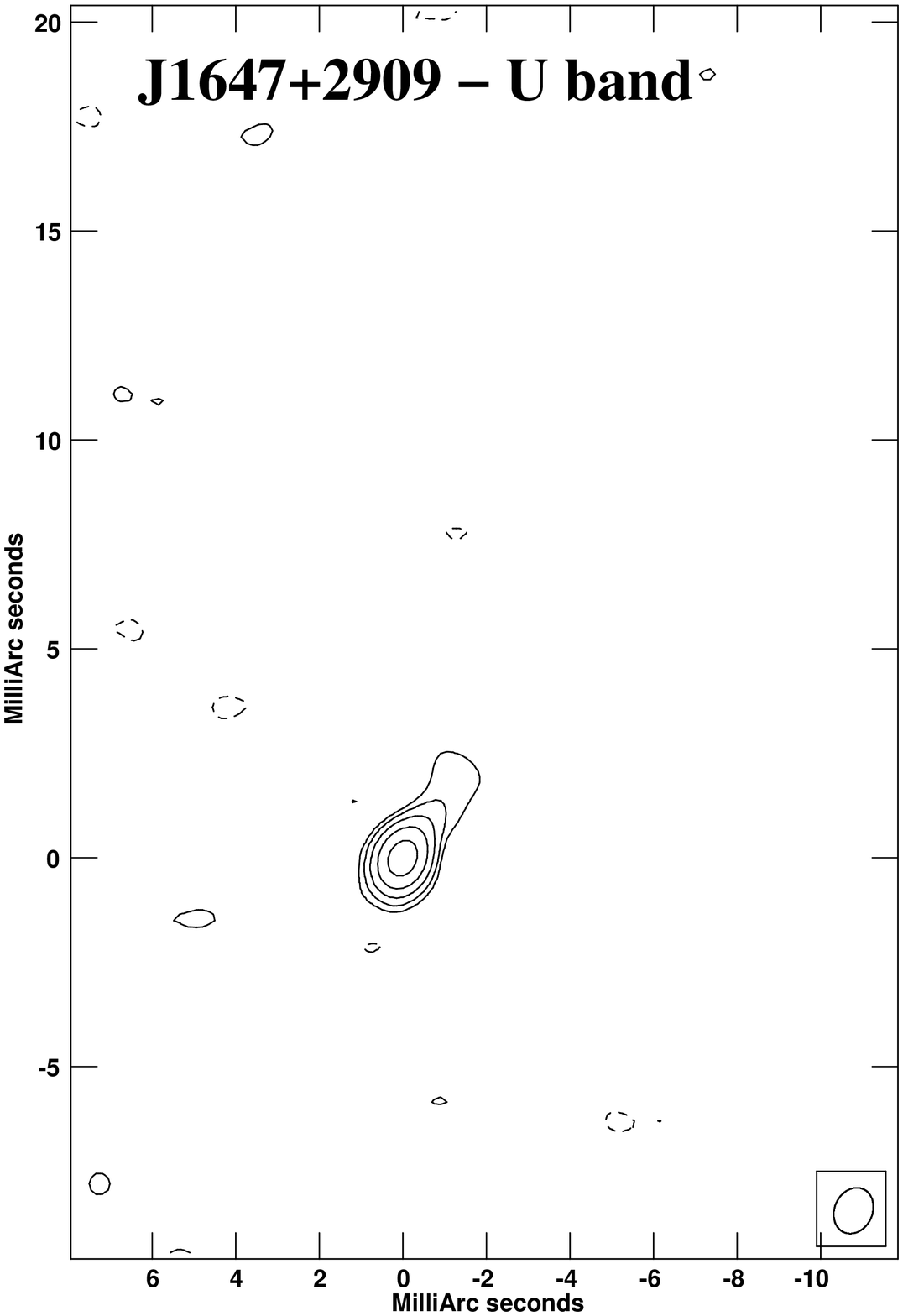}
\includegraphics[width=5.5cm, angle=0]{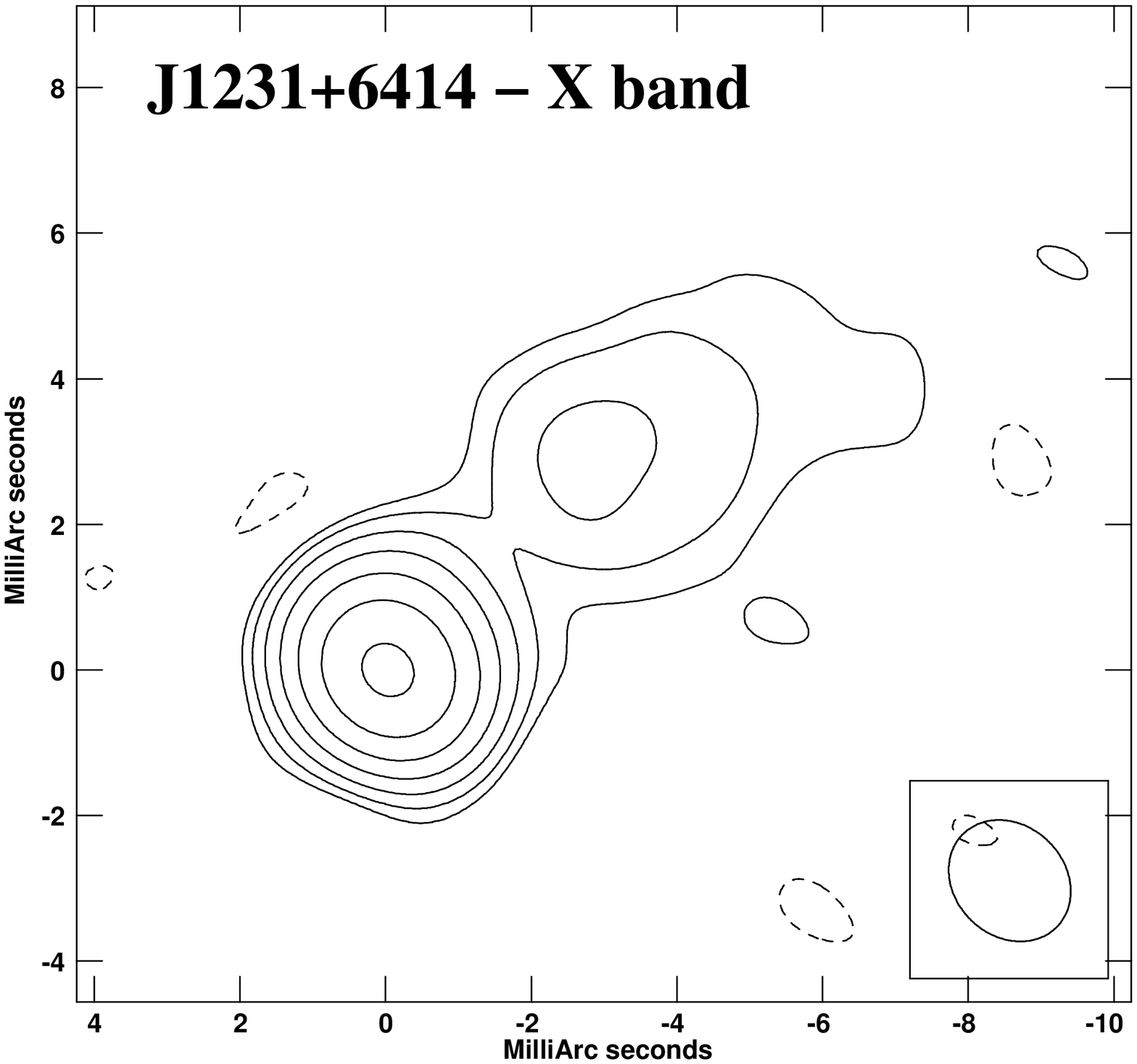}
\caption{\footnotesize ( continued).}
\label{fig_images}
\end{figure*}

\clearpage

\section{Results}\label{results}

\subsection{Detection rate} \label{rate}

In our sample, 27/42 (64 $\%$) objects show parsec-scale radio emission. In particular, 24 BL Lac objects are revealed at both frequencies, while two are detected only at 8 GHz (J1221+0821 and J1231+6414) and one only at 15 GHz (J1427+5409). 
 
We note that many of the sources revealed using phase referencing
show a flux density below the noise level of sources observed
without this technique.  This suggests that more sources could be
detected using the phase reference mode for all the observations
(e.g. J1427+5409).

Point-like morphologies are present in 8/27 (31\%) of the detected
sources at 8 GHz, and in 14/26 (54$\%$) of the revealed targets at 15
GHz (Tab. \ref{tab_parsecMorphology}). The other detected sources show
an one-sided structure.  We note that in our sample, among the 8 BL
Lacs having one-sided morphology at both frequencies, 7 are the most
luminous (and bright) BL Lacs at mas scale.

\subsection{The spectral index} \label{sect_spectral}

For each object detected at both VLBA frequencies, we estimated spectral
indexes for the core component ($\alpha_{core}$) and for the whole
source ($\alpha_{tot}$). Values are reported in Table
\ref{tab_parsecMorphology}.  For our spectral index analysis, we took
flux density measurements (Table 4) as obtained applying model-fit to the data
set with the entire {\it (u, v)} coverage. For some sources with complex parsec-scale structure (e.g. J1217+3007), we produced images at 8
and 15 GHz using the same maximum and minimum baseline in the {\it (u,
  v)} coverage. This allows a better characterization of the core emission, especially when it is well resolved at the high frequency. We applied the model-fit to these {\it (u,
  v)} data and used the flux densities extracted from these visibilities
for the estimation of the spectral indexes.  The fraction of flat
spectral indexes ($\alpha_{tot}\leq$0.6) is predominant (20/25) as
expected for BL Lac objects. 

We also attempted the study of the spectral index distribution along the source structure. However, in our observations, the {\it (u,  v)} coverage is so different between 8 and 15 GHz that a lot of data on the jet emission is lost producing the U and X band images with the same {\it (u, v)} coverage. It results that no additional information from this analysis could be derived.

\subsection{The source radio structures} \label{sect_SC}

\begin{figure}
\centering
\includegraphics[width=0.5\textwidth ]{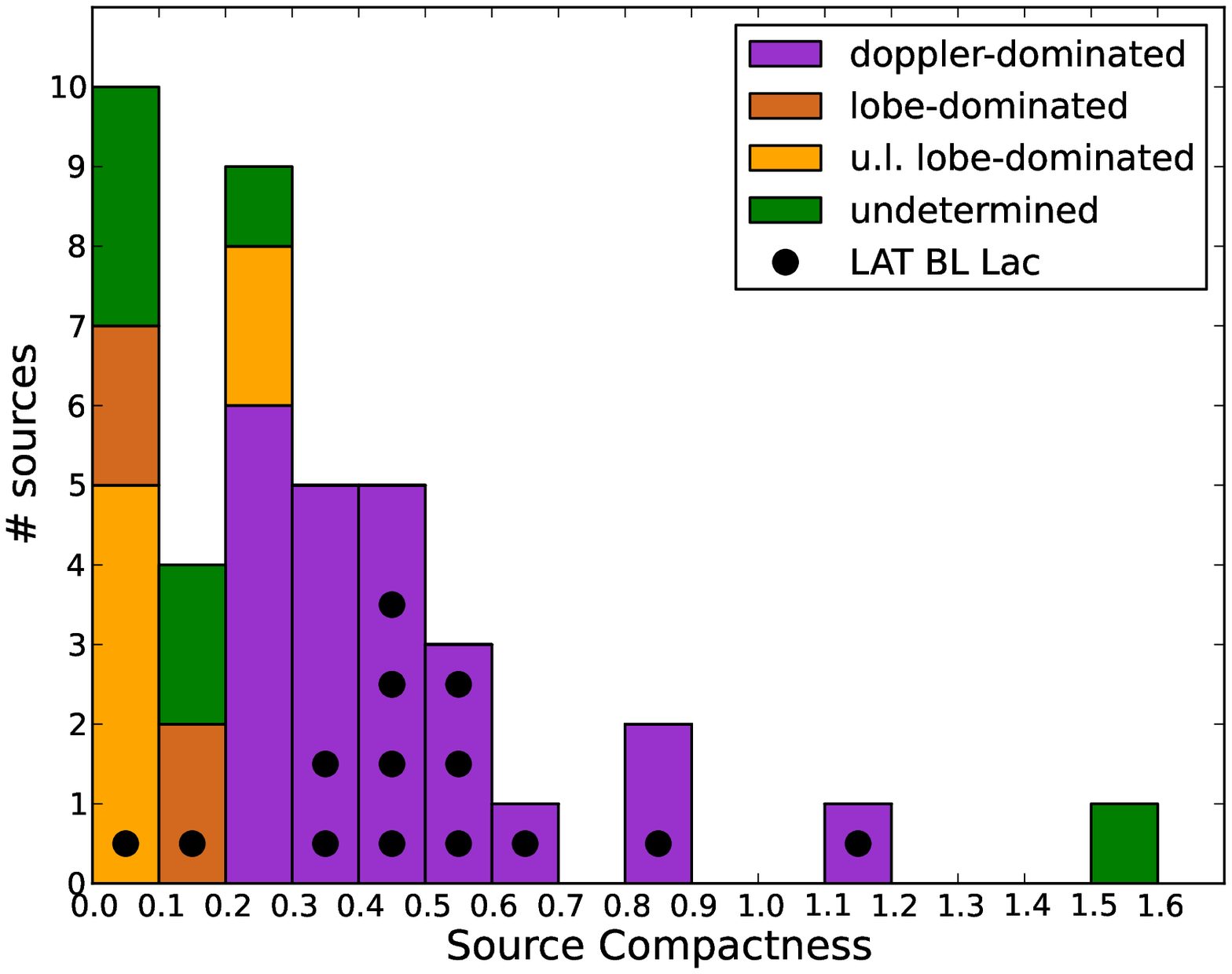}
\includegraphics[width=0.5\textwidth ]{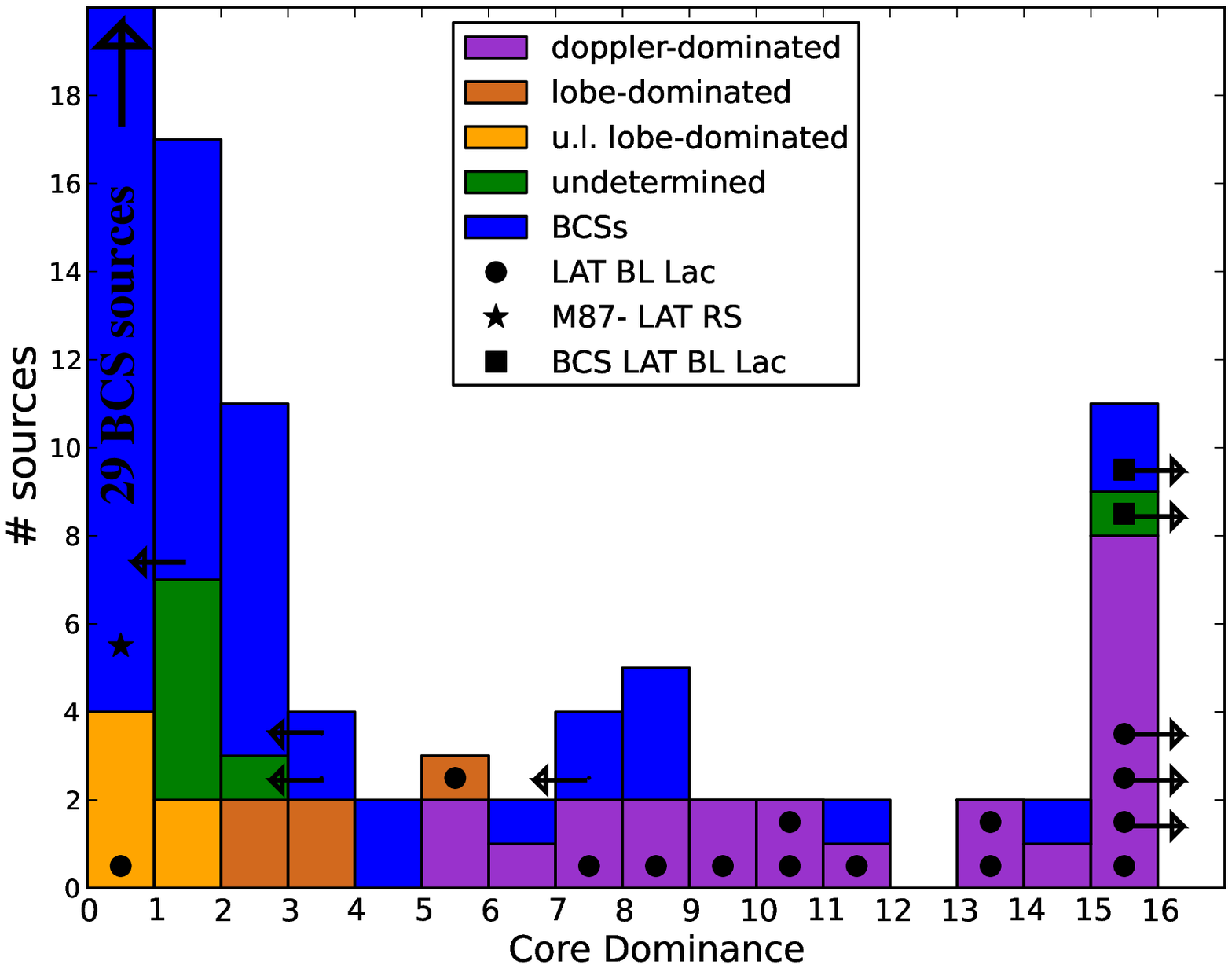}
\caption{{\bf Source compactness} ({\it on the top:}) and {\bf core dominance} ({\it on the bottom:}) {\bf distributions} for the present sample of BL Lacs using the classification given in Sect. 4.3. U.L. indicates that the corresponding values are upper limits. Gamma-ray BL Lacs are reported with black dots. In the CD histogram, we plot sources
(all radio galaxies, but 2 BL Lacs) of the Bologna Complete Sample (Liuzzo et al. 2009). Two different symbols are used to represent the three gamma-ray BCS sources: M87 (star) and 2 BL Lacs (square). Arrows indicate upper and lower limit CD values.}.\label{fig_SC}
\end{figure}

A brief description of each source is reported in the Appendix. 
We discuss here the general properties of the sample. 
We evaluated, for all objects, the source
compactness (SC) value defined as the ratio of the 8 GHz VLBA and the
NVSS total powers. SC values are reported in Table \ref{tab_SC}, while in Fig. \ref{fig_SC}, we plotted the SC distribution for our sample. 
We show SC values for both detected and non-detected sources,
assuming for the latter upper flux density limits as explained in
Sect. 3. 
Most of our sources show a parsec-scale flux density significantly lower than
expected from the kiloparsec-scale flux densities estimated
from the NVSS
and/or FIRST radio surveys. In most cases, this difference cannot be
due to the radio spectrum or to variability, and it
suggests the presence of a sub-kiloparsec radio structure lost in VLBA data
because of the lack of short baselines.

Moreover, we estimated the core dominance (CD) of each BL Lac of the
sample as defined in Liuzzo et al. (2009).  The core dominance is the
ratio between the observed core radio power and the expected core
radio power estimated from the unbeamed total radio power at low
frequency, according to the relation given in Giovannini et
al. (1994).  The core dominance is a strong indication of the Doppler
factor for each source and it allows for us to estimate constraints for the
jet velocity and orientation.  Since in most of our sample a
measure of the total radio power at low frequency is unavailable, we
estimated the total radio power at 408 MHz extrapolating the NVSS
total radio power using an average spectral index = 0.7.  For the core
radio power , we used the radio power at 8 GHz.
In Fig. \ref{fig_SC}, we show the CD distribution for
our sample. We added also the CD of radio sources (radio
galaxies and 2 BL Lacs) from the Bologna Complete Sample (BCS, see
Liuzzo et al. 2009) for a comparison. As expected, the distribution of
BL Lac sources peaks at higher core dominance according to unified
models.
  
\begin{figure} [h!]
\centering
\includegraphics[width=0.53\textwidth, angle=0]{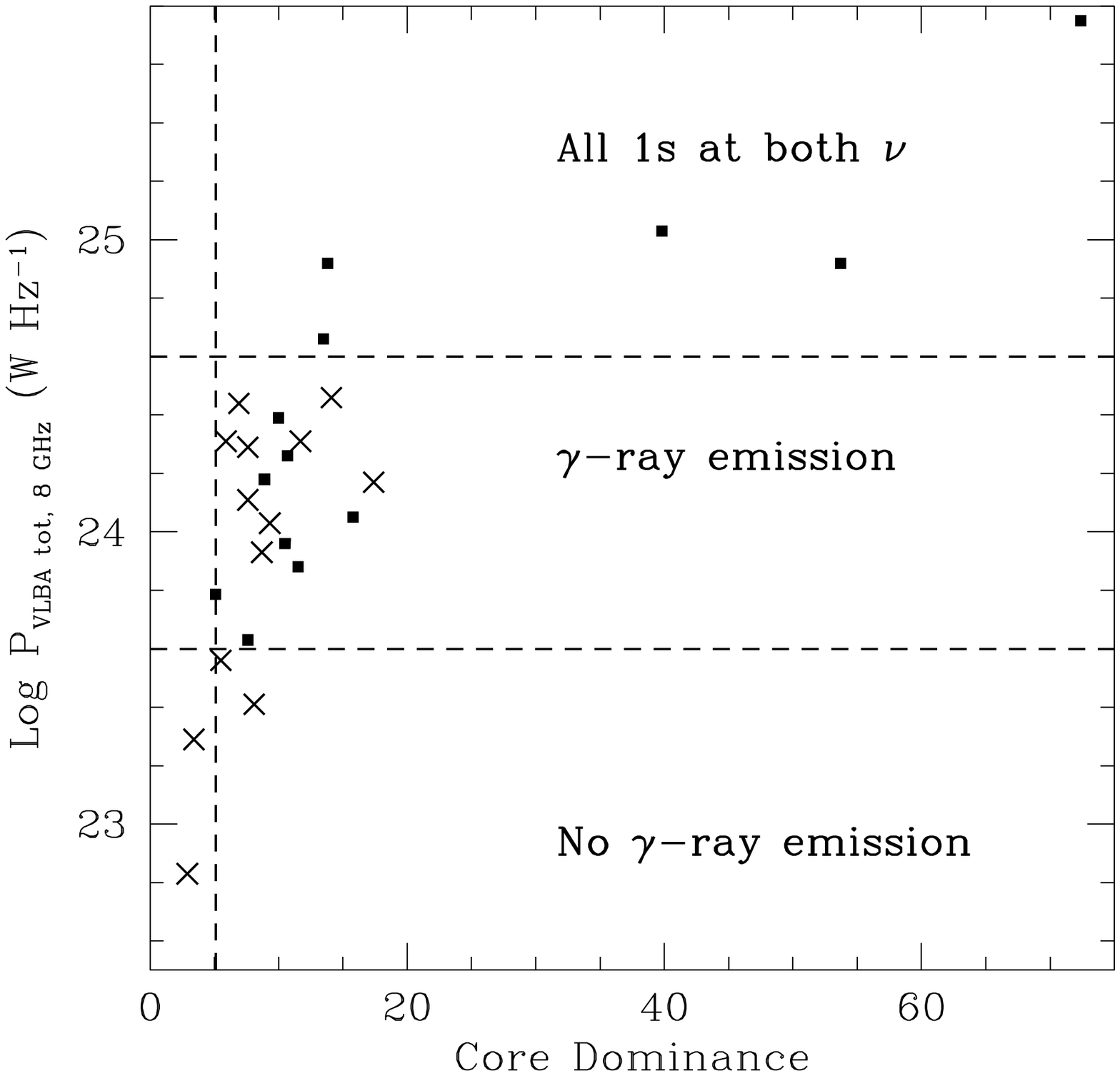}
\caption{{\bf Core Dominance} vs {\bf Log P$_{VLBA, 8 GHz}$} for LAT BL Lacs (filled squares) and non LAT BL Lacs (crosses). }\label{scpr}
\end{figure}

In Fig. \ref{scpr}, we investigated the relation between the CD parameter with the total VLBA powers.
As expected, bright BL Lacs have high core dominance, and a dependence with their high energy emission is present. We better discuss this relation in Sect. \ref{gamma}.

Looking at the derived values of source compactness and core
dominance, together with the source morphology and radio spectrum, the
BL Lac objects of our sample could be divided as follows:
\begin{itemize}
\item  {\bf Doppler Dominated (DD) sources:}
Among the 42 BL Lacs, we have 23 sources in which the emission is dominated by
the central core. They are characterized by a core dominance always larger than 5.5
 suggesting that they
are relativistic Doppler boosted objects. 
We note that a core dominance larger than 5.5 implies a jet velocity 
$\sim$ 0.8c or faster and a jet orientation with respect to the line of sight
smaller than 30$^\circ$.
 
Most of these sources display a core dominant structure and a
one-sided jet, confirming the importance of relativistic effects. The
SC is generally high, even if in some case the missing flux density is
important, suggesting the presence of a structure more complex than the
detected one-sided jet.
For comparison, we note that radio galaxies of the 
BCS (Liuzzo et al. 2009) show in most cases a core dominance smaller than 
4 and in many cases smaller than 1 (Fig. \ref{fig_SC}).

\item {\bf Lobe Dominated (LD) sources:} In 11 cases, the VLBI total
  flux densities are a small fraction of the kiloparsec-scale ones, with core
  dominance $<$ 5.5.  This group includes both undetected and detected
  sources in our VLBA maps.  In a few sources, the radio spectrum is quite
  steep, suggesting that they are dominated by an extended steep
  spectrum structure despite their classification as BL Lac objects.

\item {\bf Undetermined (U) objects:}
There are 8 sources that are not detected in our VLBA images. Their 
arcsec flux density is low, therefore we cannot establish if the lack of the detection is due to the absence of a compact core or to a sensitivity problem. Deeper observations will be necessary to 
investigate their parsec-scale properties.

\end{itemize}

\section{Gamma-ray properties.} \label{gamma}

\begin{figure} [t!]
\centering
\includegraphics[width=0.5\textwidth, angle=0]{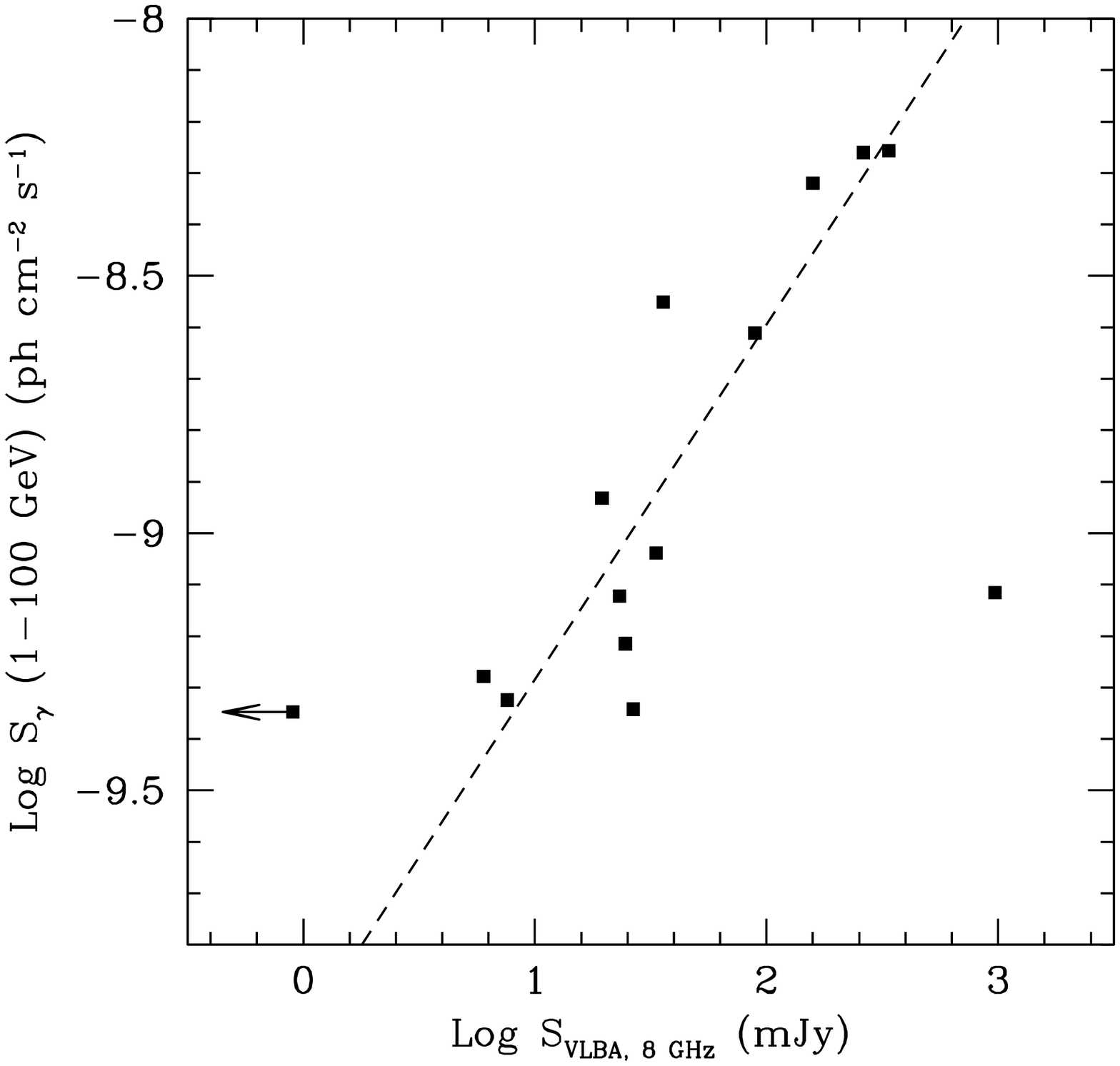}
\caption{ {\bf LogS$_{{\bf VLBA, 8 GHz}}$ vs LogS$_{{\bf \gamma}}$} for the 14 LAT BL Lacs in our sample. The arrow indicates the VLBA 8 GHz flux density upper limit for J1534+3715 which is non detected in our data (see Sect. \ref{gamma}).}\label{fig_rg}
\end{figure}

\begin{figure} [t!]
\centering
\includegraphics[width=0.5\textwidth, angle=0]{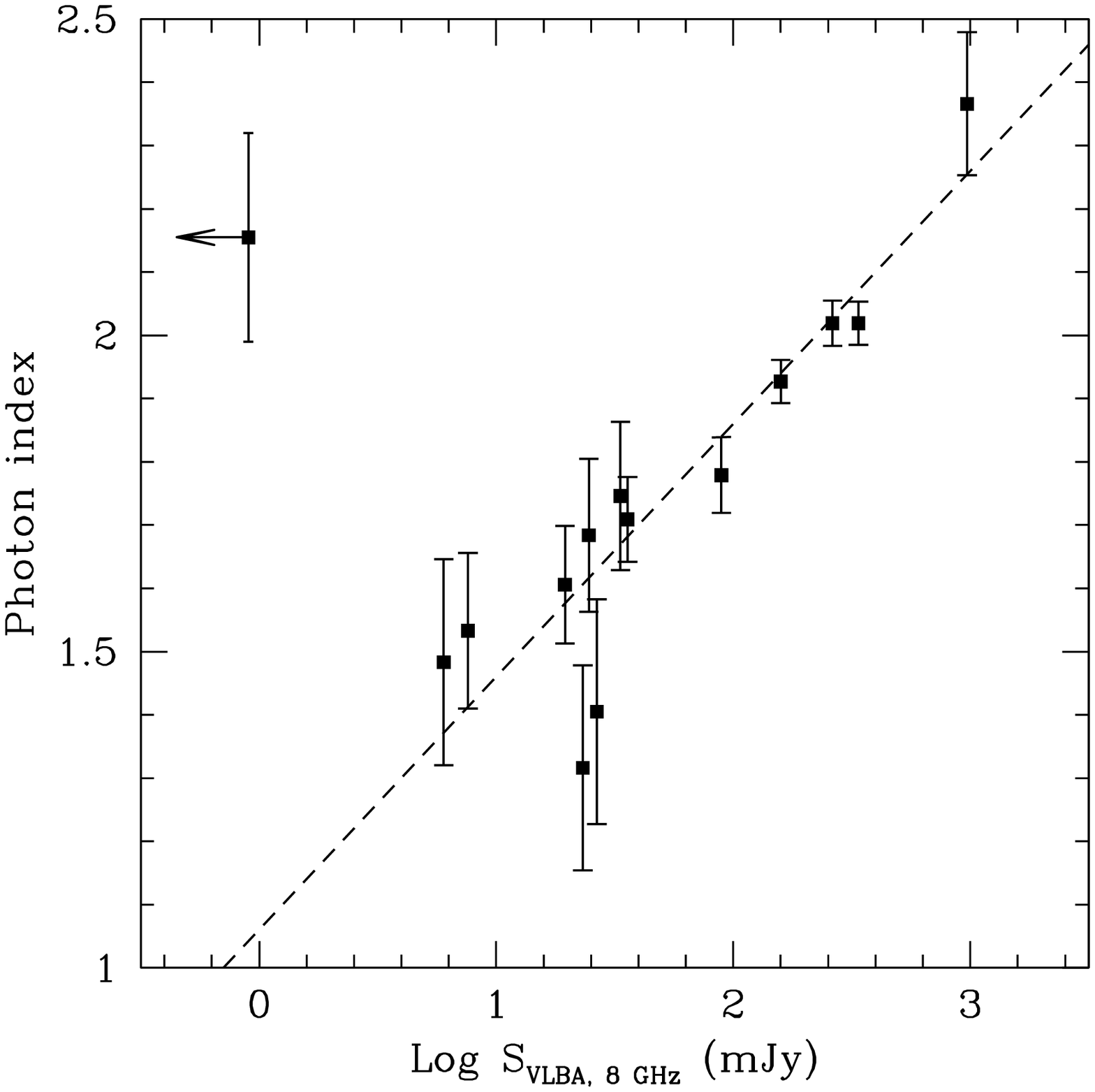}
\caption{ {\bf LogS$_{{\bf VLBA, 8 GHz}}$ vs Photon index} for the 14 LAT BL Lacs in our sample. The arrow indicates the VLBA 8 GHz flux density upper limit for J1534+3715 which is non detected in our data (see Sect. \ref{gamma}).}\label{fig_ri}
\end{figure}

We searched for gamma-ray counterparts of our BL Lac sample: 14/42
show high energy emission in the Second LAT AGN catalog (2LAC,
Ackermann et al. 2011a, 2011b).  We call them LAT BL Lacs. We collected
gamma-ray flux measurements, together with all their available
multi-frequency information in Table \ref{tab_multi}.  Among these LAT
BL Lacs, 2 (J0847+1133 and J1534+3715) are classified as LD objects,
the others are DD objects (Table \ref{tab_SC}). Moreover, the LD BL
Lac J1534+3715 is the only LAT sources not detected by our VLBA
observations: we have not considered it in our radio/gamma-ray
correlation's estimates (Fig.s \ref{scpr}$-$\ref{fig_ri}).  Comparing their parsec scale properties, LAT BL Lacs are the most luminous at mas scale respect to the non LAT objects. In particular, all LAT BL Lacs with Log P$_{VLBA, 8 GHz}$(W Hz$^{-1}$)$\geq$24.6 present resolved morphologies both at 8 and 15 GHz. On the other hand, objects with Log P$_{VLBA, 8 GHz}\leq$23.6 do not show $\gamma$ -ray emission, as sources with core dominance lower than 5.1 (Fig.\ref{scpr}). LAT BL Lac objects have also higher on average values
of the source compactness and core dominance with respect to the non LAT ones (Fig. \ref{fig_SC}). The correlation between CD parameter and total VLBA power (Fig.\ref{scpr}) is much stronger (Pearson index correlation {\it r} = 0.84) for LAT objects than for non LAT sources ({\it r} = 0.56). All these characteristics confirm the idea that LAT BL Lacs are
dominated by Doppler boosting effects (note that the two BL Lacs of the BCS sample: Mrk 421 and Mrk 501, are LAT sources and are in agreement with the properties of LAT BL Lacs in our sample).

In Fig. \ref{fig_rg}, we plot the total 8 GHz-VLBA radio flux density (S$_{VLBA, 8 GHz}$) versus
the gamma-ray fluxes: sources show a good correlation with the
exception of J1419+5423, which has a very high total radio flux density
(S$_{VLBA, 8 GHz}\sim$ 969.5 mJy), but a modest gamma-ray flux
(S$_{\gamma}\sim$7.66 10$^{-10}$ ph cm$^{-2}$ s$^{-1}$ ).  This source also
has SC=1.19, and very high core dominance, with the VLBA total flux density
greater than the kiloparsec-scale total flux density. Moreover, it is an extremely variable source (see Appendix) observed during an outburst, seen in both the Owens Valley Radio Observatory (OVRO, Richards et al. 2011) and MOJAVE monitoring programs. The likely reason for this source being an outlier on the S$_{VLBA, 8 GHz}$ - gamma-ray flux plot is its steep spectral index value in the high energy band implying that quite some gamma-ray flux is missed if a high lower cutoff (1 GeV) is selected. Thus, the photon flux at energy range [0.1-100] GeV is at the level of 10$^{-8}$ ph cm$^{-2}$ s$^{-1}$, available from the history flux extension of the 2FGL (Ackermann et al. 2011a, 2011b).

We estimated the Pearson index correlation {\it r} for Log S$_{VLBA, 8 GHz}$ and Log S$_{\gamma}$ relation: we found 
{\it r}=0.90, which means that a linear correlation is present between these quantities. We performed a linear 
fit (Log S$_{\gamma}$ = $a$ Log S$_{VLBA, 8 GHz}$ + $b$) finding $a=0.69$ and 
$b=-9.97$ (see dashed line in Fig. \ref{fig_rg}). The correlation is in agreement with results of Abdo et al. (2010) for BL Lac objects.

Fig. \ref{fig_ri} reports 8 GHz-VLBA radio total flux densities S$_{VLBA, 8
  GHz}$ vs photon indices ($\Gamma$).  The Pearson index for the
correlation involving sources detected in radio and gamma-ray bands is
{\it r}=0.90, and the linear fit (LogS$_{VLBA, 8 GHz}$=
$a\Gamma$ + $b$) gives $a=0.50$ and $b=1.06$. This result is expected
as a consequence of the presence of the 8 GHz-VLBA radio/gamma-ray flux
correlation (Abdo et al. 2010).

\section{Discussion and Conclusion} \label{discussion}

We selected a sample of nearby BL Lacs with no constraint on their
radio or high energy emission. Our aim is to investigate their
nuclear properties given that one of the surprising
results of {\it Fermi} mission is that BL Lac objects, and not FSRQs,
are the most common gamma-ray emitters in the sky. In this first
paper, we present, for the entire sample, 8 and 15 GHz VLBA images which are also available at http://www.ira.inaf.it/progetti/bllacs/ .

From VLBI observations, we found that most of the BL Lacs in our sample
(23/42) show a high core dominance ($>$ 5.5), in agreement with
expectations from unified models. We classified these sources as
Doppler Dominated objects. They show a strong evidence of an active
nuclear region affected by Doppler boosting effects, confirmed by the
{\it Fermi} detection of 12 of these sources, and by the correlation
between the radio flux density and gamma-ray flux.
However, in many sources (14/23) the SC is lower than 0.5, suggesting
that in many BL Lac sources a significant sub-kiloparsec structure is
present.

Of the remaining sources, 11/42 were classified as Lobe Dominated
sources.  In these BL Lacs, the core dominance is lower than 5.5 and 
there is no evidence of a relativistic nuclear structure or 
strong boosting effects.  This fraction is higher than
expected in a sample of BL Lac sources.  
Among these objects, 2 have been detected by {\it
  Fermi}: J0847+1133 and J1534+3715. In the case of J0847+1133, we
have a VLBI detection, and the core dominance is 5.1, therefore we can
consider it as a source with intermediate
properties. J1534+3715 was not detected in our VLBA observations and
we suggest that it could be a misaligned object. Its core dominance
is lower than 0.8. We note that, among the radio galaxies of the BCS
sample (Liuzzo et al. 2009), only M87 shows gamma-ray emission and its
core dominance is 0.9. We were unable to classify the other 8 objects
in our sample because of their low flux densities.

From a comparison of the core dominance distribution obtained
for our BL Lac sample and for the radio galaxies of the BCS sample
(Liuzzo et al. 2009), we can suggest the presence of two populations:
beamed (core dominance $\sim$ 5.5 or higher) and un-beamed objects
(core dominance lower than $\sim$ 5.5).  We suggest that BL Lacs
classified as LD objects could be intermediate or misclassified
sources, but more multi-wavelength data are necessary to investigate
this point.  The origin of gamma-ray emission in sources in the two
classes could be different: strongly related to the relativistic jets
in DD, but not in LD. More data are necessary to better investigate
this scenario and the radio structure of J1534+3715 needs to be
confirmed.

In some DD sources (6/23) and all LD objects, the SC is lower than
0.3, suggesting the presence of a relevant sub-kiloparsec
structure. The morphology and the properties of this extended radio
emitting region will be analyzed with future observations that could
help in understanding the difference between DD and LD BL Lacs and the
connection with the high energy emission.

\begin{acknowledgements}
This work was supported by contributions of European Union, Valle
D$'$Aosta Region and the Italian Minister for Work and Welfare. 
We acknowledge the financial contribution from grant PRIN-INAF-2011.
The authors wish to thank Prof. Enrico Massaro for his contribution in the
early stages of this project. This research has made use of the
NASA/IPAC Extragalactic Data Base (NED) which is operated by the JPL,
California Institute of Technology, under contract with the National
Aeronautics and Space Administration.  GBT thanks NASA for support under 
FERMI grant NNX12AO75G.
\end{acknowledgements}

\clearpage

\begin{table*}[!htbp]
\caption{ \small{ {\bf Log of the new VLBA observations} in X and U bands.}} \label{tab_oss}
\centering
\begin{tabular}{|l|l|l|l|l|l|}
\hline
Date       & Fringe &  Name &  Phase        & RA$_{core}$(J2000)  & Dec$_{core}$(J2000) \\
mm/dd/yy   & finder &       & calibrator    & h m s               & d m s              \\
\hline
11/15/2011 (X)      & 4C39.25    & {\bf J0751+1730}    & -                 & 07 51 25.08  & +17 30 51.1   \\
10/10/2011 (U)     
                    &            & {\bf J0751+2913}    & -                 & 07 51 09.57  & +29 13 35.5  \\
                    &            &{\bf J0753+2921}     & J0746+2734        & 07 53 24.61  & +29 21 21.9   \\
                    &            & {\bf J0810+4911}    & J0808+4950        & 08 10 54.60  & +49 11 03.7  \\
                    &            & {\bf J0903+4055}, * & J0904+4238        & 09 03 14.707 & +40 55 59.85  \\
                    &            &{\bf J1012+3932}     & J1023+3048        & 10 12 58.37  & +39 32 39.0      \\
                    &            & {\bf J1022+5124}    & J1041+523A        & 10 22 12.62  & +51 24 00.3      \\
\hline
 09/01/2010         & 3C345 (X)  & {\bf J1058+5628}    & -                 & 10 58 37.73  & +56 28 11.2   \\ 
                    & 4C39.25(U) & {\bf J1145-0340}    & J1136-0330        & 11 45 35.11  & -03 40 01.7   \\
                    &            & {\bf J1156+4238}, * & J1146+3958        & 11 56 46.563 & +42 38 07.50         \\
                    &            & {\bf J1215+0732}, * & J1222+0413 (U)    & 12 15 10.977 & +07 32 04.67        \\ 
                    &            & {\bf J1221+2813}    & -                 & 12 21 31.69  & +28 13 58.5                  \\
\hline
 08/29/2011         & 3C345      & {\bf J1231+6414}    & -                 & 12 31 31.40  & +64 14 18.3    \\ 
                    &            & {\bf J1257+2412}    & J1321+2216        & 12 57 31.93  & +24 12 40.1  \\
                    &            & {\bf J1427+3908}    & J1419+3821        & 14 27 45.92  & +39 08 32.3  \\
                    &            & {\bf J1510+3335}    & J1522+3144        & 15 10 41.18  & +33 35 04.5   \\
                    &            & {\bf J1534+3715}    & J1522+3144        & 15 34 47.21  & +37 15 54.6   \\
                    &            & {\bf J1604+3345}    & J1613+3412        & 16 04 46.52  & +33 45 21.8   \\
                    &            & {\bf J1647+2909}    & -                 & 16 47 26.88  & +29 09 49.6     \\
\hline
 02/25/2010         & 4C39.25    & {\bf J0809+5218}    & -                 & 08 09 49.19  & +52 18 58.2     \\
                    &            & {\bf J0916+5238}    & -                 & 09 16 51.94  & +52 38 28.5      \\
                    &            & {\bf J0930+4950}, * & J0937+5008        & 09 30 37.574 & +49 50 25.55      \\
                    &            & {\bf J1053+4929}    & -                 & 10 53 44.10  & +49 29 55.9        \\
                    &            & {\bf J1120+4212}, * & J1126+4516        & 11 20 48.062 & +42 12 12.46       \\
                    &            & {\bf J1136+6737}, * & J1143+6633        & 11 36 30.086 & +67 37 04.34 \\
                    &            & {\bf J1217+3007}    &-                  & 12 17 52.08  & +30 07 00.5\\
                    &            & {\bf J1221+3010}    &-                  & 12 21 21.94  & +30 10 37.1\\
                    &            & {\bf J1419+5423}    &-                  & 14 19 46.60  & +54 23 14.8\\
                    &            & {\bf J1427+5409}, * & J1429+5406        & 14 27 30.279 & +54 09 23.71\\
                    &            & {\bf J1436+5639}, * & J1429+5406        & 14 36 57.719 & +56 39 24.86\\
\hline
10/17/2009         & 3C 345      & {\bf J0754+3910}    &-                  & 07 54 37.08  & +39 10 47.6\\
                   &             & {\bf J0809+3455}    &-                  & 08 09 38.87   & +34 55 37.1\\
                   &             & {\bf J0847+1133}    & J0851+0845 (U)    & 08 47 12.94   & +11 33 50.1\\
                   &             & {\bf J0850+3455}    & J0850+3747 (U)    & 08 50 36.18   & +34 55 22.8\\
                   &             & {\bf J1201-0007}    &-                  & 12 01 06.22   & -00 07 01.8\\
                   &             & {\bf J1201-0011}    & J1201+0028        & 12 01 43.66   & -00 11 14.0\\
                   &             & {\bf J1221+0821}    &-                  & 12 21 32.06   & +08 21 44.1\\
                   &             & {\bf J1253+0326}    &-                  & 12 53 47.03   & +03 26 30.4 \\
                   &             & {\bf J1341+3959}, * & J1359+4011        & 13 41 05.107  & +39 59 45.42\\
                   &             & {\bf J1428+4240}    &-                  & 14 28 32.62   & +42 40 21.2\\
                   &             & {\bf J1442+1200}    &-                  & 14 42 48.27   & +12 00 40.3\\
                   &             & {\bf J1516+2918}    &-                  & 15 16 41.60   & +29 18 09.5\\
\hline
\end{tabular}
\tablefoot{Col. 1: date of the VLBA observations; col. 2: fringe finder of the scheduling block; col. 3 name of the observed target (stars indicate sources for which we obtained VLBA absolute core positions); col. 4: phase calibrator if used (when the phase calibrator is followed by a parenthesis with a 
specified frequency (X and/or U band), it means that phase referencing was performed only at that frequency); col.s 5 and 6: RA and DEC in J2000 of the core.}
\end{table*}

\begin{table*}[ht!]
\caption{{\bf Images parameters} of 8 GHz and 15 GHz observations.}\label{tab_vlbaImage}
\centering
\begin{tabular}{|l|lcc|lcc|}
\hline
 &  & {8 GHz}  &  &  & 15 GHz  &   \\
\cline{2-7}
Name        & HPBW                   & rms                   & S$_{peak}$                &  HPBW                  & rms                    & S$_{peak}$\\
IAU         & mas$\times$mas, deg    & mJy beam$^{-1}$  & mJy beam$^{-1}$    & mas$\times$mas, deg    & mJy beam$^{-1}$  & mJy beam$^{-1}$ \\
\hline
J0751+1730, $^{1, 2}$ & -                      &-       &-      & -                      & -        &-    \\
J0751+2913, $^{3}$    &  -                     & -      &-      & 1.2$\times$0.8,-23.2   & 0.6      &-     \\
J0753+2921            & 1.8$\times$1.1,  28.3  & 0.3    &-      & 0.7$\times$0.5,  4.5   & 0.4      &-     \\
J0754+3910            & 1.7$\times$0.9, -33.0  & 0.2    & 23.3  & 1.3$\times$0.5,-30.8   & 0.8      & 13.7  \\
J0809+3455            & 2.0$\times$1.0, -32.2  & 0.2    & 49.9  & 1.2$\times$0.8,-29.9   & 0.4      & 45.8 \\
J0809+5218            & 1.8$\times$1.6, -12.8  & 0.2    & 61.5  & 2.1$\times$1.6, 39.7   & 0.2      & 61.8 \\
J0810+4911            & 1.0$\times$0.8, -0.9   & 0.2    &-      & 0.5$\times$0.5,-28.9   & 0.6      &-     \\
J0847+1133, $^{1}$    & ----                   &-       & -     & 2.1$\times$1.9,-31.1   & 1.1      & -    \\
J0850+3455            & 2.3$\times$1.3, -29.8  & 0.1    & 22.1  & 1.7$\times$1.7,-86.2   & 0.2      & 21.1 \\
J0903+4055            & 1.2$\times$1.0, -30.1  & 0.2    & 8.4   & 2.0$\times$1.5, 20.1   & 0.3      & 5.9  \\
J0916+5238            & 2.1$\times$1.6, 13.7   & 0.3    & 25.3  & 3.2$\times$2.2, 7.4    & 0.6      & 16.8 \\
J0930+4950            & 2.2$\times$1.6, -24.5  & 0.3    & 5.0   & 1.6$\times$1.6, 37.1   & 0.2      & 4.4  \\
J1012+3932            & 1.2$\times$0.9, 38.5   & 0.3    &-      & 1.0$\times$0.5, -46.9  & 0.5      &-     \\
J1022+5124            & 1.2$\times$0.9,-46.5   & 0.2    &-      & 1.0$\times$0.4, -59.7  & 0.6      &-     \\
J1053+4929            & 1.6$\times$1.5, -42.2  & 0.2    & 25.7  & 2.4$\times$1.8, -78.5  & 0.5      & 23.2 \\
J1058+5628            & 1.9$\times$1.7,-78.7   & 0.1    & 129.3 & 0.8$\times$0.7, 84.8   & 0.3      & 95.0 \\
J1120+4212            & 2.1$\times$1.3, -4.5   & 0.2    & 16.2  & 1.9$\times$1.4, 28.6   & 0.2      & 11.6  \\
J1136+6737            & 2.2$\times$1.1,43.4    & 0.3    & 22.8  & 1.4$\times$0.9, -58.6  & 0.2      & 16.0 \\
J1145-0340            & 1.8$\times$0.9, 3.7    & 0.4    &-      & 1.0$\times$0.4, -0.2   & 0.3      &-     \\
J1156+4238            & 1.6$\times$1.5, -26.0  & 0.2    & 2.2   & 1.0$\times$0.9, 6.5    & 0.1      & 1.3  \\
J1201-0007, $^{1, 2}$ &-            &          & -              & -                      &-         &-     \\
J1201-0011            & 3.4$\times$1.5, -22.4  & 0.2    &-      & 4.5$\times$0.8, -27.4  & 1.0      &-     \\
J1215+0732            & 1.9$\times$1.4, -5.7   & 0.3    & 27.3  & 1.3$\times$0.8, -4.4   & 0.3      & 13.7\\
J1217+3007            & 2.8$\times$2.8, -37.6  & 0.2    & 232.9 & 1.1$\times$0.9, -7.3   & 0.3      & 187.0 \\
J1221+0821, $^{2}$    & 3.3$\times$1.6, -12.0  & 0.1    & 15.4  &-                       &-         &-     \\
J1221+2813            & 1.8$\times$1.4, -5.73  & 0.2    & 200.6 & 1.0$\times$0.7, -11.1  & 1.0      & 168.3  \\
J1221+3010            & 1.4$\times$0.9, -3.8   & 0.2    & 27.9  & 3.0$\times$0.6,-30.6   & 0.2      & 28.2 \\
J1231+6414, $^{1, 2}$ & 1.8$\times$1.5, 45.8   & 0.1    & 28.8  & -                      & -        & -  \\
J1253+0326            & 3.4$\times$1.2,-19.5   & 0.2    & 34.3  & 1.7$\times$0.5 -20.3   & 0.4      & 23.9 \\
J1257+2412            & 1.4$\times$0.9,  0.0   & 0.3    & -     & 0.8$\times$0.5, -4.4   & 0.3      &-      \\ 
J1341+3959            & 3.4$\times$2.0, -62.5  & 0.2    & 5.6   & 1.0$\times$0.5, -22.4  & 0.3      & 4.8  \\
J1419+5423            & 2.4$\times$2.3, 20.9   & 0.3    & 802.4 & 1.1$\times$0.8, -8.6   & 1.2      & 556.6\\
J1427+3908            & 1.3$\times$1.0, -13.4  & 0.2    &-      & 0.6$\times$0.5, 5.43   & 0.3      & -    \\
J1427+5409 $^{1}$     & -                      &-       &-      & 0.5$\times$0.4,-7.8    & 0.4      & 5.1  \\
J1428+4240            & 2.1$\times$1.0, -30.3  & 0.2    & 21.4  & 1.5$\times$0.4, -19.9  & 1.8      & 13.3 \\
J1436+5639            & 2.6$\times$1.5, -50.4  & 0.1    & 4.6   & 0.6$\times$0.5, 10.8   & 0.2      & 3.6  \\
J1442+1200            & 2.8$\times$1.0, -22.8  & 0.2    & 25.4  & 3.3$\times$0.8,-20.69  & 0.8      & 19.5 \\ 
J1510+3335            & 1.2$\times$0.9, -5.7   & 0.3    & -     & 0.6$\times$0.5, -6.4   & 0.3      & -     \\
J1516+2918            & 2.4$\times$1.0, -24.6  & 0.2    & 27.7  & 1.8$\times$0.7, -26.4  & 0.4      & 25.8 \\
J1534+3715            & 1.2$\times$1.0,-8.0    & 0.3    & -     & 0.6$\times$0.5, -1.1   & 0.3      &-     \\
J1604+3345            & 1.5$\times$1.0, -33.2  & 0.2    &-      & 0.9$\times$0.5, -30.9  & 0.4      &-      \\
J1647+2909            & 2.1$\times$1.2, -26.1  & 0.3    & 50.1  & 1.2$\times$0.9, -22.1  & 0.8      & 46.5 \\
\hline
\end{tabular}
\tablefoot{In col. 1, we report the source names: $^{(1,2)}$ is for undetected sources which were not observed in phase referencing mode at 8 and 15 GHz, respectively; $^{(3)}$ indicates the source that is not observed during all observational blocks in the project. We also list the half-power beam width (HPBW, col.s 2 and 5), the rms (col.s 3 and 5), and the flux density peaks (col.s 4 and 7) for images at 8 and 15 GHz, respectively.}
\end{table*}

\begin{table*}[!htbp]
\footnotesize
\caption{ {\bf Parsec-scale morphologies} and {\bf spectral indexes}.} \label{tab_parsecMorphology}
\centering
\begin{tabular}{|cccrr|}
\hline
Name  & pc scale &structure& $\alpha_{core}$  &   $\alpha_{tot}$ \\
IAU & (8 GHz)   & (15 GHz) &                   & \\
\hline
J0754+3910 & 1s & p   & 0.37$\pm 0.04$  & 0.66$\pm 0.04$ \\
J0809+3455 & 1s & 1s  & 0.00$\pm 0.03$  & 0.27$\pm 0.03$ \\
J0809+5218 & 1s & 1s  & 0.28$\pm 0.03$  & 0.06$\pm 0.03$\\
J0850+3455 & 1s & 1s  & 0.14$\pm 0.03$  & 0.09$\pm 0.03$             \\
J0903+4055 & p  & p   & 0.59$\pm 0.07$  & 0.59$\pm 0.07$ \\ 
J0916+5238 & 1s & p   & 0.21$\pm 0.04$  & 0.53$\pm 0.04$\\
J0930+4950 & p  & p   & 0.60$\pm 0.08$  & 0.60$\pm 0.08$  \\
J1053+4929 & 1s & p   & -0.23$\pm 0.04$ & 0.58$\pm 0.04$ \\
J1058+5628 & 1s & 1s  & 0.03$\pm 0.03$  & 0.33$\pm0.03$ \\
J1120+4212 & 1s & 1s  & 0.1$\pm 0.04$   & 0.54$\pm 0.04$  \\
J1136+6737 & p  & 1s  & 0.29$\pm 0.03$  & 0.25$\pm 0.04$\\
J1156+4238 & p  & p   & -1.16$\pm 0.12$ & -1.16$\pm 0.12$   \\
J1215+0732 & 1s & p   & 0.05$\pm 0.04$  & 0.91$\pm 0.04$\\ 
J1217+3007 & 1s & 1s  & 0.11$\pm 0.03$  & 0.30$\pm0.03$\\
J1221+0821 & p  & -   & -               & -  \\
J1221+2813 & 1s & 1s  & 0.00$\pm0.04$   & 0.25$\pm 0.04$\\
J1221+3010 & 1s & p   & 0.17$\pm0.05$   & 0.17$\pm 0.05$\\
J1231+6414 & 1s & --   & 0.80$\pm 0.05$  & 1.53$\pm 0.04$\\
J1253+0326 & 1s & p   & 0.14$\pm 0.07$  & 0.47$\pm 0.07$ \\ 
J1341+3959 & p  & p   & 0.13$\pm 0.03$  & 0.13$\pm 0.03$ \\
J1419+5423 & 1s & 1s  & -0.15$\pm 0.3$  & 0.27$\pm 0.74$ \\
J1427+5409 & -  & p   & -               & -\\
J1428+4240 & p  & 1s  & 0.05$\pm 0.04$  & -0.07$\pm 0.05$ \\
J1436+5639 & p  & p   & 0.83$\pm 0.08$  & 0.83$\pm 0.10$ \\\
J1442+1200 & 1s & p   & 0.04$\pm 0.04$  & 0.32$\pm 0.04$\\
J1516+2918 & 1s & p   & -0.03$\pm 0.03$ & 0.09$\pm 0.03$ \\
J1647+2909 & 1s & 1s  & 0.03$\pm 0.03$  & 0.19 $\pm 0.03$  \\
\hline
\end{tabular}
\tablefoot{Col. 1: source name; col.s 2 and 3: parsec-scale structure at 8 and 15 GHz respectively, ``p'' indicates point-like morphology and ``1s'' one-sided morphologies; col.s 4 and 5: core and total spectral indexes from 8 and 15 GHz VLBA measurements (see Sect. \ref{sect_spectral}).}
\end{table*}

\addtocounter{table}{1}

\begin{table*}[!h]
\caption{{\bf Kiloparsec and parsec-scale radio properties} of the whole sample.} \label{tab_SC}
\centering
\begin{tabular}{|ccrrrrcrc|}
\hline
Name   & Log$P_{NVSS}$     & Log$P_{core, 8.4 GHz}$& Log$P_{tot, 8.4 GHz}$& Log$P_{core, 15 GHz}$& Log$P_{tot, 15 GHz}$& CD & SC  & Notes\\
IAU    &   W Hz$^{-1}$ &  W Hz$^{-1}$     & W Hz$^{-1}$     &  W Hz$^{-1}$    & W Hz$^{-1}$    &                    &     &\\
\hline 
J0751+1730 & 23.9  & $\leq$24.12 & $\leq$24.12 & $\leq$24.10 & $\leq$24.1 & $\leq$15.5  & $\leq$1.6  & U\\
J0751+2913 & 24.08 & N.O.        & N.O.        & $\leq$23.24 & $\leq$23.24 & -          &-           & U\\
J0753+2921 & 23.41 & $\leq$22.76 & $\leq$22.76 & $\leq$22.89 & $\leq$22.89 & $\leq$2.6  & $\leq$0.23 & U\\ 
J0754+3910 & 24.10 & 23.69       & 23.41       & 23.59       & 23.59       & 8.1        & 0.45       & DD \\
J0809+3455 & 24.56 & 23.79       & 24.03       & 23.79       & 23.92       & 9.3        & 0.27       & DD\\
J0809+5218 & 24.93 & 24.42       & 24.66       & 24.35       & 24.60       & 13.5       & 0.49       & DD  \\
J0810+4911 & 23.53 & $\leq$22.23 & $\leq$22.23 & $\leq$22.76 & $\leq$22.76 & $\leq$0.6  & $\leq$0.06 & LD\\
J0847+1133 & 24.53 & 23.79 $^{*}$& 23.79 $^{*}$& $\leq$22.49 & $\leq$22.49 & 5.1        & 0.18       & LD \\
J0850+3455 & 24.25 & 24.06       & 24.17       & 24.04       & 24.15       & 17.4       & 0.83       & DD\\
J0903+4055 & 24.51 & 23.91       & 24.29       & 23.75       & 23.75       & 7.6        & 0.25       & DD\\
J0916+5238 & 25.15 & 24.35       & 24.31       & 24.19       & 24.19       & 8.4        & 0.21       & DD\\
J0930+4950 & 24.28 & 23.83       & 23.93       & 23.67       & 23.67       & 8.7        & 0.36       & DD  \\
J1012+3932 & 24.15 & $\leq$22.82 & $\leq$22.82 & $\leq$23.04 & $\leq$23.04 & $\leq$1.0  & $\leq$0.05 & LD\\ 
J1022+5124 & 23.44 & $\leq$22.47 & $\leq$22.47 & $\leq$22.95 & $\leq$22.95 & $\leq$1.3  & $\leq$0.11 & U\\
J1053+4929 & 24.50 & 23.98       & 24.18       & 24.05       & 24.05       & 8.9        & 0.51       & DD\\
J1058+5628 & 25.06 & 24.64       & 24.92       & 24.63       & 24.81       & 53.7       & 0.70       & DD\\
J1120+4212 & 23.94 & 23.74       & 23.88       & 23.45       & 23.64       & 11.5       & 0.83       & DD \\
J1136+6737 & 24.31 & 24.05       & 24.05       & 23.88       & 24.07       & 15.8       & 0.55       & DD\\
J1145-0340 & 24.12 & $\leq$22.92 & $\leq$22.92 & $\leq$22.80 & $\leq$22.80 & $\leq$1.3  & $\leq$0.06 & LD\\
J1156+4238 & 24.07 & 23.29       & 23.29       & 23.60       & 23.60       & 3.4        & 0.17       & LD\\
J1201-0007 & 24.68 & $\leq$24.02 & $\leq$24.02 & $\leq$24.00 & $\leq$24.00 & 2.7        & $\leq$0.22 & LD \\
J1201-0011 & 24.27 & $\leq$22.61 & $\leq$22.61 & $\leq$23.30 & $\leq$23.30 & $\leq$0.5  & $\leq$0.02 & LD\\
J1215+0732 & 24.80 & 23.98       & 24.31       & 23.90       & 23.90       & 5.9        & 0.26       & DD \\
J1217+3007 & 25.38 & 24.98       & 25.03       & 24.84       & 24.95       & 39.8       & 0.45       & DD   \\
J1221+0821 & 24.89 & 23.87       & 23.83       & $\leq$23.33 & $\leq$23.33 & 4.0        & 0.10       & LD \\
J1221+2813 & 25.26 & 24.64       & 24.92       & 24.64       & 24.85       & 13.8       & 0.47       & DD \\
J1221+3010 & 24.78 & 24.20       & 24.39       & 24.43       & 24.43       & 10.0       & 0.50       & DD \\
J1231+6414 & 24.60 & 24.24       & 24.46       & $\leq$23.98 & $\leq$23.98 & 14.1       & 0.63       & DD\\
J1253+0326 & 24.03 & 23.47       & 23.56       & 23.43       & 23.43       & 5.5        & 0.34       & DD\\
J1257+2412 & 23.80 & $\leq$22.64 & $\leq$22.64 & $\leq$22.64 & $\leq$22.64 & $\leq$1.1  & $\leq$0.07 & U\\
J1341+3959 & 24.80 & 23.67       &    22.83    & 23.64       & 23.64       & 2.9        & 0.07       & LD\\
J1419+5423 & 25.68 & 25.62       & 25.75       & 25.40       & 25.68       & 72.4       & 1.19       & DD \\
J1427+3908 & 23.68 & $\leq$22.61 & $\leq$22.61 & $\leq$22.79 & $\leq$22.79 & $\leq$1.2  & $\leq$0.09 & U\\
J1427+5409 & 24.08 & $\leq$23.61 & $\leq$23.61 & 23.20       & 23.20       & $\leq$2.0  & $\leq$0.34& LD\\ 
J1428+4240 & 24.37 & 23.97       & 23.96       & 23.83       & 24.00       & 10.5       & 0.40       & DD \\
J1436+5639 & 24.06 & 23.63       & 23.63       & 23.46       & 23.46       & 7.6        & 0.37       & DD\\ 
J1442+1200 & 24.69 & 24.17       & 24.26       & 24.16       & 24.16       & 10.7       & 0.39       & DD \\
J1510+3335 & 23.36 & $\leq$22.44 & $\leq$22.44 & $\leq$22.47 & $\leq$22.47 & $\leq$1.3  & $\leq$0.12 & U\\
J1516+2918 & 24.75 & 24.06       & 24.11       & 24.07       & 24.07       & 7.6        & 0.23       & DD \\
J1534+3715 & 24.02 & $\leq$22.65 & $\leq$22.65 & $\leq$22.65 & $\leq$22.65 & $\leq$0.8  & $\leq$0.04 & LD\\
J1604+3345 & 23.75 & $\leq$22.68 & $\leq$22.68 & $\leq$22.98 & $\leq$22.98 & $\leq$1.3  & $\leq$0.09 & U\\
J1647+2909 & 25.22 & 24.32       & 24.44       & 24.20       & 24.37       & 6.9        & 0.23       & DD\\
\hline
\end{tabular}
\tablefoot{In col. 1 is the source name; in col. 2 the NVSS total power; in col.s 3, 4, 5, and 6 VLBA core and total powers at 8 and 15 GHz are listed, N.O. indicates source non observed with VLBA, $^{*}$ indicates that we took the flux density at 8 GHz detected by Bourda et al. 2010, see Appendix). In col. 5, SC indicates source compactness. In Col. 6, LD is for Lobe Dominated sources, DD for Doppler Dominated, and U for Undetermined objects (see details in Sect. \ref{sect_SC}).}
\end{table*}

\onecolumn

\begin{table*}
\caption{{\bf Multiwavelength properties} of the sample.}\label{tab_multi}
\centering
\scriptsize
\begin{tabular}{|c|c|c|c|c|c|c|c|}
\hline
Name & z  & $S_{NVSS}$& $S_{FIRST}$ &$S_{VLBA-8 GHz}$& $S_{VLBA-15 GHz}$ & $F_{\gamma, (1-100 GeV)}$& 2FGL \\
       &  & (mJy)     & (mJy)        & (mJy)         & (mJy)             & ($10^{-10}$ ph cm$^{-2}$ s$^{-1}$)& name \\
\hline
\hline
J0751+1730 & 0.185 & 9.7   & 10.96  & $\leq$15.3 & $\leq$14.7 & -     & - \\
J0751+2913 & 0.194 & 12.4  & 8.92   & N.O.      & $\leq$1.8  & -      & - \\
J0753+2921 & 0.161 & 4.0   & 4.49   & $\leq$0.9 & $\leq$1.2  & -      & - \\
J0754+3910 & 0.096 & 57.8   & 49.26  & 27.0      & 17.8      & -      & - \\
J0809+3455 & 0.083 & 227.4 & 169.12 & 62.3      & 52.5       & -      & -  \\
J0809+5218 & 0.138 & 183.8 & 187.05 & 89.1      & 85.7       & 24.49  & 2FGL J0809.8+5218 \\
J0810+4911 & 0.115 & 10.8  & 10.91  & $\leq$0.6 & $\leq$1.8  & -      & - \\
J0847+1133 & 0.199 & 33.0  & 33.66  & 6$^{*}$ & $\leq$3.0    & 5.27   & 2FGL J0847.2+1134 \\
J0850+3455 & 0.145 & 34.5  & 30.86  & 28.6      & 27.1       & -      & - \\
J0903+4055 & 0.188 & 35.8  & 29.75  & 9.1       & 6.3        & -      & - \\
J0916+5238 & 0.190 & 139.0 & 108.93 & 29.5      & 16.8       & -      & - \\
J0930+4950 & 0.187 & 21.3  & 15.08  & 7.6       & 5.2        & -      & - \\
J1012+3932 & 0.171 & 19.0     & 20.11  & $\leq$0.9 & $\leq$1.5 & -    & - \\
J1022+5124 & 0.142 & 5.6   & 2.69   & $\leq$0.6 & $\leq$1.8  & -      & - \\
J1053+4929 & 0.140 & 65.5  & 62.61  & 33.4      & 23.2       & 9.15   & 2FGL J0847.2+1134 \\
J1058+5628 & 0.143 & 229.5 & 219.45 & 158.6     & 128.6      & 47.88  & 2FGL J1058.6+5628  \\
J1120+4212 & 0.124 & 23.5  & 24.56  & 19.5      & 11.8       & 11.69  & 2FGL J1121.0+4211 \\
J1136+6737 & 0.136 & 45.2  &-       & 24.6      & 26.3       & 6.10   & 2FGL J1136.3+6736  \\
J1145-0340 & 0.167 & 18.7  & 10.48  & $\leq$1.2 & $\leq$0.9  & -      & - \\
J1156+4238 & 0.172 & 15.6  & 14.38  & 2.6       & 5.4        & -      & - \\
J1201-0007 & 0.165 & 69.5  & 67.57  & $\leq$15.3 & $\leq$14.7 & -     & - \\
J1201-0011 & 0.164 & 28.0  & 23.47  & $\leq$0.6 & $\leq$3    & -      & - \\
J1215+0732 & 0.136 & 138.8 & 81.80  & 36.6      & 17.7       & -      & -  \\
J1217+3007 & 0.130 & 587.8 & 466.45 & 262.3     & 217.4      & 54.92  & 2FGL J1217.8+3006  \\
J1221+0821 & 0.132 & 178.4 & 162.53 & 17.3      & $\leq$14.7 & -      & - \\
J1221+2813 & 0.102 & 739.0 & 921.26 & 337.2     & 288.6      & 55.39  & 2FGL J1221.4+2814 \\
J1221+3010 & 0.182 & 72.0  & 62.49  & 35.8      & 32.2       & 28.12  & 2FGL J1221.3+3010 \\
J1231+6414 & 0.163 & 59.3  &-       & 37.6      & $\leq$14.7 & -      & - \\
J1253+0326 & 0.066 & 107.4 & 79.21  & 36.5      & 27.1       &-       & - \\
J1257+2412 & 0.141 & 13.1  & 10.32  & $\leq$0.9 & $\leq$0.9  & -      & - \\
J1341+3959 & 0.172 & 85.6  & 57.85  & 6.3       & 5.8        &-       & - \\
J1419+5423 & 0.153 & 818.2 & 581.55 & 969.5     & 820.0      & 7.66   & 2FGL J1420.2+5422 \\ 
J1427+3908 & 0.165 & 7.0  & 4.79   & $\leq$0.6 & $\leq$0.9   & -      & - \\
J1427+5409 & 0.106 & 44.8  & 29.79  & $\leq$15.3 & 5.9       & -      & - \\
J1428+4240 & 0.129 & 57.5  & 42.72  & 23.2      & 24.2       & 7.54   & 2FGL J1428.6+4240 \\
J1436+5639 & 0.150 & 20.7  & 17.11  & 7.6       & 4.6        & 4.74   & 2FGL J1437.1+5640 \\ 
J1442+1200 & 0.163 & 68.0  & 69.97  & 26.6      & 21.8       & 4.55   & 2FGL J1442.7+1159 \\
J1510+3335 & 0.114 & 7.36   & 4.10   & $\leq$0.9 & $\leq$0.9 & -      & - \\ 
J1516+2918 & 0.130 & 136.5 & 73.96  & 30.6      & 28.9       & -      & - \\
J1534+3715 & 0.143 & 21.0  & 21.57  & $\leq$0.9 & $\leq$0.9  & 4.49   & 2FGL J1535.4+3720 \\ 
J1604+3345 & 0.177 & 7.1   & 5.84   & $\leq$0.6 & $\leq$1.2  & -      & - \\
J1647+2909 & 0.132 & 394.7 & 275.79 & 64.5      & 55.2       & -      & - \\
\hline
\end{tabular}
\tablefoot{Col. 1: source name; col. 2: redshift; col. 3: NVSS flux density; col. 4 FIRST flux density; col.s 5 and 6: 8 GHz and 15 GHz VLBA flux densities (N.O means that the source is not observed, $^{*}$ is as Table 4); col. 7: gamma-ray flux; col. 8: 2FGL name. }
\end{table*}

\longtab{4}{
\begin{longtable}{|c|crrrrrr|}
\caption{{\bf Model-fit results} for detected sources. Col. 1: source name, col. 2: the shape of the fitted components: elliptical (e), circular (c) and $\delta$; col.3: the flux density S  in mJy; col.s 4 and 5: polar coordinates (r) and ($\theta$) referred to the pointing center, with polar angle measured from the north through east; col. 6: the major axis $a$; col. 7: the ratio $b/a$ of minor $b$ and major $a$ axes of the fitted component; col. 8:  the position angle $\Phi$ of the major axis measured from north to east.} \label{tab_modelfit}\\
\hline
Name & CC & $S$ & r & $\sigma$& $a$ &$b/a$& $\phi$  \\
& & (mJy)& (mas) & (deg)& (mas) &  & (deg)\\
\hline
\endfirsthead
\caption{continued.}\\
\hline\hline
Name & CC & $S$ & r & $\sigma$& $a$ &$b/a$& $\phi$  \\
& & (mJy)& (mas) & (deg)& (mas) &  & (deg)\\
\hline
\hline
\endhead
\hline
\endfoot
\hline
\hline
J0754+3910 (X)  & $\delta$ & 22.5 & 0.00  & 00.0 & 0.00  & 1.00  & 00.0  \\
                &  e       &  4.5 & 1.39  & 32.0 & 1.72  & 0.42  & -51.3 \\     
J0754+3910 (U)  & $\delta$ & 17.8 & 0.00  & 00.0 & 0.00  & 1.00  & 00.0  \\
\hline
J0809+3455 (X)  & $\delta$  & 38.3 & 0.00 & 00.0 & 0.00  & 1.00  & 00.0\\
                &  e        & 17.4 & 0.39 & 18.5 & 1.93  & 0.10  &-14.9\\
                &  e        & 6.6  & 5.32 & 26.4 & 10.12 & 0.05  &-37.6\\
J0809+3455 (U)  & $\delta$  & 38.2 & 0.00 & 00.0 & 0.00  & 1.00  & 00.0\\
                &  e        & 14.3 & 0.50 & 06.1 & 1.37  & 0.18  & -12.7\\
\hline
J0809+5218 (X)  & $\delta$ & 56.6  & 0.00 & 00.0  & 0.00  & 1.00  & 00.0\\
                & $\delta$ & 16.0  & 0.75 & -42.9 & 0.00  & 1.00  & 00.0\\
                &  e       & 16.6  & 2.65 & -24.0 &  5.35 & 0.06  &  -22.1   \\
J0809+5218 (U)  & $\delta$ & 47.6  & 0.00 & 00.0  & 0.00  & 1.00  & 00.0\\
                & e        & 33.1  & 0.91 & -57.7 & 2.26  & 0.32  & 55.1   \\
                & $\delta$ & 2.9   & 4.72 & -35.5 & 0.00  & 1.00  & 00.0  \\    
                & $\delta$ & 2.1   & 5.46 & -35.7 & 0.00  & 1.00  & 00.0  \\
\hline
J0850+3455 (X)  &  c  & 22.8  & 0.00 & 00.0  & 0.34 & 1.00   &-32.3 \\
                &  e  & 5.8   & 2.85 & 89.2  & 5.99 & 0.29   &-79.8 \\
J0850+3455 (U)  &  c  & 21.7  & 0.00 & 00.0  & 0.19 & 1.00   &-23.4 \\ 
                &  c  & 5.4   & 2.30 & 10.6  & 1.33 & 1.00   &-39.6 \\  
\hline
J0903+4055 (X)  & $\delta$ & 9.1   & 0.00 & 00.0 & 0.00 & 1.00   & 00.0  \\
J0903+4055 (U)  & $\delta$ & 6.3   & 0.00 & 00.0 & 0.00 & 1.00   & 00.0  \\
\hline
J0916+5238 (X)  & $\delta$ & 24.1  & 0.00 & 00.0  & 0.00 & 1.00   & 00.0\\
                & $\delta$ & 5.4   & 2.42 & -12.7 & 0.00 & 1.00   & 00.0\\
J0916+5238 (U)  & $\delta$ & 16.8  & 0.00 & 00.0  & 0.00 & 1.00   & 00.0\\
\hline
J0930+4950 (X)  & $\delta$ & 7.6   & 0.00 & 00.0 & 0.00 & 1.00   & 00.0\\
J0930+4950 (U)  & $\delta$ & 5.2   & 0.00 & 00.0 & 0.00 & 1.00   & 00.0\\ 
\hline
J1053+4929(X)   & $\delta$ & 20.1  & 0.00 & 00.0 & 0.00 & 1.00   & 00.0 \\
                &  e  & 13.3  & 0.50 & 71.9 & 2.57 & 0.18   & -63.4\\
J1053+4929(U)   & $\delta$ & 23.2  & 0.00 & 00.0 & 0.00 & 1.00   & 00.0\\
\hline
J1058+5628 (X)  & $\delta$ & 87.6  & 0.00 & 00.0  & 0.00 & 1.00   & 00.0\\
                &  e       & 57.0  & 0.41 & -71.3 & 0.93 & 0.59   & -65.4 \\
                &  e       & 14.4  & 2.72 & 75.7  & 6.05 & 0.28   & 89.3\\
J1058+5628 (U)  & $\delta$ & 85.6  & 0.00 & 00.0  & 0.00 & 1.00   & 00.0\\
                &  e       & 43.0  & 0.53 & -81.0 & 1.00 & 0.48   &-60.6\\
\hline
J1120+4212 (X)  & $\delta$ & 14.8  & 0.00 & 00.0  & 0.00 & 1.00   & 00.0\\
                &  e       & 4.7   & 1.08 & -26.3 & 1.92 & 0.75   & -7.0\\
J1120+4212 (U)  & $\delta$ & 7.6   & 0.00 & 00.0  & 0.00 & 1.00   & 00.0 \\
                & $\delta$ & 4.2   & 0.83 & -0.7  & 0.00 & 1.00   & 00.0 \\
\hline
J1136+6737 (X)  &  c  & 24.6  & 0.00 & 00.0 & 0.17 & 1.00   & -49.1\\
J1136+6737 (U)  &  c  & 16.7  & 0.00 & 00.0 & 0.24 & 1.00   & -49.1\\
                &  c  & 9.6   & 1.05 & 27.1 & 0.38 & 1.00   & -40.5\\
\hline
J1156+4238 (X)  & $\delta$ & 2.6   & 0.00 & 00.0 & 0.00 & 1.00   & 00.0 \\
J1156+4238 (U)  &  e       & 5.4   & 0.00 & 00.0 & 2.42 & 0.56   & 42.8 \\        
\hline
J1215+0732 (X)  & $\delta$ & 21.3  & 0.00 & 00.0 & 0.00 & 1.00   & 00.0\\
                &  e       & 15.3  & 0.44 &-85.7 & 2.00 & 0.07   &-85.0\\
J1215+0732 (U)  &  e       & 17.7  & 0.00 & 00.0 & 0.53 & 0.43   &-84.4\\
\hline
J1217+3007 (X)  &  e       & 230.5 & 0.00 & 00.0 & 0.45 & 0.22   & -30.7\\
                &  c       & 15.2  & 2.32 & 49.1 & 0.29 & 1.00   & 47.5\\
                &  e       & 16.6  & 6.30 & 52.2 & 7.18 & 0.9    & 5.0\\
J1217+3007 (U) & $ \delta$ & 169.5 & 0.00 & 00.0 & 0.00 & 1.00   & 0.0\\
               &  c        & 33.9  & 0.35 &-6.5  & 0.11 & 1.00   & 48.0\\
               &  e        & 14.0  & 1.22 &-8.1  & 0.82 & 0.35   & -25.6\\ 
\hline
J1221+0821 (X) &  e   & 17.3  & 0.00 & 00.0 & 1.51 & 0.20   &-5.8\\
\hline
J1221+2813 (X) &$ \delta$  & 178.0 & 0.00 & 00.0 & 0.0  & 1.00   & 0.0\\
               &  c        & 35.8  & 0.45 & 47.0 & 0.23 & 1.00   & 24.0\\
               &  e        & 123.4 & 5.96 & 52.5 & 11.92& 0.30   &-53.9 \\
J1221+2813 (U) &  c        & 178.3 & 0.00 & 00.0 & 0.14 & 1.00   & -41.6\\
               &  c        & 29.8  & 0.46 &-65.3 & 0.22 & 1.00   & -4.8\\  
               &  e        & 80.5  & 6.38 &-65.4 & 10.77&  0.17  & -61.5\\       
\hline
J1221+3010 (X) & $\delta$  & 18.9  & 0.00 & 00.0 & 0.00 & 1.00   & 00.0 \\
               &  e        & 16.9  & 0.28 & 87.3 & 1.02 & 0.60   & 81.2 \\ 
J1221+3010 (U) &  e        & 32.2  & 0.00 & 00.0 & 0.49 & 0.60   &-33.7 \\
\hline
J1231+6414 (X) & $\delta$  & 26.1  & 0.00 & 00.0 & 0.00 & 1.00    & 00.0\\
               & $\delta$  & 4.0   & 0.63 & 00.5 & 0.00 & 1.00    & 00.0\\
               &  e        & 7.6   & 3.03 & 00.6 & 4.55 & 0.40    & -58.7\\
\hline
J1253+0326(X)  & $\delta$  & 29.5  & 0.00 & 00.0 & 0.00 & 1.00     & 0.0  \\
               & $\delta$  & 7.0   & 0.67 & 07.1 & 0.00 & 1.00     & 00.0\\
J1253+0326(U)  & $\delta$  & 27.1  & 0.00 & 00.0 & 0.00 & 1.00     & 00.0\\
\hline
J1341+3959 (X) & $\delta$  & 6.3   & 0.00 & 00.0 & 0.0  & 1.00     & 0.0  \\
J1341+3959 (U) & $\delta$  & 5.8   & 0.00 & 00.0 & 0.0  & 1.00     & 0.0  \\
\hline
J1419+5423 (X) &  c        & 716.3 & 0.00 & 00.0 & 0.16 & 1.00     & -55.0\\
               &  c        & 134.4 & 5.03 & 55.1 & 0.35 & 1.00     & 131.6\\
               &  e        & 94.8  & 0.88 & 53.2 & 2.65 & 0.69     & -63.7\\                              
               &  c        & 18.3  & 14.42& 53.4 & 5.26 & 1.00     & -62.6\\
               & $\delta$  & 5.7   & 30.61& 59.1 & 0.00 & 1.00     & 0.0\\
J1419+5423 (U) &  c        & 434.9 & 0.00 & 00.0 & 0.09 & 1.00     & -34.5\\
               &  c        & 290.5 & 0.25 & 39.6 & 0.24 & 1.00     & -62.4\\
               &  c        & 49.3  & 1.34 & 45.8 & 0.48 & 1.00     & 43.3\\
               &  c        & 45.3  & 5.07 & 51.6 & 1.85 & 1.00     & -72.6\\
\hline
J1427+5409 (U) & $\delta$  & 5.9   & 0.00 & 00.0 & 0.00 & 1.00     & 00.0\\
\hline
J1428+4240(X)  &  c        & 23.2  & 0.00 & 00.0 & 0.16  & 1.00     &-44.3\\
J1428+4240 (U) & $\delta$  & 16.7  & 0.00 & 00.0  & 0.00  & 1.00     & 0.0  \\
               & $\delta$  & 7.5   & 0.53 & -16.7 & 0.00  & 1.00     & 0.0  \\
\hline
J1436+5639 (X) &  e   & 7.6   & 0.00 & 00.0 & 2.21 & 0.10     & 51.6  \\
J1436+5639 (U) &  c   & 4.6   & 0.00 & 00.0 & 0.44 & 1.00     & -31.2\\
\hline
J1442+1200 (X) & $\delta$ & 22.4   & 0.00 & 00.0 & 0.00 & 1.00     & 00.0\\
               & $\delta$ & 4.2    & 0.38 & -60.6& 0.00 & 1.00     & 00.0   \\
J1442+1200 (U) & $\delta$ & 21.8   & 0.00 & 00.0 & 0.00 & 1.00     & 00.0 \\
\hline
J1516+2918 (X) & $\delta$ & 28.4   & 0.00 & 00.0  & 0.00 & 1.00     & 00.0\\
               & $\delta$ & 2.2    & 2.76 & -18.8 & 0.00 & 1.00     & 00.0\\
J1516+2918 (U )&  e       & 28.9   & 0.00 & 00.0  & 0.66 & 0.23     & -14.1\\
\hline
J1647+2909 (X) & $\delta$ & 49.2   & 0.00 & 00.0 & 0.00 & 1.00     & 00.0\\
               & $\delta$ & 7.4    & 1.49 & 25.7 & 0.00 & 1.00     & 00.0\\
               &  e       & 7.9    & 5.42 & 26.8 & 6.87 & 0.12     & -18.9\\
J1647+2909 (X) & $\delta$ & 37.2   & 0.00 & 00.0 & 0.00 & 1.00     & 00.0\\
               & $\delta$ & 10.8   & 0.46 & 16.0 & 0.00 & 1.00     & 00.0\\
               &  e       & 7.2    & 1.60 & 23.5 & 1.75 & 0.18     & -19.5\\
\hline
\end{longtable}}

\twocolumn

\appendix

\subsection{{\bf Notes on single sources.}} \label{app}

\begin{itemize}

\item {\bf J0751+1730.} This source is unresolved at kiloparsec-scale with a total flux density of 
$\sim$ 10 mJy in NVSS maps. 
From VLA data archive (AB878) in B configuration, 
at 8 GHz, we obtained a slightly resolved image ($\sim$ 1.4\arcsec with a 
HPBW = 0.7\arcsec) which implies $\alpha_{1.4}^{8.4} \sim$0.9. 
It is undetected in our VLBA images but with an upper limit higher than NVSS
flux density (for technical issues it was not observed in phase reference 
mode), for this reason we classify this source among the U sources, even if
the steep spectrum from VLA data suggests the presence of a lobe dominated
structure.

\item {\bf J0751+2913.} This BL Lac is unresolved in FIRST images with
  flux density of $\sim$ 9 mJy.  Not observed for technical reasons at
  8 GHz, it is not detected at 15 GHz above 1.8 mJy beam$^{-1}$.  Because of
  the low flux density at 1.4 GHz we put it among undetermined
  objects.

\item {\bf J0753+2921.} This source is unresolved in FIRST images (flux density = 4.5 mJy), and is not 
detected in archival VLA data (AE110) at 5 GHz (HPBW = 5 \arcsec) suggesting 
a steep spectral index and a resolved structure, in agreement with the non
detection above 0.9 mJy beam$^{-1}$ and 1.2 mJy beam$^{-1}$ at 8 and 15 GHz, 
respectively. We classify it as an U object because of
the low flux density.

\item {\bf J0754+3910.} This source shows a faint one-sided structure
  in NVSS images, but it appears unresolved in FIRST. In our 8 GHz
  VLBA map, it shows a core-jet morphology with a total flux density
  of $\sim$ 27.0 mJy. At 15 GHz, it appears point-like with a total
  flux density of 17.8 mJy.  Bourda et al. (2010) found a VLBI flux
  density = 31 mJy (2 GHz) and 26 mJy (8 GHz).  No gamma-ray
  emission is found associated to this target. We classify this
  source as DD.

\item {\bf J0809+3455.} In the FIRST image, this source (also known as
  B2 0806+35) reveals a two-sided large-scale emission, and a nuclear
  component with a bright one-sided jet to the south. The NVSS image
  shows a northern extension (more than 4\arcsec in size) confirming
  the asymmetric brightness of the kiloparsec-scale source.  In our
  VLBA image at 8 GHz, we detect a one-sided structure in agreement
  with the FIRST bright jet and the VLBA map at 5 GHz by Bondi et
  al. (2004). The VLBA total flux density is 62.3 mJy and 52.5 mJy at
  8 and 15 GHz respectively. The core has a flat spectrum and the jet
  flux density is $\sim$ 24 mJy and 14.3 mJy at lower and higher
  frequency respectively.  No gamma-ray emission is found associated
  to this target. We classify this BL Lac as DD.

\item {\bf J0809+5218.} This BL Lac shows a point-like structure in
  NVSS (total flux density $\sim$ 183.8 mJy) and FIRST images, as well
  with the VLA in A configuration (Giroletti et al. 2004). Our VLBA
  maps display, at both frequencies, a collimated jet oriented at NE,
  but moving after a few mas in N direction, with a large PA change,
  moving to the same PA visible in 5 GHz VLBA images (Giroletti et al
  2004b; MOJAVE Survey, Lister et al. 2009a, 2009b), and VLA images (Helmboldt
  et al. 2007).  Total flux densities are 89.1 $\sim$ mJy and $\sim$
  85.7 mJy at 8 and 15 GHz, respectively.  This BL Lac shows gamma-ray
  emission (see Table \ref{tab_multi} ).  We classify it as DD.

\item {\bf J0810+4911.} In NVSS and FIRST images this source appears
  symmetrically extended in PA $\sim$ 135 deg, with a flux density
  $\sim$ 10 mJy.  The source is undetected above $\sim$0.6 mJy beam$^{-1}$
  and 1.8 mJy beam$^{-1}$ at 8 and 15 GHz respectively. SC is $\leq$0.06.  No
  counterpart in the 2LAC is found.  We classify this BL Lac as LD.

\item {\bf J0847+1133.} This source is unresolved on kiloparsec-scale
  with flux density $\sim$ 33 mJy.  In our VLBA images, due to observational problems, it is undetected at
  both frequencies, with an upper limit of $\sim$15.3 mJy beam$^{-1}$ at 8
  GHz and 3 mJy beam$^{-1}$ at 15 GHz.  Bourda et al. (2010) revealed a VLBI flux density of 6
  mJy in X band. We used this measurement to discuss properties of
  this source.  This source is a LAT BL Lac showing a faint gamma-ray
  emission (Table \ref{tab_multi}). We classify this BL Lac as
  LD.

\item {\bf J0850+3455.} At kiloparsec-scales, this source is unresolved
  with a total flux density of 34.5 mJy in NVSS images. In our VLBA
  data, it shows a resolved morphology at both frequencies with total
  flux density of 28.6 and 27.1 mJy at 8 and 15 GHz respectively, in
  agreement with measurements of Bourda et al. (2010).  No gamma-ray
  emission is found associated to this object. We classify this BL
  Lac as DD.

\item {\bf J0903+4055.} This source is unresolved on kiloparsec-scales
  with a flux density of $\sim$ 35.8 mJy in the NVSS image. In our VLBA
  data, it shows a point-like structure with a flux density of $\sim$
  9.1 mJy at 8 GHz and 6.3 mJy at 15 GHz.  No gamma-ray emission is
  associated to this object. We classify it as DD.

\item {\bf J0916+5238.} On kiloparsec-scales, this BL Lac shows a
  core-jet structure with total flux density $\sim$140 mJy in NVSS
  data and some evidences of extended emission. In our 8 GHz VLBA
  maps, it has a one-sided morphology: the jet is elongated to the
  north with a flux density of $\sim$ 5.4 mJy while the
  central unresolved component shows a $\sim$ 24.1 mJy flux
  density. At 15 GHz, only the core is revealed with a flux density of
  16.8 mJy.  No gamma-ray counterpart is found in the 2 LAC for this
  source. We classify it as DD.

\item {\bf J0930+4950.} This source appears clearly point-like both at
  arcsecond scale with a NVSS total flux density of $\sim$21.3 mJy and
  in VLBA images (7.6 mJy at 8 GHz). No gamma-ray emission is found
  associated to this target. We classify it as DD.

\item {\bf J1012+3932.} On kiloparsec-scales, this object is
  point-like with NVSS total flux density of $\sim$ 20 mJy.  In our
  VLBA images, it is not detected above $\sim$0.9 mJy beam$^{-1}$ and 1.5
  mJy beam$^{-1}$ at 8 and 15 GHz, respectively. SC is $\leq$0.05. No
  gamma-ray counterpart is found in the 2LAC. We classify this
  target as LD.

\item {\bf J1022+5124.}  This target shows a kiloparsec-scale
  point-like structure with a low flux density in the NVSS image of $\sim$
  5.6 mJy. On parsec scale, it is not detected above $\sim$ 0.6 mJy beam$^{-1}$ and 1.8 mJy beam$^{-1}$ at 8 and 15 GHz
  respectively. No gamma-ray emission is found associated to this BL
  Lac.  We classify it as U.

\item {\bf J1053+4929.} This BL Lac is unresolved on kiloparsec-scales
  with a total flux density of $\sim$65.5 mJy in NVSS data. In our
  VLBA maps, it is one-sided at 8 GHz and point-like at 15 GHz.
  Its $\alpha_{core}$ is $\sim$-0.23 and it has an SC of
  $\sim$0.51. No counterpart in the 2LAC catalog is found. We
  classify this target as DD.

\item {\bf J1058+5628.} This BL Lac object has been studied by Nesci
  et al. (2010) in the B band: it is a quasi-periodic BL Lac, with an
  approximate period of 2590 days. In the radio band, it shows a
  kiloparsec unresolved structure with a total flux density in NVSS
  image of $\sim$ 230 mJy. At mas scale, this object has a one-sided
  morphology at both frequencies to the western direction with 8 GHz
  core flux density of $\sim86.7$ mJy, $\alpha_{core}\sim$ 0.03, 8 GHz
  total flux density of $\sim$ 158.6 mJy and $\alpha_{tot}\sim$0.33.
  The jet direction is in agreement with that found by the MOJAVE
  survey (Lister et al. 2009a, 2009b). SC is $\sim$0.70.  This target has a
  counterpart in the 2LAC (Table \ref{tab_multi}). We classify
  it as DD.

\item {\bf J1120+4212.} This source has a point-like morphology on
  arcsecond scale with a NVSS total flux density of $\sim$ 23.5
  mJy. Our VLBA maps revealed a one-sided structure at both
  frequencies, with a total 8 GHz flux density of $\sim$ 14.8 mJy and
  8 GHz jet flux density of $\sim$4.7 mJy. The $\alpha_{tot}$ is
  $\sim$ 0.54, while SC is $\sim$0.83. Moreover, it is a LAT BL Lac
  (Table \ref{tab_multi}). We classify it as DD.

\item {\bf J1136+6737.} This BL Lac has a kiloparsec-scale point-like
  structure, and is also point-like in our 8 GHz VLBA data with an SC
  $\sim$ 0.55.  At 15 GHz, it is resolved into a one-sided jet
  morphology with $\alpha_{core}\sim$0.29, $\alpha_{tot}\sim$ 0.25, and NE
  jet flux density of $\sim$ 9.6 mJy.  Gamma-ray emission is found
  associated with this object (Table \ref{tab_multi}). We
  classify it as DD.

\item {\bf J1145-0340.} This object is a part of a galaxy cluster
  (Piranomonte et al. 2007). On arcsecond scales, it appears
  unresolved with a total flux density of $\sim$ 18.7 mJy in NVSS
  data. In our VLBA images, the source is not detected above $\sim$1.2 mJy beam$^{-1}$
  and $\sim$0.9 mJy beam$^{-1}$ at 8 and 15 GHz, respectively.  SC
  is $\leq$0.06. No gamma-ray emission is found associated to this
  target in the 2LAC. We classify it as LD.

\item {\bf J1156+4238.} On arcsecond scales, this object is unresolved
  with a total flux density of $\sim$ 15.6 mJy in the NVSS image. In
  our VLBA data, it shows a point-like structure with a total flux
  density of $\sim$ 2.6 mJy at 8 GHZ and $\alpha_{core, tot}\sim$
  -1.16. We measure SC for this BL Lac to be $\sim$0.17. No gamma-ray
  emission is found associated to this source. We classify it as LD.

\item {\bf J1201-0007.} At kiloparsec-scales, this source is
  unresolved with a total flux density of 69.5 mJy in the NVSS
  image. The FIRST image shows that this source is resolved into a
  double structure in N-S direction with flux density peaks of 8.7 mJy
  Northern peak) and 17.5 mJy (Southern peak).  In our VLBA data, it
  is undetected at both frequencies above 15.3 mJy beam$^{-1}$ at 8 GHz and
  14.7 mJy beam$^{-1}$ at 15 GHz in agreement with the resolved arcsecond
  structure. No gamma-ray emission is found associated with this
  target.  We classify this BL Lac as LD.

\item {\bf J1201-0011.} On kiloparsec-scales, this source is unresolved
  with a flux density of 28 mJy in the NVSS image. In our VLBA data, it
  is undetected at both frequencies above $\geq$ 0.6 mJy beam$^{-1}$ at 8 GHz
  and $\geq$ 3  mJy beam$^{-1}$ at 15 GHz. We measure SC to be $\leq$0.22. No gamma-ray
  emission is found associated to this target. We classify this BL
  Lac as LD.

\item {\bf J1215+0732.} On kiloparsec-scales, a diffuse halo is
  present both in NVSS and FIRST images where a total flux density of
  $\sim$138.8 mJy and $\sim$81.8 mJy is revealed in NVSS and FIRST
  respectively. On parsec-scales, this object shows at 8 GHz a jet
  extending to the east (in agreement with the observations of
  Giroletti et al. 2004b, Rector et al. 2003 at 5 GHz) with a total
  flux density of $\sim$36.6 mJy. At 15 GHz, the source is point-like
  with $\alpha_{tot}\sim$0.9.  SC is measured to be $\sim$0.26. No
  gamma-ray emission is found associated to this BL Lac. We classify
  it as DD.

\item {\bf J1217+3007.}  This object is especially variable in the
  optical and polarized in optical and radio bands (Kollgaard et
  al. 1992, Fan \& Lin 1999, Giroletti et al. 2004). NVSS data reveal
  a point-like structure with a total flux density of $\sim$588
  mJy. At 8 GHz, the parsec-scale jet is formed by a more collimated
  component connected to the central core by a separated, diffuse
  component starting at a distance of about 7 mas from the core and
  extending for 10 mas south-east. This extended emission is also prominent in
  the 5 GHz VLBA image obtained by Bondi et al. (2004). At 15 GHz,
  the VLBA map shows mainly the first jet component ($\sim$ 48 mJy),
  even if an hint of the same diffuse structure can also be noted here
  at a distance of $\sim$ 14 mas from the nucleus. The jet structure
  is in agreement with that observed in the MOJAVE and VIPS Surveys
  (Lister et al. 2009a, 2009b, Helmboldt et al. 2007) where the jet detection
  extends up to 40 mas from the core. The total flux density in our
  VLBA images is similar at the two frequencies ($\sim$262 mJy at 8
  GHz and $\sim$ 217 mJy at 15 GHz). However, the NVSS flux density, on the
  kiloparsec-scale, is more than twice the 8 GHz parsec-scale flux density,
  suggesting a spectrum steeper than $-$0.4.
  This target is also a LAT BL Lac (Table \ref{tab_multi}). We
  classify it as DD.

\item {\bf J1221+0821.} This source has a point-like kiloparsec
  structure with a total flux density of $\sim$178 mJy from NVSS
  maps.  In our 8 GHz VLBA image, it appears unresolved with a total
  flux density of $\sim$17.3 mJy, while at 15 GHz it is not detected
  above 14.7 mJy beam$^{-1}$.  SC is $\sim$0.1. It is a non LAT BL Lac. We
  classify it as LD.

\item {\bf J1221+2813.} This source is known to be an extremely
  variable object, in infrared especially (Fan $\&$ Lin 1999), lying
  in the center of an elliptical galaxy (Weistrop et al. 1985). It was
  initially identified as the variable star W Comae. Arcsecond scale
  radio images (Kollgaard et al. 1992, Perlam et al. 1994) indicate
  a faint extended emission south-west of the core, which is also
  revealed in the FIRST image.  The total flux density is $\sim$
  739 mJy in NVSS images.  On parsec-scales, our VLBA observations show
  a long south-west jet-like feature at both frequencies. At 8 GHz it
  shows a straight jet extended more than 20 mas, where a large bend
  in PA (to the South) is present. This fainter resolved jet is
  marginally visible here but confirmed in lower resolution images
  (e.g. Helmboldt et al.  2007) while at 15 GHz it stops at about 10
  mas from the nucleus, which has a flux density $\sim$ 178.0 mJy and
  is flat spectrum. The jet structure contains numerous blobs of
  emission, especially at 15 GHz where it has a flux density of
  $\sim$37.9 mJy.  The same was found by Kellermann et al. (1998) in
  their 15 GHz VLBA observations and in the MOJAVE Survey (Lister et
  al. 2009a, 2009b).  Proper motions have been studied by Gabuzda et
  al. (1994) and Jorstad et al. (2001) with multi-epoch VLBI
  observations at 5 and 22 GHz respectively, which have shown
  significant, often non-radial, proper motions.  The SC for this target is
  $\sim$0.47. It has a gamma-ray counterpart in the 2LAC (Table
  \ref{tab_multi}). We classify this BL Lac as DD.

\item {\bf J1221+3010.}  The source is compact on kiloparsec-scales
  with a NVSS flux density of $\sim$ 72 mJy.  Previous VLBA
  observations of Giroletti et al. (2004b) revealed a 10 mas
  jet emerging at P.A. $\sim$ 90$\deg$. The total VLBA correlated flux density is
  57 mJy and dominates the total flux density of the source. Our VLBA
  maps at 8 GHz detect a one-sided structure aligned with the 5 GHz
  jet with $\alpha_{core}\sim$ 0.17 and jet flux density of 16.9
  mJy. In our 15 GHz map, the source is unresolved, even if a 10 mJy
  jet is detected in the MOJAVE Survey (Lister et al. 2009a, 2009b) in the
  same direction of the 8 GHz one.  Its SC is $\sim$ 0.5. This target
  is also a LAT BL Lac (Table 6). We classify it as DD.

\item {\bf J1231+6414.}
The kiloparsec-scale structure shows an almost unresolved core with a
total flux density of 59.3 mJy in the NVSS  (see also Perlman 1994). A
faint emission from a halo-like feature is also observed.  On
parsec-scales at 8 GHz this source shows a one-sided jet pointing
north-west.  A similar structure has been found by Giroletti et
al. (2004b) at 5 GHz.  The 8 GHz total flux density is $\sim$ 37.6
mJy. At 15 GHz it is undetected.  SC for this object is 0.63. No
association in the 2 LAC is found (Table 6).  We classify this BL Lac as DD.

\item {\bf J1253+0326.} This BL Lac shows an unresolved morphology
  both in NVSS and FIRST images with a total flux density of $\sim$107
  and $\sim$79.2 mJy, respectively. On parsec-scales, at 8 GHz it
  appears one-sided with a core flux density of $\sim$ 29.5 mJy and a
  jet flux density of $\sim$7 mJy, while at 15 GHz it has a point-like
  structure with a total flux density of $\sim$ 27.1 mJy. SC is
  $\sim$0.34. No gamma-ray emission is found associated to this
  target. We classify it as DD.

\item {\bf J1257+2412.} This extremely weak object is only marginally
  resolved with the VLA with a total flux density of $\sim$ 13.1 mJy
  in NVSS data. It was barely detected by the VLBA ($\sim$3 mJy core)
  at 5 GHz (Giroletti et al. 2004b). In our VLBA images at 8 and 15
  GHz, no emission is revealed above $\sim$ 0.9 mJy beam$^{-1}$ at both
  frequencies. SC is $\leq$0.07. No counterpart is found in the 2 LAC.  We
  classify this source as LD.

\item {\bf J1341+3959.} This source is unresolved both in NVSS and
  FIRST maps with a total flux density of $\sim$ 85.6 mJy and
  $\sim$57.9 mJy, respectively. At parsec-scales, it shows a
  point-like structure with a core flux density of $\sim$ 6.3 mJy at 8
  GHz and $\sim$ 5.8 mJy at 15 GHz. SC is $\sim$0.07.  It does not
  have a counterpart in the 2 LAC. We classify this BL Lac as LD.

\item {\bf J1419+5423.}  This is the most powerful object in the
  sample, and it is also the best studied, being in the BL Lac 1 Jy
  sample (Kuehr et al. 1981).  It is an extremely peculiar (see Sect. \ref{gamma}) and variable ,
  both on long and short timescales (Massaro et al. 2004) and at all
  frequencies. Variability in the optical and radio bands has shown
to be 
  correlated (Massaro et al. 2004), as has polarization (Gabuzda
  et al. 1996). Multi-epoch studies have shown that the flux density
  variability affects are related to the core component, rather than
  the extended south-east jet: for example, an increase of a factor 3
  was observed in the core flux density in the period 1999-2001
  (Massaro et al. 2004).  Its radio structure on kiloparsec-scales is
  point-like with a total flux density of $\sim$ 818 mJy in the NVSS. 
Our VLBA maps show that at 8 GHz the source has a $\sim$25
  mas jet with about 25\% of the total flux density, which is
  $\sim$970 mJy. At 15 GHz, the jet is much shorter, around 4 mas,
  and makes up 47\% of the total flux density of $\sim$820 mJy. The jet
  structure is aligned with that observed by the MOJAVE Survey (Lister
  et al. 2009a, 2009b).  It has a counterpart in the 2LAC (Table
  \ref{tab_multi}) . We classify this target as DD.

\item {\bf J1427+3908.}

On arcsecond scales, this source is unresolved with an NVSS total flux
density of $\sim$7 mJy. In our VLBA data, no radio emission is
revealed above $\sim$0.6 mJy beam$^{-1}$ and 0.9 mJy beam$^{-1}$ at 8 and 15 GHz
respectively. SC is $\leq$0.09. No counterpart is found in the
2LAC. We classify this object as U.

\item {\bf J1427+5409.} The source is unresolved at kiloparsec-scales
  with a total flux density of $\sim$44.8 mJy in NVSS data.  In our 8
  GHz VLBA image it is not detected above $\sim$ 15.3 mJy while in the
  phase referencing 15 GHz VLBA data it shows a point-like structure
  with a flux density of $\sim$ 5.9 mJy. No gamma-ray counterpart is
  found associated with this object. We classify it as LD.

\item {\bf J1428+4240.}  In the NVSS images the source is unresolved
  with a total flux density of $\sim$ 57.5 mJy.  From VLA observations
  with $\sim$2 \arcsec resolution, this BL Lac shows some evidence of
  a diffuse radio halo around the core and it has  relatively weak
  radio and optical polarization for a BL Lacertae object
  (Laurent-Muehleisen et al. 1993; Jannuzi er al. 1994;
  Giroletti et al. 2004b). Previous VLBI images did not find polarized
  emission (Kollgaard et al. 1994). In our VLBA map at 8 GHz it
  appears unresolved with a total flux density of $\sim$ 23.2 mJy,
  while at 15 GHz it shows a one-sided structure with a core flux
  density of $\sim$ 16.7 mJy and jet flux density of $\sim$ 7.5 mJy.
  Gamma-ray emission is also found associated to this object (Table \ref{tab_multi}). We classify it as DD.

\item {\bf J1436+5639} This BL Lac shows a point-like structure both
  at arcsecond and milliarcsecond-scales. In the NVSS it has a total flux
  density of $\sim$20.7 mJy, while in our 8 GHz VLBA data the
  unresolved central component has a flux density of 7.6 mJy and $\alpha_{tot}\sim$ 0.83.
  It is a LAT BL Lac (see Table \ref{tab_multi}). We classify this
  object as DD.

\item {\bf J1442+1200.} In the NVSS the source appears unresolved with
  a total flux density of $\sim$ 68 mJy.  VLA images obtained by
  Giovannini et al. (2004b) with subarcsecond resolution show a nuclear
  flat spectrum emission surrounded by a steep-spectrum halo
  structure $\sim$8\arcsec in size at 8.4 GHz moderately elongated in
  the southwest direction.  Previous VLBA images (Giroletti et
  al. 2004b) revealed only a faint (15 mJy) core, possibly extended to
  the west. In our 8 GHz VLBA maps, we confirm the jet extension to
  the west with a flux density of $\sim$ 4.2 mJy, while the core has a
  flux density of $\sim$22.4 mJy. At 15 GHz, this BL Lac is unresolved
  with a total flux density of $\sim$ 21.8 mJy. It has a 2 LAC
  counterpart (Table \ref{tab_multi}). We classify this target as
  DD.

\item {\bf J1510+3335.} This source has an unresolved kiloparsec-scale
  morphology with a total flux density of $\sim$7.4 mJy in the NVSS.
 In our VLBA maps it is not detected above $\sim$ 0.9 mJy at
  both frequency.  SC is $\leq$0.12. No gamma-ray emission is found
  for this BL Lac. We classify it as U.

\item {\bf J1516+2918.} At kiloparsec-scales, this source has
  a jet elongated in the north-west direction. The total flux density is
  $\sim$ 137 mJy in the NVSS map, and $\sim$74 mJy in FIRST data.
  In our VLBA images, it is one-sided at 8 GHz with a NE jet (flux
  density of $\sim$ 2.2 mJy) and total flux density of $\sim$ 30.6
  mJy.  At 15 GHz, it has a $\sim$ 28.9 mJy point-like structure. It
  is not a LAT BL Lac. We classify it as DD.

\item {\bf J1534+3715.} This source is unresolved on kiloparsec-scales
  with a total flux density of $\sim$21 mJy in NVSS images.  On
  parsec-scales, no radio emission is revealed above $\sim$0.9 mJy beam$^{-1}$
  at both frequencies.  SC is $\leq$ 0.04. A gamma-ray counterpart is
  found in the 2LAC (see Table \ref{tab_multi}).  We classify this BL
  lac as LD.  The non detection by the VLBA could be due to an unsuccessful
  phase referencing observation.  Deeper investigation and new VLBI
  observations will be necessary to investigate its nuclear
  properties, and to confirm the lack of a compact emission despite 
  the high frequency properties.

\item {\bf J1604+3345.} This source is unresolved with a total flux
  density of $\sim$ 7.1 mJy in NVSS data. In our
  VLBA maps, it is not detected above $\sim$0.6 mJy beam$^{-1}$ and 1.2 mJy beam$^{-1}$
   at 8 and 15 GHz, respectively. SC is $\leq$0.09. No
  gamma-ray counterpart is found associated to this BL Lac. We
  classify it as U.

\item {\bf J1647+2909.} On arcsecond scales, this BL Lac shows some
  diffuse halo-like structure around the central core component. The
  total flux density in the NVSS image is $\sim$ 395 mJy and 276 mJy in
  the FIRST image.  In our VLBA images, it shows a one-sided
  morphology in the north direction (aligned with that revealed by the
  VIPS survey, Helmboldt et al. 2007) with a total 8 GHz flux density
  of 64.5 mJy, $\alpha_{core}\sim$ 0.45, and SC $\sim$0.23. No
  gamma-ray counterpart is found associated to this target. We
  classify it as DD.

\end{itemize}

\end{document}